\documentclass[aps,reprint,onecolumn,superscriptaddress, 
               showkeys, showpacs, longbibliography, notitlepage]{revtex4-1}

\usepackage{graphicx}
\usepackage{upgreek}  %upright greek symbols
\usepackage{colordvi} %color in the text
\usepackage{textcomp} % \textmu
\usepackage{amsmath} %  \text
\usepackage{amssymb}
\usepackage{bbm}
\usepackage{afterpage}
\usepackage{versions}
\includeversion{arXiv}
\excludeversion{journal}

\newcommand{\bnabla}{\boldsymbol{\nabla}}
\newcommand{\bcdot}{\boldsymbol{\cdot}}
\newcommand{\rb}{\right)}
\newcommand{\lb}{\left(}

\newcommand{\diff}{\ensuremath{\;\mathrm d}}
\newcommand{\D}{\mbox{d}}
\newcommand{\totdiff}[2]{\ensuremath{\frac{\D#1}{\D#2}}}
\newcommand{\substdiff}[1]{\frac{\mathrm D #1}{\mathrm D t}}
\newcommand{\substdiffText}[1]{\mathrm D #1/\mathrm D t}
\newcommand{\mum}{\upmu \textrm{m}}
\newcommand{\mus}{\upmu \textrm{s}}
\newcommand{\tx}[1]{\text{#1}}
\newcommand{\rmax}{{R_\tx{max}}}

\begin{document}

\title{
 Jet formation from bubbles near a solid boundary in a compressible liquid. Numerical study of distance dependence} 

\author{Christiane Lechner}
\email[Corresponding author: ]{christiane.lechner@tuwien.ac.at}
\affiliation{\mbox{Drittes Physikalisches Institut, Universit\"at
G\"ottingen, Friedrich-Hund-Platz 1, 37077 G\"ottingen, Germany}}
\affiliation{\mbox{Institute of Fluid Mechanics and Heat Transfer, 
TU Wien, Getreidemarkt 9, 1060 Vienna, Austria}}

\author{Werner Lauterborn}
\author{Max Koch}
\author{Robert Mettin}

\affiliation{\mbox{Drittes Physikalisches Institut, Universit\"at
G\"ottingen, Friedrich-Hund-Platz 1, 37077 G\"ottingen, Germany}}

\date{\today}

\begin{abstract}
A small, spherical bubble of high internal pressure is inserted into water 
at constant ambient pressure as a model of a laser-induced bubble.  
Its subsequent dynamics near a flat solid boundary is studied in 
dependence on the distance of the bubble to the boundary by numerically 
solving the Navier-Stokes equations with the help of the open source software 
environment OpenFOAM. Implemented is the finite volume method for 
discretization of the equations of motion and the volume of fluid method 
for capturing the interface between the bubble interior and exterior. 
The bubble contains a small amount of non-condensable gas that is treated 
as an ideal gas. The liquid is water obeying the Tait-equation. 
Surface tension is included where necessary. 
The evolution of the bubble shape and a selection of pressure and 
velocity fields are given for normalized distances $D^* = D/R_{\rm max}$ 
between 0 and 3 ($D$ = initial distance of the bubble centre to the boundary, 
$R_{\rm max}$ = maximum radius the bubble 
would attain without any boundary). $R_{\rm max} = $ 500~$\upmu$m is chosen for 
the study. 
Normal axial jet formation ($\sim$100\,m\,s$^{-1}$) by axial flow focusing 
is found for $0.24 \le D^* \le 3$ and the 
change to a different type of axial jet formation ($\sim$1000\,m\,s$^{-1}$) 
by annular-liquid-flow collision for bubbles 
very near to the solid boundary ($0 \le D^* \le 0.2$). 
The transition region ($0.2 < D^* < 0.24$) is characterized 
by additional inbound and outbound annular jets. 
Remarkably, the inclusion of the viscosity of the water 
is decisive to get the fast jets. 
\end{abstract}

\maketitle

\section{Introduction}\label{sec:intro} 
It was the erosion problem on ship propellers that, more than hundred years ago, has drawn attention to bubbles in liquids
\citep{Brennen-1995}. It took some time, however, until the first ideas came up
\citep{Kornfeld-1944} of how bubbles could destroy obviously any material. Experiments with bubbles near boundaries had to be conceived and a theoretical description  of the respective bubble motion had to be developed
\citep{Rayleigh-1917, Naude-1961, Benjamin-1966}. Progress for experiments 
came from high-speed instrumentation, in particular high-speed photography and holography 
(see the reviews by \citet{Lauterborn-1985,Lauterborn-1986} and the book chapter by
\citet{Lauterborn-2018a}). 
Progress for theory came from (at the time) high-speed 
numerical simulations in combination with the development of numerical codes
\citep{Plesset-1971}, later in particular from the boundary integral method 
\citep{Blake-1986, Blake-1987b, Best-1992, Best-1993, Zhang-1993, Wang-2014}.
Nowadays, a large variety of codes is available for bubble dynamics and jet 
formation  near boundaries (see, e.g., \citep{Koch-2016} and the references 
therein, and
\citep{Tiwari-2015, Denner-2018, Beig-2018,  
Fuster-2018} for recent developments).

Besides high-speed instrumentation and computation suitable bubbles had to be provided for investigation. To prepare single bubbles or bubble configurations for experiments needs some efforts. For single, at generation highly spherical bubbles near boundaries (first mainly solid boundaries) laser-induced bubbles
\citep{Lauterborn-1974, Lauterborn-1976, Lauterborn-1979, Lauterborn-1982}
became the method of choice. 
Thus it were laser-induced bubble shapes that were first compared by 
\citet{Lauterborn-1975} 
with the respective shapes from numerical simulations of jet-forming bubbles near a flat solid boundary done by 
\citet{Plesset-1971}.
A reasonable agreement was obtained concerning the development of the liquid jet towards the solid boundary by involution of the top of the bubble and with respect to jet velocities. The experimental work contains some more examples of jet formation and jet velocities of bubbles near a solid boundary 
beyond the numerical calculations,
feasible then, that stop at jet impact onto the opposite bubble wall. 

In the sequel, the laser-bubble technique has been employed to study various hitherto unreachable aspects of bubble dynamics. 
\citet{Vogel-1988a, Vogel-1988b}
studied shock wave emission and the flow field that leads to jet formation. 
Moreover, \citet*{Vogel-1989}
presented images of bubbles in top and side view with jet formation and torus-bubble collapse for different normalized distances $D^*$ of the bubble to a solid boundary. $D^*$ is defined as  
\begin{equation} 
D^* = \frac{D}{R_{\rm max}} \ ,
\end{equation}
where $D$ is the  initial distance of the bubble centre to the boundary 
and $R_{\rm max}$ the maximum radius the bubble would attain without any boundary. 
Further examples of the use of laser-induced bubbles were given by 
\citet{Tomita-1990}, \citet*{Testud-1990}, \citet{Ward-1991} and \citet*{Ohl-1995}.
\citet{Tomita-1990}
conducted a series of experiments with laser-induced bubbles in different configurations from bubbles near solid boundaries (plane, concave, convex), near a free surface and an elastic boundary to two-bubble interaction in free liquid and near a solid boundary. 
In all these studies, bubble shapes are 
given to build a view on how jets develop and proceed under various circumstances.  
\citet*{Testud-1990}
presented streak images from single bubble oscillation with shock-wave radiation as well as from two-bubble interaction with attraction and repulsion according to the relative phases of oscillation. 
\citet{Ward-1991}
measured the pressure distribution around a laser-induced bubble ($R_{\rm max} = 1.14$~mm) collapsing near a solid boundary ($D^* = 1.7$).
They experimentally confirmed the pressure hump that develops on the distal side of the bubble from the boundary as predicted by 
\citet*{Blake-1986}
for an empty bubble in an incompressible liquid.  
\citet*{Ohl-1995}
extended the time resolution to 20 million frames per second (50~ns only between consecutive frames) and thus could resolve the bubble collapse phase 
better. They gave an example of torus-bubble splitting at collapse of a bubble near a solid boundary, a phenomenon 
also observed before upon first \citep{Tomita-1990}, 
second and third collapse \citep{Lauterborn-1974}  
and later in the interaction of two laser-induced bubbles 
\citetext{\citealp{Lauterborn-1984}; see also \citealp{Han-2015}}.

As to the erosion problem from bubbles, 
\citet*{Isselin-1998} and \citet{Philipp-1998}
made a thorough experimental investigation on single bubble dynamics in dependence on the normalized distance to a solid boundary for evaluating the damage potential of bubbles collapsing nearby. Similar studies had been done before with spark-induced bubbles 
\citep{Tomita-1986}.
It seems that only the direct contact of a bubble with a nearby solid boundary leads to observable damage, either by jet impact or (torus) bubble collapse. 
However, the knowledge still was not sufficient for a detailed experimental or theoretical picture of bubble collapse and the induced erosion.
Therefore, research proceeded with both experiments and numerics. 
\citet{Tong-1999}
experimentally and numerically studied a laser-induced bubble at $D^* =$ 1.2 and 0.92 and described a special form of splash, the Blake splash 
\citep*{Blake-1998}.
It occurs for bubbles  near to a solid boundary ($D^* \sim 1$) by the collision of the outwards flow underneath the bubble (between bubble and boundary) from the jet that has hit the lower (opposite) bubble wall and the inflow of liquid along the solid surface from the collapsing bubble. 
When both flows meet underneath the bubble an upwards flow leads to an 
indentation of the lower bubble wall and to a surface wave running up 
inside the bubble along the outer bubble wall. 
Splitting of the bubble may occur, when the splash gets as extended as to reach the opposite bubble wall. More results on the Blake splash can be found in 
\citet{Brujan-2002} and \citet{Lauterborn-2018b}.

 Whereas high-speed photography is best in delivering bubble shapes (including jets) for comparison with theory 
\cite{Benjamin-1966, Lauterborn-1975, Vogel-1989, Tomita-1990, Ohl-1995,
Lindau-2003, Koukouvinis-2016a, Koukouvinis-2016b, Supponen-2017,
Pishchalnikov-2019},
numerical simulations are able to also provide, for instance, pressure and velocity fields besides the bubble shapes 
\citep[e.g.,][]{Blake-1998, Tong-1999}. 
There have been and are, however, efforts to also visualize the flow field around bubbles experimentally 
\citetext{see the early review by \citet{Lauterborn-1984};
further \cite{Vogel-1985, Vogel-1988b, Vogel-1989}}.
More recent works are from 
\citet{Kroeninger-2010} and \citet{Reuter-2017a}. 
Particle tracking velocimetry  was used by \citet{Kroeninger-2010}
to measure the flow field in the vicinity of a collapsing laser-induced bubble including jet formation near a solid boundary. In this method, the liquid is seeded with micro-particles, and velocity vectors are determined from their displacement between two frames of a high-speed photographic series. Reasonable agreement with simulations done with the boundary integral method was obtained. A novel digital presentation method for flow fields from experiments was introduced by 
\citet{Reuter-2017a}
that they call Lagrangian ink mapping. In this visualization method fluid elements are (digitally) colour coded and their displacement with time from the experimental measurements is followed.
The method is similar to experiments with analogue monochrome ink, but in the digital version with more colours and with extended possibilities. This visualization technique may be transferred to numerical results to better present the motion of flows for human perception.

As to the pressure in the liquid, the shock waves from bubble collapse were early measured to high precision, first for acoustic cavitation bubbles 
\citep{Kuttruff-1962,Kuttruff-1969},
then also for laser-induced bubbles 
\citep{Hentschel-1982, Vogel-1988a},
whereas numerical simulations were lacking behind. 
Upon bubble collapse, a sequence of shock waves may be emitted 
\citep{Lauterborn-1974}.
Several authors have looked into this topic in more detail 
\citep{Ward-1991, Ohl-1999, Shaw-2000, Lindau-2003, Supponen-2017}.
The following picture emerged for bubbles collapsing near a flat solid
boundary: 
A first pressure wave (the laser-induced shock wave not counted) 
is emitted, when the jet hits the opposite bubble wall. 
As the now toroidal bubble is still collapsing, at least a second shock wave 
(a torus shock wave) is emitted at bubble minimum. 
But the bubble may break up into several parts and then more shock waves 
are emitted. 
Recently, this picture has been augmented for $D^* \le 0.2$ by an additional shock wave resulting from the self-impact of an inbound annular jet, before the axial jet hits the opposite bubble wall 
\citep{Lechner-2019}.

It is only recently that the pressure waves from bubbles collapsing near solid boundaries 
could also be calculated and compared with experimental findings 
\cite{Johnsen-2008, Johnsen-2009, Muller-2010, 
Ochiai-2011, Lauer-2012,
Chahine-2014, Hsiao-2014, Koch-2016, Lechner-2017, Lauterborn-2018b,
Lechner-2019}.
In these works, very different bubble models and initial conditions are studied from (initially) spherical, 
cylindrical or planar bubbles 
\citep*[e.g.,][]{Muller-2010},
axisymmetric bubbles 
\citep[e.g.,][]{Ochiai-2011, Pishchalnikov-2019} 
and three-dimensional bubbles \citep{Lauer-2012, Beig-2018}
in a constant external pressure field to shock-induced bubble collapse near boundaries as in lithotripsy 
\citep{Johnsen-2008,Johnsen-2009}.
Studies that also calculate the influence on the solid surface of the shock waves emitted and of the jets are those from 
\citet{Chahine-2014} and \citet{Hsiao-2014}.
They present pit depths and widths for different materials from numerical simulations. 

The main results on bubble dynamics obtained with laser-induced bubbles (among others) up to 2010 can be found in the reviews by 
\citet{Lauterborn-1999} and \citet{Lauterborn-2010}.
A review on the shock waves encountered in connection with laser-induced bubbles has been given in a book chapter by 
\citet{Lauterborn-2013}.

The dynamics of bubbles near solid boundaries is dominated by the phenomenon of jet formation. Almost any of the already cited works contain information on jet formation from bubbles near solid surfaces. 
However, jet formation not only occurs with solid boundaries, but for a larger, in fact huge, class of bubbles. 
For instance, there exists a large body of studies on jet formation near boundaries other than solid ones. 
These lie outside the range of the present work. 
Just to mention the efforts to unite different jet-inducing mechanisms, including free and solid boundaries
\citep{Obreschkow-2011,Supponen-2016}, 
by the concept of the Kelvin impulse set forth by 
\citet{Benjamin-1966}
and developed further by Blake and coworkers 
\cite{Blake-1988, Blake-2015}.
Moreover, bubbles are used for cleaning of contaminated surfaces with acoustic waves 
\citep[see, e.g.,][]{Olaf-1957, Prabowo-2011, Reuter-2017b}. 
In this context, jets in sound fields 
were studied, again since long time 
\citep[see, e.g.,][]{Hund-1969, Crum-1979, Blake-1999, Rossello-2018}.  
Owing to the stochastic nature of bubble formation in acoustic fields, however, 
studies with single, acoustic bubbles are difficult. 
Laser-induced bubbles helped also in this case with their flexibility in generation and placing into the liquid 
\citep{Ohl-2006, Reuter-2016}.
Jet and shock-wave formation together with induced flows 
along the surface were singled out as the main actors. 
Temperature rise by jet impact and bubble collapse seems not to be an issue
\citep*{Beig-2018}. 
The present study can be considered a contribution to the problem with a model of laser-induced bubbles.

The intention of the present study is to explore the increased numerical possibilities available by the recent progress in software development to solve the Navier-Stokes equations together with computational speed to overcome prohibitive time limitations. Aim is a numerical code and a data base for the prediction of the collapse behaviour of bubbles near solid boundaries and for comparison with experimental data. Hopefully then, a consistent picture and understanding  emerges of the astounding destructive power of bubbles near boundaries. 

The work is organized as follows. 
The equations of motion and the equations of state 
 for the bubble contents and the liquid are given in Section \ref{sec:model}. 
 The numerical implementation of the equations is briefly sketched in Section \ref{sec:implementation}. 
The results are presented 
in Section \ref{sec:resultsdistdep} with bubble shapes, flow fields and pressure fields in dependence on the normalized distance $D^*$ in the range [0, 3]. 
A discussion of the results is given in Section \ref{sec:discussion} followed by the conclusions in Section \ref{sec:conclusions}.

\section{Bubble model}\label{sec:model}
A bubble model for a cold liquid (a liquid far from its 
boiling point, \citep[cf.][]{Brennen-1995}) with the following properties is used. 
The bubble contains a constant, small amount of non-condensable gas  
to comply with experiments 
\citep{Fujikawa-1980, Akhatov-2001, Akhatov-2002}.
The vapour pressure is small compared to the ambient pressure and is neglected.   
The liquid is Tait-compressible for inclusion of pressure waves up to weak shock waves 
\citep{Koch-2016}.
Thermodynamic effects and mass exchange through the bubble wall are neglected. 
%

%++++++++++++++++++++++++++++++++++++++++++++++++++++++++++++++++++++
\subsection{Equations of motion\label{subsec:eqmotion}}
 The equations of motion are formulated for a ``single fluid'', i.e., with one
density field $\rho(\boldsymbol x, t)$, one velocity field $\boldsymbol U(\boldsymbol x, t)$, 
and one pressure field $p(\boldsymbol x,t)$, satisfying the Navier-Stokes equation (\ref{eq:momentumEq}) 
and the continuity equation (\ref{eq:contEq}):
 \begin{eqnarray}
 \frac{\partial (\rho \, \boldsymbol U)}{\partial t} + \bnabla \bcdot (\rho \, \boldsymbol U
 \otimes \boldsymbol U) & = & - \bnabla p  + \rho \, \boldsymbol{g} + 
\bnabla \bcdot  \mathbbm{T} 
+ {\boldsymbol f}_\sigma \, , \label{eq:momentumEq} \\ \nonumber \\
     \frac{\partial \rho}{\partial t}  + 
          \bnabla \bcdot (\rho \, \boldsymbol U) & = & 0 \, .
        \label{eq:contEq} 
\end{eqnarray} 
$\bnabla$ denotes the gradient, $\bnabla \bcdot$ is the divergence, and $\otimes$
the tensorial product. $\boldsymbol{g}$ is the gravitational acceleration. 
$\mathbbm{T}$ is the viscous stress tensor of a Newtonian fluid:  
\begin{equation}
\mathbbm{T} =  \mu  \left( \bnabla \boldsymbol U + (\bnabla \boldsymbol U)^T - \frac{2}{3} 
 (\bnabla \bcdot \boldsymbol U) \, \mathbbm{I} \, \right), 
\label{eq:stressNewtonian}
\end{equation}
with the viscosity field $\mu(\boldsymbol x,t)$ 
and $\mathbbm{I}$ the unit tensor. 

Surface tension is treated as a force density field ${\boldsymbol
f}_\sigma(\boldsymbol x, t)$
\citep[see, e.g.,][]{Tryggvason-2001}. 
In the surface integral
\begin{equation} 
{\boldsymbol f}_\sigma(\boldsymbol x, t) = \int\limits_{S(t)} \sigma \, \kappa(\boldsymbol x')  \, {\boldsymbol n}(\boldsymbol x') 
\delta (\boldsymbol x  - \boldsymbol x') {\rm d}S', \label{eq:surfaceEq} 
\end{equation}
$\sigma$ represents 
the surface tension coefficient (taken as constant, $\sigma = 0.0725$\,N\,m$^{-1}$ for water), 
$\kappa$ is twice the mean curvature of the interface, 
and  ${\boldsymbol n}$ the unit vector normal to the interface, taken to  
point from the gas into the liquid. 
$\delta(\boldsymbol x - \boldsymbol x')$ denotes the Dirac delta in three dimensions,
with $\boldsymbol x' \in
S(t)$ a point on the interface
and $\boldsymbol x$ the point, at which the equation is evaluated.

The influence of gravity was found to be
negligible 
for the small bubbles of maximum radius $R_{\rm max}$ = 500\,$\upmu$m studied. 
Therefore gravity was omitted in the equations of motion. 
Surface tension is included, except for larger distances of the
bubble from the boundary, where it does not play a significant role. 
The viscosity of the liquid and of the gas is included, because it was found essential, in particular for bubbles in close vicinity to the solid boundary. 

In order to distinguish between  liquid ($l$) and gas ($g$), 
volume fraction fields
$\alpha_l(\boldsymbol x,t)$ and $\alpha_g(\boldsymbol x,t)$ are introduced with $\alpha_l = 1$ in the liquid
phase, $\alpha_l=0$ in the gas phase, and  $\alpha_g(\boldsymbol x,t)=1-\alpha_l(\boldsymbol x,t)$.
The position of the interface is then given implicitly by the transition of 
$\alpha_l$ from 1 to 0. 
The viscosity field $\mu(\boldsymbol x,t)$ can
be written as 
$\mu(\boldsymbol x,t) = \alpha_l(\boldsymbol x, t) \mu_l + \alpha_g (\boldsymbol x,t)\mu_g $ 
\citep[see, e.g.,][]{Gopala-2008}. 
The dynamic viscosities $\mu_l$ of the liquid and $\mu_g$ of the gas are taken to be constant 
($\mu_l = 1.002 \times 10^{-3}\,{\rm kg\ s^{-1}\,m^{-1}}$, $\mu_g = 1.7 \times 10^{-5}\,{\rm kg\ s^{-1}\,m^{-1}}$).  
The density field $\rho(\boldsymbol x,t)$ 
is given by 
$\rho(\boldsymbol x,t) = \alpha_l(\boldsymbol x, t) \rho_l(\boldsymbol x,t) + \alpha_g (\boldsymbol x,t) \rho_g(\boldsymbol x,t)$ 
with $\rho_l$ and $\rho_g$ the densities of the liquid and gas, respectively.
As there is no mass transfer between bubble interior (gas) and exterior (liquid), 
the respective phase-fraction density fields $\alpha_l\rho_l$ and $\alpha_g\rho_g$ 
separately obey the continuity equation:
\begin{equation}
\frac{\partial (\alpha_i \rho_i)}{\partial t}   + 
          \bnabla \bcdot (\alpha_i \rho_i \boldsymbol U)   = 0 \, , \qquad 
  { i=l,g} \, . 
        \label{eq:contEqlg}
\end{equation}

\subsection{Equations of state}\label{subsec:eos}
The equations of motion are closed by the equations of state for the gas and the liquid. 
For the gas in the bubble, the change of state is assumed to be adiabatic: 
\begin{equation}
\rho_g(p)  = \rho_{gn}{\lb\displaystyle \frac{p}{p_n} \rb^{1/\gamma_g}} ,
\label{eq:eosgas}
\end{equation} 
with $p_n$ and $\rho_{gn}$ the pressure and the density of the gas in the bubble at normal conditions, 
respectively, and 
$\gamma_g = 1.4$ the ratio of the specific heats of the gas (air).

For the liquid, the Tait equation of state for water is used 
\citep[see, e.g.,][]{Fujikawa-1980}:

\begin{equation}
\rho_l(p) = \rho_\infty\lb\frac{p+B}{p_\infty+B}\rb^{1/n_T},
\label{eq:tait}
\end{equation}
with $p_\infty$ the atmospheric pressure, $\rho_\infty$ the equilibrium
density, the Tait exponent $n_{\rm T} = 7.15$ and the Tait pressure $B = 305\,$MPa.

\section{Numerical implementation of the bubble model}\label{sec:implementation}

 OpenFOAM (Open Field Operation And Manipulation) has been selected to perform 
 the numerical calculations. 
 OpenFOAM 
\citep{Weller-1998}
is an open source computational fluid dynamics  
software 
based on the finite volume method (FVM) 
for discretization of the equations of motion. 
 The volume of fluid (VOF) method is used for capturing and locating the 
 gas--liquid interface 
\cite{Hirt-1981, Miller-2013, Klostermann-2013}.
 It is a robust method for following evolving, topologically complex interfaces 
\citep{Fuster-2009},
as encountered in the present case, for instance, 
 with jet formation and disintegration, bubble splitting and merging 
 upon torus-bubble collapse and rebound, and nanojet and droplet formation 
\citep{Lechner-2017}.
The pressure-based two-phase solver 
 {\tt compressibleInterFoam} of {\tt foam-extend-3.2}
 (respectively
 {\tt 4.0})  is 
 adapted for solving the Navier-Stokes equations 
 for a bubble near a flat solid boundary. 
The numerical implementation of the equations
 (\ref{eq:momentumEq}) --
(\ref{eq:tait}) 
is briefly sketched here. 
The reader is referred to 
\cite{Miller-2013} and \cite{Koch-2016}
for more details. 

\subsection{{Reformulation of the equations}}
Equations (\ref{eq:contEqlg}), (\ref{eq:momentumEq}), and (\ref{eq:contEq}) 
are reformulated as evolution equations for the variables  
$\alpha_l,$ $\boldsymbol U$, and $p$ 
\citep{Weller-2008}.
To this end, the derivatives of the
densities 
$\rho_l$ and $\rho_g$ are replaced, making use of the respective equations
of state $\rho_i(p)$, i.e., 
\begin{equation}
  \diff \rho_i = \totdiff{\rho_i}{p} \, \diff p =: \psi_i \diff p \, , 
  \quad i=l,g \, .
  \label{eq:rhoDeriv}
\end{equation}
The quantities $\psi_i$, given by
\begin{equation}
\psi_g(p)  = \frac{\rho_g(p)}{\gamma_g \, p}, \qquad
\psi_l(p)  =\frac{\rho_l(p)}{n_T\lb p +B\rb} \, ,
\label{eq:eosCompressibility}
\end{equation}
are the adiabatic compressibilities multiplied by the density 
and are related to the speeds of sound, $c_i$, by  $\psi_i = 1/c_i^2$. 
In the following, the quantities $\psi_i$ will
be called compressibilities for short. 

Equations (\ref{eq:contEqlg}) in
 non-conservative form read, after division by $\rho_i$: 
\begin{equation}
  \substdiff{\alpha_i} + \frac{\alpha_i \psi_i}{\rho_i} \substdiff{p} +
 \alpha_i \nabla \cdot \boldsymbol U = 0,  \quad {i = l,g}\, ,
 \label{eq:contEqlgNonConservative}
\end{equation}
where $\substdiffText{}: = \partial /\partial t + \boldsymbol U \cdot \nabla$
denotes the substantial derivative and Eq.~(\ref{eq:rhoDeriv}) has been used to
replace $\substdiffText{\rho_i}$ in the second term. Summing up the
phase-fraction equations for liquid and gas yields the continuity equation 
(\ref{eq:contEq}) in non-conservative form: 
\begin{equation}
  \left( \frac{\alpha_l \psi_l}{\rho_l} + \frac{\alpha_g \psi_g}{\rho_g}\right)\substdiff{p} + \nabla \cdot \boldsymbol U = 0
 \,.
  \label{eq:pEqn}
\end{equation}

Equation (\ref{eq:pEqn}) is used to eliminate the divergence of
$\boldsymbol U$ in Eq.~(\ref{eq:contEqlgNonConservative}) such that the transport
equation for the liquid phase fraction can be written as
\begin{equation}
  \substdiff{\alpha_l} = \alpha_l \alpha_g \left( \frac{\psi_g}{\rho_g} -
 \frac{\psi_l}{\rho_l}\right) \substdiff{p} \label{eq:ContEql} \ .
\end{equation}
Summarizing, Eqs  (\ref{eq:momentumEq}),
(\ref{eq:pEqn}), and (\ref{eq:ContEql}) are solved for $\boldsymbol U, p$, and
$\alpha_l$.
The phase fraction of the gas is given by
$\alpha_g = 1 - \alpha_l$. 
The densities $\rho_l$ and $\rho_g$ are computed from the equations of state
(\ref{eq:eosgas}) and (\ref{eq:tait}), respectively, and the compressibilities $\psi_l$ and $\psi_g$ from 
their derivatives (\ref{eq:eosCompressibility}), giving the density field 
$\rho = \alpha_l \rho_l + \alpha_g \rho_g$ and the viscosity field $\mu = \alpha_l \mu_l + \alpha_g \mu_g$.

\subsection{Liquid-phase-fraction field  $\alpha_l$} \label{subsec:alphafield}
The position of the interface is captured implicitly by solving
the transport equation (\ref{eq:ContEql}) for $\alpha_l$. No geometric
reconstruction of the interface is necessary. To counteract the
numerical diffusion of the interface, 
\citet{Weller-2008}
introduced a ``compression
term'', the third term on the left hand side of Eq.~(\ref{eq:alphaEqn}): 
\begin{equation}
 \frac{\partial \alpha_l}{\partial t} + \bnabla \bcdot (\alpha_l \boldsymbol U) +  
 \bnabla \bcdot (\boldsymbol U_r \, \alpha_l \, \alpha_g) 
  = \alpha_l \alpha_g (\frac{\psi_g}{\rho_g} 
- \frac{\psi_l}{\rho_l})\substdiff{p} 
   + \alpha_l \bnabla \bcdot \boldsymbol U \, .
  \label{eq:alphaEqn}
\end{equation}
The compression term, only active in the interface
region where $\alpha_l \, \alpha_g > 0$, can be motivated from a two-fluid
formulation, where $\boldsymbol U_r$ denotes the
relative velocity of the two fluids \citep[see, e.g.,][]{Berberovic-2009}.
${\boldsymbol U}_r$ is taken to be normal to the iso-lines of $\alpha_l$, 
i.e., normal to the interface. At the face centres $f$ of a computational cell, $\boldsymbol U_r$ is
computed as 
\begin{equation}
  \boldsymbol U_{r,f} = c_{\alpha} 
   \frac{| \phi_f|}{|\boldsymbol S_f|} \boldsymbol n_f 
    \quad \textrm{for } c_{\alpha} \le 1,
\end{equation}
with $\boldsymbol S_f$ the surface area vector,  $\phi_f$ the volume 
flux through the face, and $c_{\alpha}$ a parameter that is used to 
adjust the strength of the compression of the interface, set to unity
here, $c_{\alpha} = 1$. The vector 
$\boldsymbol n_{f}$ denotes the unit normal to the interface, 
determined from the liquid phase fraction
$\alpha_l$ by
 $\boldsymbol n_{f} = \bnabla \alpha_{l,f}/|\bnabla
\alpha_{l,f} + \delta_n |$, whereby $\delta_n = 10^{-8}/\bar V^{1/3}$ ($\bar V$ denoting the average
cell volume). 
Equation (\ref{eq:alphaEqn}) is solved explicitly with the
MULES (multidimensional universal limiter with explicit solution)
scheme in several sub-cycles within a time step.

\subsection{Momentum, surface-tension and pressure field} \label{subsec:momfield}

Following \citet*{Brackbill-1992},
the surface integral ${\boldsymbol f}_\sigma$ in 
Eq. (\ref{eq:momentumEq}) is approximated by 
$\sigma \kappa_V \bnabla {\alpha_l}$, so that the momentum equation reads 
\begin{equation}
   \frac{\partial (\rho \, \boldsymbol U)}{\partial t} +  \bnabla \bcdot (\rho \,
 \boldsymbol 
 U \otimes \boldsymbol U) = - \bnabla p  +  \bnabla \bcdot \mathbbm{T} + \sigma \, \kappa_V \,
 \bnabla \alpha_l \, , \label{eq:UEqn}
\end{equation}
with
\begin{equation}
\kappa_V = -\bnabla \bcdot (\bnabla \alpha_l/(|\bnabla \alpha_l| +
\delta_{n}))
\label{eq:kappaV}
\end{equation}
and $\mathbbm{T}$ given by Eq. (\ref{eq:stressNewtonian}). 
In contrast to the 
curvature $\kappa(\boldsymbol x')$ in Eq. (\ref{eq:momentumEq}), which is only defined 
on the interface, 
$\kappa_V$ is defined on the whole computational domain, deviating from zero
only in the interface region, where $\alpha_l$ is not constant.

Equations (\ref{eq:UEqn}) and (\ref{eq:pEqn}) are solved using a PISO 
(pressure implicit with splitting of operators \cite{Issa-1986})-like algorithm.

\subsection{Initial conditions} \label{subsec:initcond}
\begin{figure}%[!h]  %Figur 1  sketch initial condition with boundary
\begin{journal}
\centerline{
\includegraphics[width=0.4\textwidth]{figure1a.eps}
\hspace{7ex}
\includegraphics[width=0.4\textwidth]{figure1b.eps}
}
\end{journal}
\begin{arXiv}
\centerline{
\includegraphics[width=0.4\textwidth]{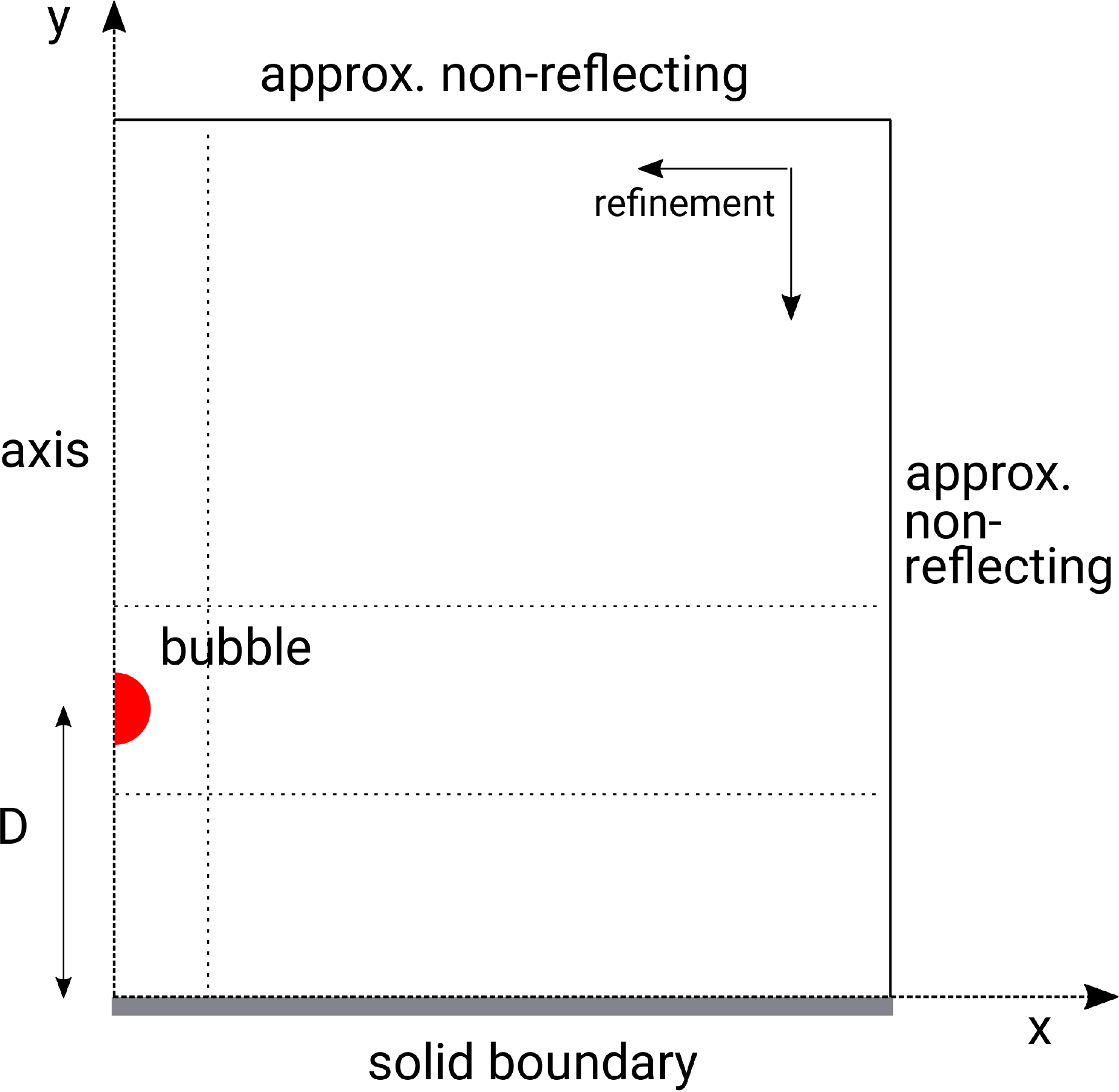}
\hspace{7ex}
\includegraphics[width=0.4\textwidth]{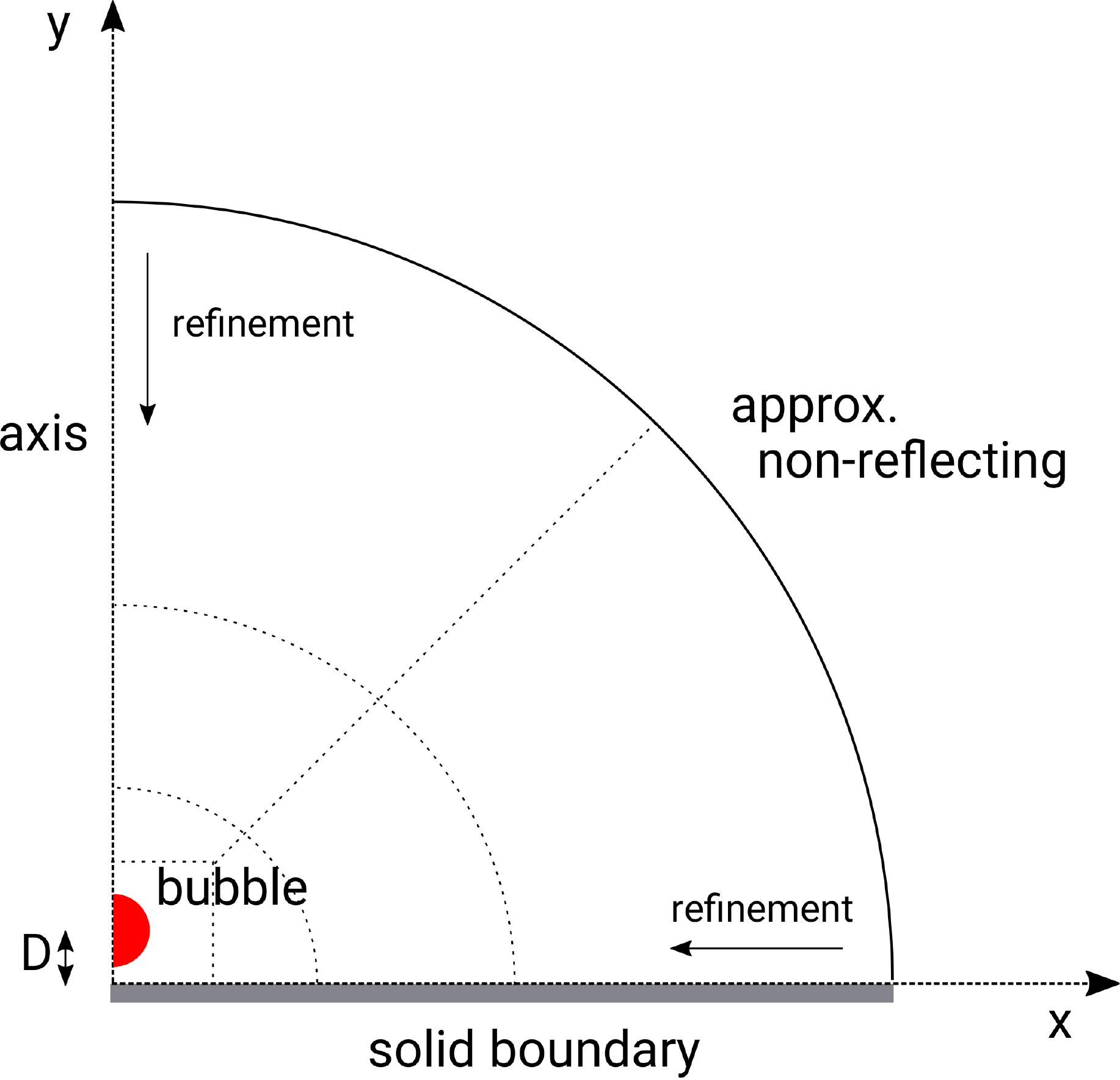}
}
\end{arXiv}

\caption{(colour online) 
Sketch of the geometry (not to scale) and the two different
meshes used for the simulations of bubble dynamics near a flat solid boundary. 
A small, initially spherical bubble of high internal pressure is inserted 
into the liquid at a distance $D$ from the boundary (realizable by a laser-induced bubble). 
Left: Cartesian mesh (large $D^* = D/R_{\rm max}$). Right: inner Cartesian region
matched to a polar grid (small $D^*$). A few grid lines are indicated. 
}
\label{fig:inited}
\end{figure}

The initial conditions can be inferred from figure\,\ref{fig:inited}.
For the simulations, a small,
spherical bubble with high internal pressure is placed at an initial distance $D$ 
from the solid boundary in a still liquid with constant ambient pressure
$p_\infty$ (energy-deposit bubble \cite{Lauterborn-2018b}).
The gas in the bubble is assumed to be adiabatically
compressed from its equilibrium state with radius $R_{\rm n,init}$. 
$R_{\rm n,init}$ is determined in a series of numerical simulations in
spherical symmetry such that the bubble expands to a maximum volume with
radius $R_{\max} = 500$\,$\upmu$m. 
Part of the gas content is removed before collapse 
of the bubble in order to allow for a strong enough first collapse. 
This is done by altering the bubble radius at rest, $R_{\rm n}$, from $R_{\rm n,init}$ to $R_{\rm n,final}$ (see below). 
This operation is necessary to correct for the condensing water vapour that is diminishing the internal bubble mass
\citep{Akhatov-2001, Koch-2016}.
The remaining gas content, corresponding to
an equilibrium radius $R_{\rm n,final}$, is chosen such that
the maximum rebound radius of a bubble in an unbounded liquid is 
in agreement with experimental data \citep{Koch-2016}.
For a bubble with 
a chosen initial radius $R_{\rm init} = 20$\,$\upmu$m,
placed in water under normal ambient conditions 
($p_{\infty} = 1$ bar, $T_{\infty} = 293.15$ K), the above procedure leads to
an initial internal pressure of $1.1 \times 10^9$ Pa,  
corresponding to an equilibrium radius of $R_{\rm n,init} = 184.1$ $\upmu$m. 
$R_{\rm n,final}$ is determined to
be $64$\,$\upmu$m.

\subsection{Mesh and time steps} \label{subsec:mesh} 
The mesh organization is given in figure\,\ref{fig:inited}.
As the computations are done in axial symmetry, a wedge
shaped mesh with small opening angle is constructed, 
which is one cell wide in azimuthal direction. 
Appropriate symmetry conditions at the wedge boundaries  
ensure axial symmetry. The essentially two-dimensional computational domain
is bounded by the axis of symmetry, the flat solid boundary, and approximate non-reflecting boundaries at
a distance of more than $80\,R_{\rm max}$ in both directions.
This large distance has been found necessary 
to simulate a bubble in a (semi-) unbounded liquid.
The bubble is placed with its centre at the axis of symmetry. 
Rotation around this axis gives a three-dimensional view on the bubble. 
The distance $D$ 
of the centre of the bubble from the solid boundary is varied to give normalized distances 
$D^* = D/R_{\rm max}$ in the range [0, 3].
At the solid boundary no-slip boundary conditions are imposed. Furthermore, 
the liquid volume fraction is set to 1, thus enforcing a (thin) liquid film 
between the bubble and the solid for small $D^*$.
This boundary condition is chosen for the sake of simplicity as it avoids
potential issues with a moving contact line. 
It does not affect
the results of this paper.
Two different grids are used depending on $D^*$. For large
$D^*$, the grid is Cartesian with uniform grid spacing $\Delta x=\Delta y \,
= \Delta x_{\min}$ in a region covering
the bubble 
during its translational motion towards the solid
boundary. 
The cell size increases progressively at larger distances from the bubble (see
figure\,\ref{fig:inited}, left).
For small $D^*$, a central uniform Cartesian region 
with an extension of $150\, \mum$ is matched to a polar grid
(figure\,\ref{fig:inited}, right). 
This arrangement
of grid cells takes advantage of the
fact that bubbles with small $D^*$ expand to a roughly hemispherical shape
with the center at the solid boundary.
Thus, grid cells are aligned with the bubble wall during a major part of
the bubble evolution. 
The cells of the polar grid have an
aspect ratio of one in a region covering the bubble, and larger radial progression 
further outwards. Unless otherwise stated, $\Delta x_{\min} = 1\, \mum$ for most of
the bubble evolution, with a refinement to $\Delta x_{\min} = 0.5\, \mum$,  
when 
thin liquid jets develop. 
Grid convergence for the present problems is demonstrated in
Appendix A.
The time-step is adjusted such that the maximum Courant
numbers built with the flow velocity and the velocity of the interface do
not exceed the values $0.2$ and $0.08$, respectively. 
During the stages where the compressibility of the liquid is important, 
the acoustic Courant number is well below 1. 
The von Neumann number 
is well below unity.

\subsection{Validation} \label{subsec:val} 
\begin{figure}  %fig.2
\begin{journal}
    \centerline{
\includegraphics[width=0.9\textwidth]{figure2.eps} 
}
\end{journal}
\begin{arXiv}
    \centerline{
\includegraphics[width=0.9\textwidth]{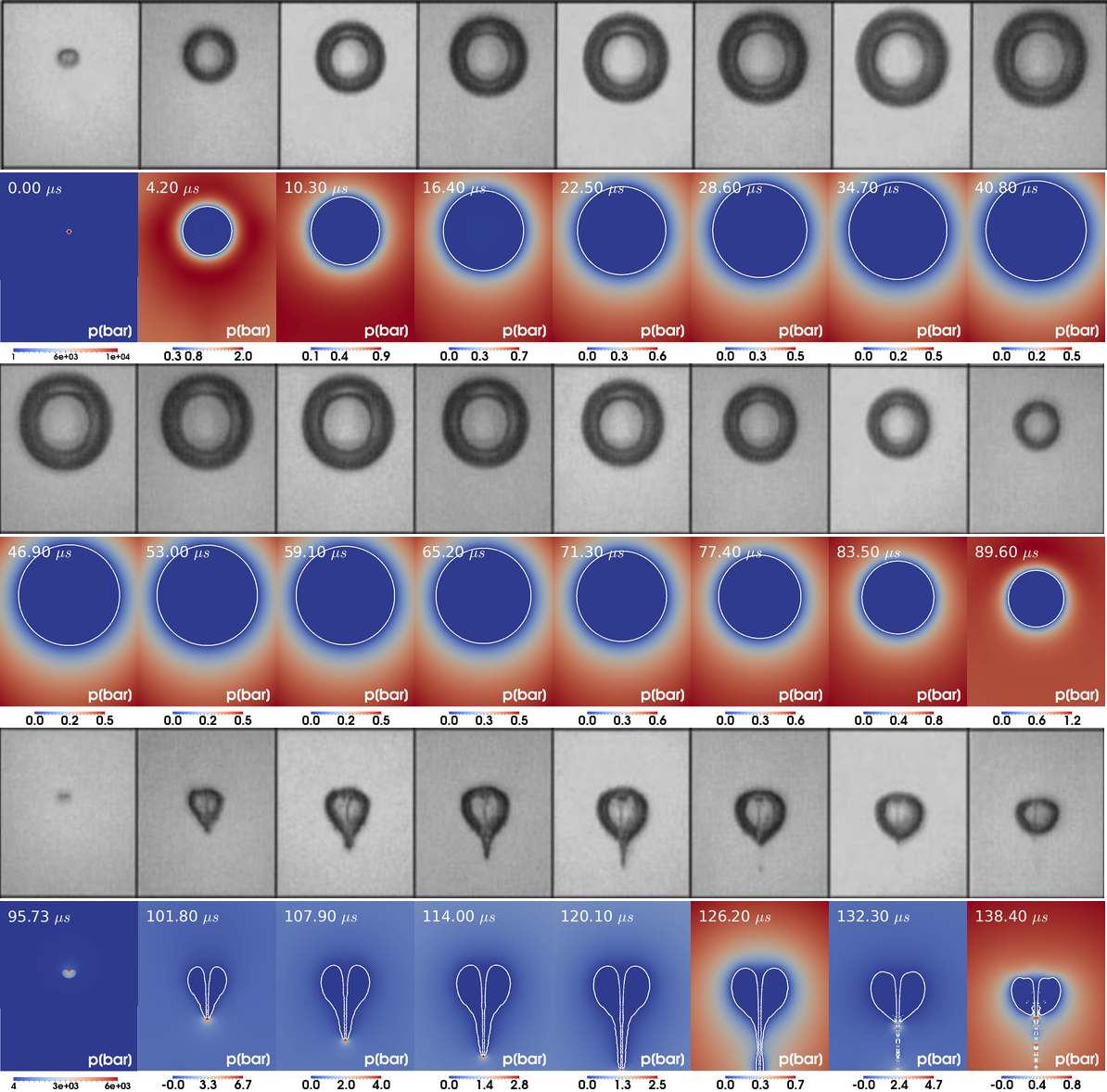} 
}
\end{arXiv}
\caption{(colour online) 
    Comparison of experiments \citep[][part of
         figure\,2a]{Philipp-1998}
  and numerical simulations (bubble shapes and pressure fields) for $D^* =$ 3. Bubble expansion, first collapse with jet formation, and first rebound. 
    Experiment: High-speed pictures in gray scale. Time between frames $\Delta t_{\rm exp} = 17.7\ \upmu$s, $R_{\rm max,exp} = 1450\ \upmu$m, frame width = 3.9 mm. 
    Numerical simulation: Time between frames 6.1 $\upmu$s, $R_{\rm max,sim } = 500\ \upmu$m, frame size 1.345 $\times$
  1.659 mm (width $\times$ height). The solid boundary is located 414.7 $\upmu$m below the lower border of each frame. 
}
\label{fig:valcompD3}
\end{figure}

There exist high-speed photographs of bubble dynamics near a flat solid boundary  
\citep[e.g.,][]{Philipp-1998}
that may be compared with the present simulations for validating the code. 
Figure \ref{fig:valcompD3} shows the comparison with simulations for $D^* = 3$, 
a case with a strong collapse and a pronounced jet. 
In the three double rows the respective upper rows are photographs of the bubble in backlight 
and the respective lower rows are from a simulation run, where the pressure field is presented 
and the bubble--liquid interface is highlighted by a white line. 
The maximum radius of the bubble in the experiments is 
$R_{\rm max,exp} = 1450\,\upmu$m and in the present simulation 
$R_{\rm max,sim} = 500\,\upmu$. Consequently, the frame sizes and 
times in the upper and lower rows of figure\,\ref{fig:valcompD3} differ by a factor of 
$R_{\rm max,sim}/R_{\rm max,num}$. 
This scaling for comparison is valid, because viscosity and surface tension 
do not play a significant role in this case. 
%
%-----------------------

A visual comparison frame by frame reveals an excellent agreement in expansion, collapse, and rebound. The jet becomes visible in the space and time resolution of the series only during rebound 
(see frame 101.80\,$\upmu$s and subsequent frames). 
Note that the bubble is photographed in side view, whereas the simulations show a central cut through the bubble. 
In the experiments, the jet therefore appears as a dark line along the axial centre line in the bright centre of the bubble. 
The overall agreement in bubble shape is excellent even for the gaseous jet hull in the rebound phase. For instance, the dissolution of the jet and its gaseous hull starts at the same frame 
(see frame at 126.20\,$\upmu$s). 

\begin{figure} % fig 3  
\begin{center}
 \begin{journal}
 \includegraphics[width=0.6\textwidth]{figure3.eps}
 \end{journal}
 \begin{arXiv}
 \includegraphics[width=0.6\textwidth]{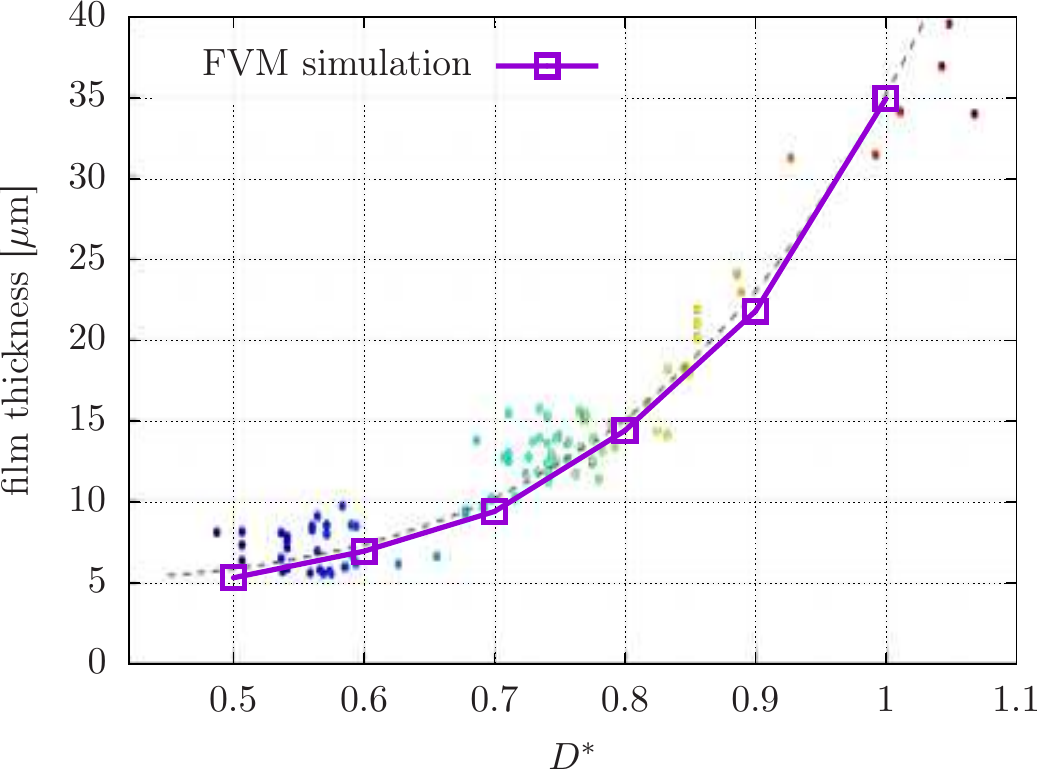}
 \end{arXiv}
 \caption{ 
(colour online) Thickness of the liquid film between the bubble and
the solid boundary at the moment the jet impacts onto the lower bubble
wall as a function of $D^*$. Comparison of experiments \citep[][Fig.~11]{Reuter-2019} 
and numerical simulations. Experiment: average maximum bubble radius 
$R_{\rm max, exp} = 410\ \upmu$m. Numerical simulation: $R_{\rm max, sim} =
500\ \upmu$m, $\sigma = 0$. 
Shown are 
the experimental data (filled circles), a polynomial fit to the experimental
data ($-\, \!\! -\, \!\! -$) and data from the present numerical simulation rescaled
by $R_{\rm max, exp}/R_{\rm max, sim}$ ({\small $\square$}).
\label{fig:reuter2019_Fig11_exp_num}
}
\end{center}
\end{figure}

A more quantitative comparsion with experimental data can be found in 
figure \ref{fig:reuter2019_Fig11_exp_num}. The recently developed method of 
total internal reflection shadowmetry by \citet{Reuter-2019} 
allowed them to measure the thickness of the liquid film between the bubble and
the solid boundary for small $D^*$ values ($D^* \lesssim 1$). Figure
\ref{fig:reuter2019_Fig11_exp_num} shows the film thickness at the moment
the liquid jet impacts onto the lower bubble wall as a function of $D^*$. 
Excellent agreement is
found between the experimental data and the data obtained from the numerical
simulation.

There also exist experiments as to the shock-wave emission from laser-induced bubbles 
\citep*{Vogel-1996}
and numerical calculations as to shock wave emission from collapsing bubbles 
\citep{Hickling-1964}.
The code has been validated with respect to shock-wave emission and propagation in these studies by 
\citet{Koch-2016}.
All three comparisons---jet formation in
figure\,\ref{fig:valcompD3}, film thickness at jet impact in
figure\,\ref{fig:reuter2019_Fig11_exp_num} and shock-wave propagation in 
\citet{Koch-2016}---give confidence for the extension to other parameters, here to cover the range of normalized distances $D^* = [0.05,3]$ and the special case $D^* = 0$.

\section{Results on distance dependence}\label{sec:resultsdistdep}
Bubbles that expand and collapse at different distances from a flat solid boundary experience 
a strongly different fate. 
Very far from the boundary ($D^* > 10$) they collapse essentially spherically 
\cite{Ohl-1998, Ohl-1999, Ohl-2002},
provided no instabilities are developing 
\cite{Koch-2016} as with large bubbles. 
When the bubble expands and collapses nearer to the solid boundary, a liquid jet towards the solid is formed 
with changing properties in dependence on the distance. 
Essentially two types of jets are found in the region $0 \le D^* \le 3$ investigated. 
The two jet-forming mechanisms are demonstrated with examples in this Section.

\subsection{Bubble shape dynamics}\label{subsec:shapes}
A survey is given of typical bubble shapes upon collapse, when the bubble distance to the solid boundary is varied 
from larger distances ($D^* =$ 3) to very small ones ($D^* = 0.05$). 
%
%-----------------------
\begin{figure}  %fig4
\begin{journal}
    \centerline{
\includegraphics[width=0.45\textwidth]{figure4a.eps}
\includegraphics[width=0.45\textwidth]{figure4b.eps}
}%\vspace{-2ex}
    \centerline{
\includegraphics[width=0.45\textwidth]{figure4c.eps}
\includegraphics[width=0.45\textwidth]{figure4d.eps}
}%\vspace{-2ex}
\end{journal}
\begin{arXiv}
    \centerline{
\includegraphics[width=0.45\textwidth]{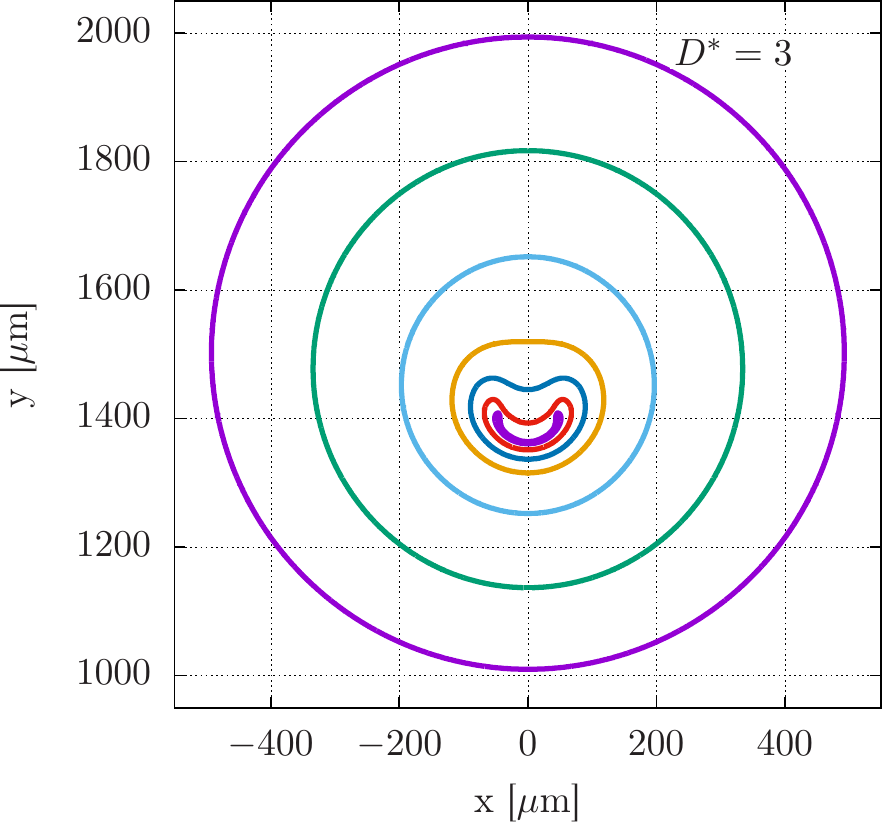}
\includegraphics[width=0.45\textwidth]{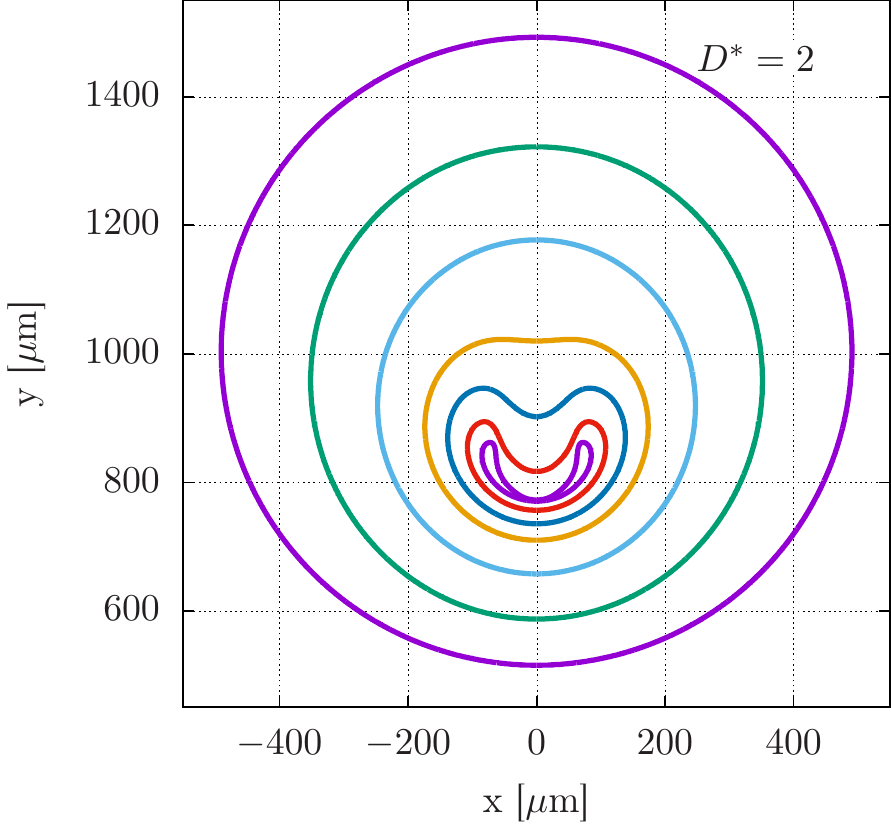}
}%\vspace{-2ex}
    \centerline{
\includegraphics[width=0.45\textwidth]{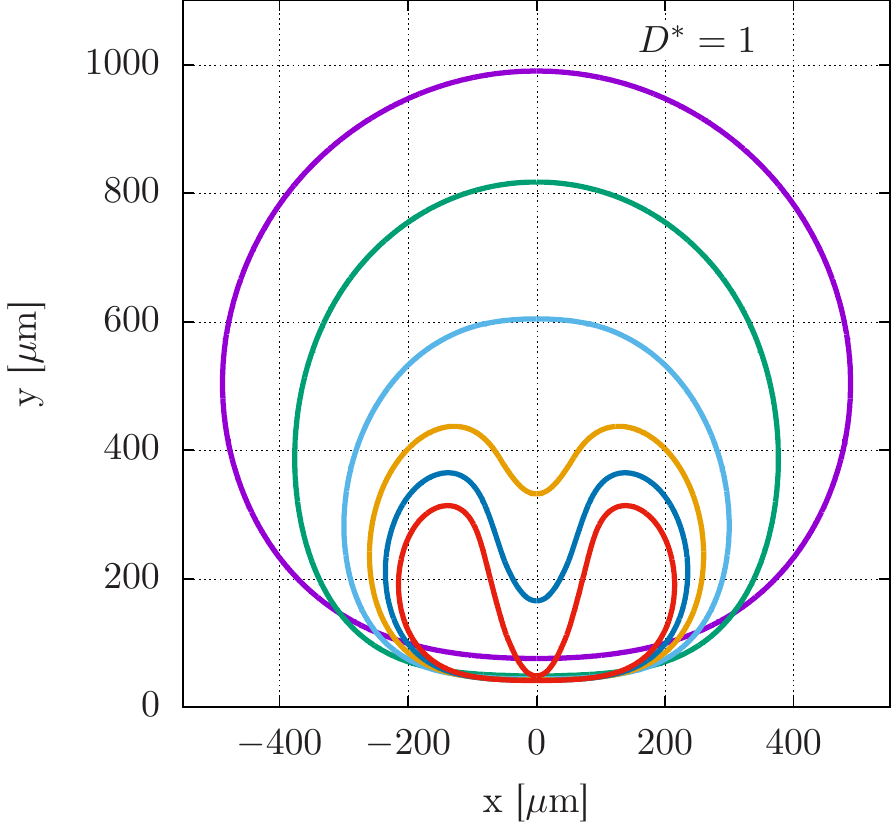}
\includegraphics[width=0.45\textwidth]{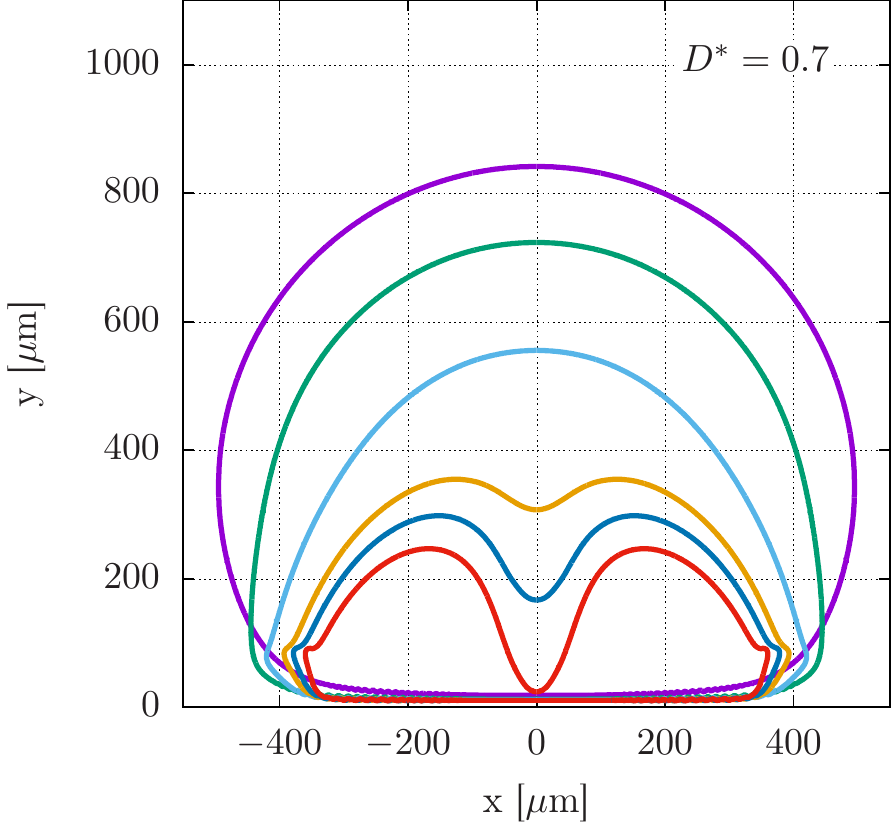}
}%\vspace{-2ex}
\end{arXiv} 
\caption{(colour online) 
Typical bubble shapes upon collapse after expansion to maximum volume (outer to inner curves) for different normalized distances $D^*$. Simulations with vanishing surface tension ($\sigma = 0$), as it has a negligible influence. 
Upper left: $D^* =$ 3, $t$ = 47.7, 85, 93, 95, 95.4, 95.6, and 95.7 $\upmu$s.  
Upper right: $D^* =$ 2, $t$ = 48.6, 85, 93, 96.2, 97.2, 97.8, and 98.1 $\upmu$s.  
Lower left: $D^* =$ 1, $t$ = 51, 85, 95, 99, 101, and 102.4 $\upmu$s. 
Lower right: $D^* =$ 0.7, $t$ = 53, 80, 92, 98, 100, and 102\,$\upmu$s.   
}
\label{fig:shapes} 
\end{figure} 
The special case $D^* = 0$ is covered in Section \ref{subsec:D0}.
In figures \ref{fig:shapes} and \ref{fig:shapes1}, the shape development upon collapse of the respective, expanded bubble is given for altogether eight distances $D^* =$ 3, 2, 1, 0.7, 0.3, 0.2, 0.1, and 0.05 by the outline of the shape in a central cut through the bubble at different instants in time. 
The shape sequences start from the maximum volume to cover the collapse phase 
up to the impact of the jet onto the lower bubble wall.  
The further collapse scenario is not included for the sake of
readability of the diagrams.

At large distances, jet formation and impact onto the opposite bubble wall occur by a gradual and smooth involution of the top of the bubble with impact onto the opposite bubble wall very late in the collapse phase
(figure\,\ref{fig:shapes}, $D^* =$ 3). The jet is broad in relation to the bubble size at jet impact (almost of bubble size) and leads to a flat-bubble appearance at impact and further collapse (seen from aside, along the solid surface). 
The bubble ($R_{\rm max} = 500\,\upmu$m) is moving a considerable distance ($>$100 $\upmu$m) towards the solid boundary. 
As the influence of the solid wall gets larger and larger when the 
bubble 
is positioned
nearer and nearer to the boundary, the jet is formed earlier 
and earlier in the collapse phase (figure\,\ref{fig:shapes}, $D^* =$ 2, 1, and 0.7). Thereby the jet gets broader, but thinner in relation to the actual bubble shape at jet impact onto the lower bubble wall, and the bubble stays larger at jet impact. The motion of the bubble centre towards the boundary gets also larger ($>$200 $\upmu$m at $D^* = 2$, $>$300 $\upmu$m at $D^* = 1$), whereby at $D^* = 1$ the lower wall of the bubble near the solid boundary is hardly moving. Below $D^* =1$, the bubble shape at maximum volume changes more and more to a hemispherical one. For $D^* = 0.7$, 
the bubble gets flatter in shape near the boundary compared to $D^* = 1$ and develops 
a sharp curvature at the outer rim, sharp curvatures not being present at $D^* = 1$. 

%
%-----------------------
\begin{figure}  %fig5
\begin{journal}
%\vspace{-4ex}
    \centerline{
\includegraphics[width=0.45\textwidth]{figure5a.eps} 
\includegraphics[width=0.45\textwidth]{figure5b.eps}
}\vspace{1ex}%\vspace{-8ex}
    \centerline{
\includegraphics[width=0.45\textwidth]{figure5c.eps}
\includegraphics[width=0.45\textwidth]{figure5d.eps}
}%\vspace{-4ex}
\end{journal}
\begin{arXiv}
%\vspace{-4ex}
    \centerline{
\includegraphics[width=0.45\textwidth]{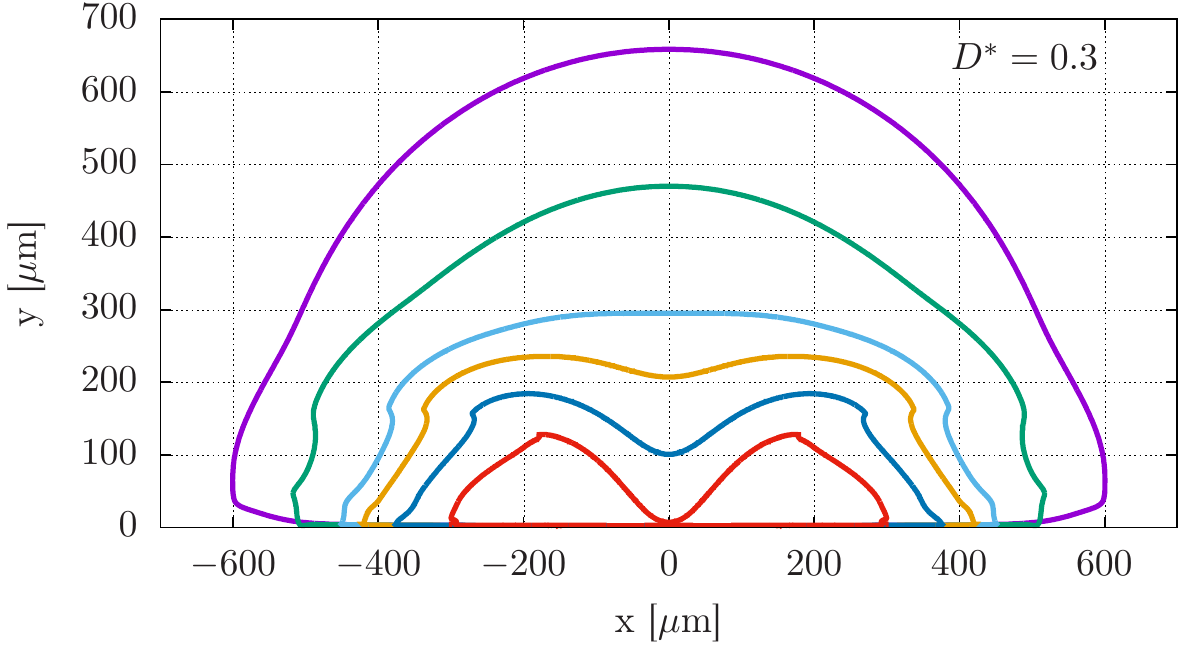} 
\includegraphics[width=0.45\textwidth]{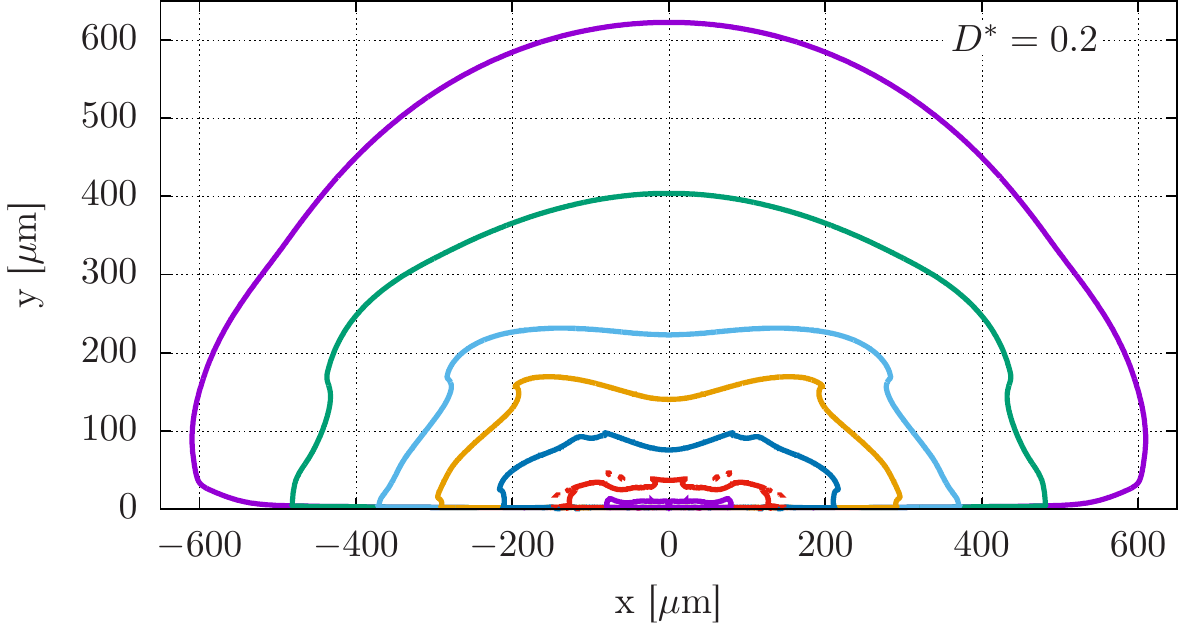}
}\vspace{1ex}%\vspace{-8ex}
    \centerline{
\includegraphics[width=0.45\textwidth]{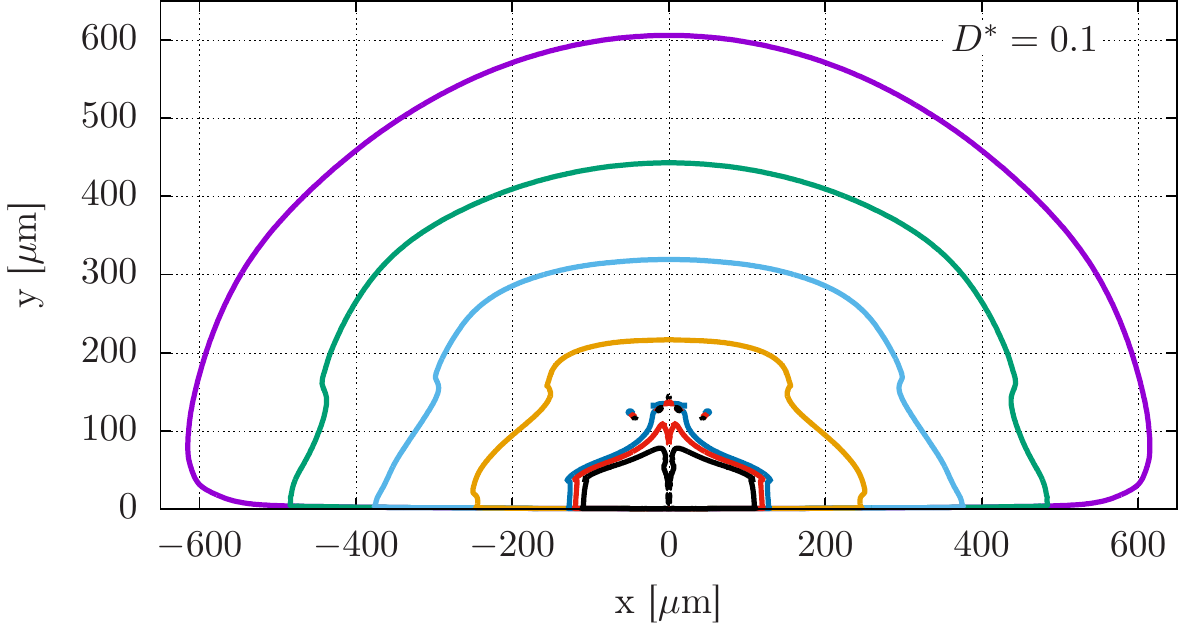}
\includegraphics[width=0.45\textwidth]{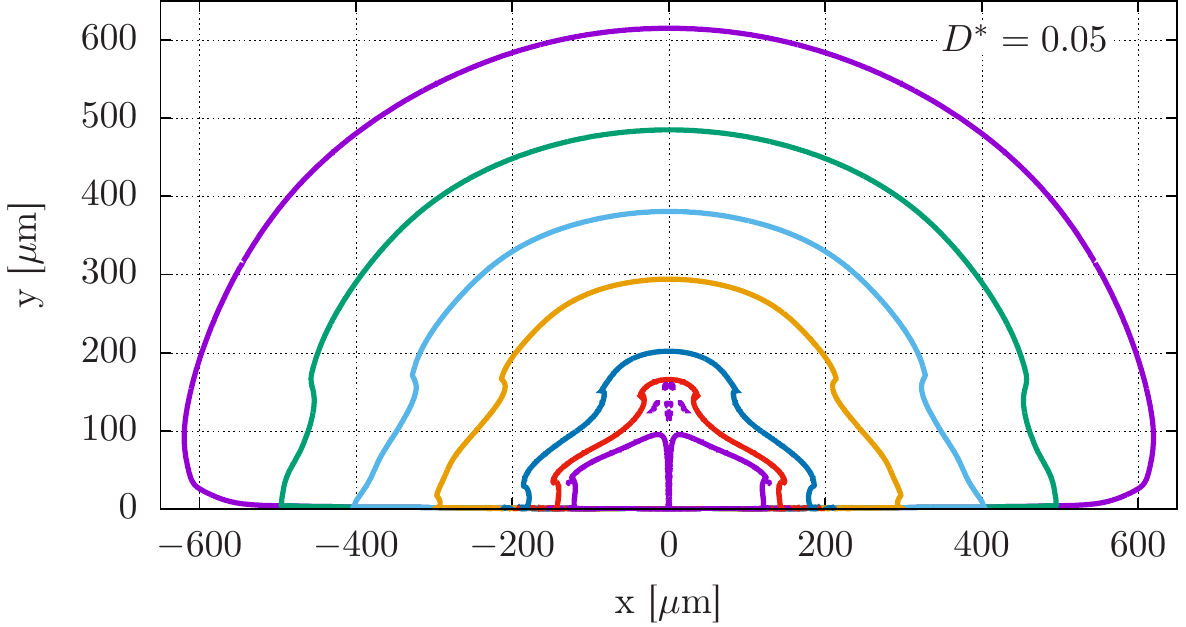}
}%\vspace{-4ex}
\end{arXiv}
\caption{(colour online)
Typical bubble shapes upon collapse after expansion to maximum volume (outer to inner curves) 
for different normalized distances $D^*$. 
$\sigma = 0.0725$\, N\,m$^{-1}$ (water).
Upper left: $D^* =$ 0.3, $t$ = 55, 90, 100, 103, 106, and 108.5 $\upmu$s. 
Upper right: $D^* =$ 0.2, $t$ = 55, 95, 105, 108, 110, 111, and 111.2 $\upmu$s.
Lower left: $D^* =$ 0.1, $t$ = 56.2, 95, 105, 110, 112.2, 112.28, and 112.37 $\upmu$s. 
Lower right: $D^* =$ 0.05, 
$t$ = 57, 95, 105, 110, 113, 113.6, and 113.85 $\upmu$s. 
}
\label{fig:shapes1} 
\end{figure} 

The bubble shapes upon collapse up to jet impact are quite smooth above $D^*$ about 0.7. 
The situation, however, becomes more complex for the range of $D^*$ below this value. 
Four examples are given in figure\,\ref{fig:shapes1} for $D^* = $ 0.3, 0.2, 0.1, and 0.05. 
The diagrams are calculated including surface tension, as the sharp curvatures appearing influence
more and more the bubble shape dynamics, albeit altogether only
slightly.
Upon expansion, the bubble gets almost attached to the solid boundary and gets a shape not looking too far away from a hemispherical bubble. The radial extension of the (almost) hemispherical bubble, however, is larger than 600 $\upmu$m for bubbles very near to the solid boundary, the radius of the spherical bubble without boundary only being 500 $\upmu$m at its maximum. 
Thus the bubble is stretching out along the solid surface upon expansion. 
An understanding of the stretching may be found in a volume argument.   
When comparing a spherical bubble of radius $R$ with a hemispherical one of the same volume, 
the relation $2^{1/3} R$ holds, i.e., for a bubble with $R = \rmax = 500$ $\upmu$m a value of about 630 $\upmu$m is obtained. 
This goes favourably with the about 620 $\upmu$m at $D^*=0.05$ with its approximately hemispherical shape at maximum volume. 

The influence of viscosity 
impresses a characteristic deformation on the shape of the bubble, 
notably introducing surface areas of sharper curvatures of the rim of the bubble near the solid boundary. 
Now, there are strongly different curvatures along the surface of the bubble. 
Nevertheless, at $D^* = 0.3$ (figure\,\ref{fig:shapes1}), a clearly visible, axial liquid jet is still formed from the top of the bubble that will hit the opposite bubble wall. 
However, when the bubble expands ever more closely to the solid boundary, a value of $D^*$ is reached, where no simple axial jet seems to be formed. 
An example is given in figure\,\ref{fig:shapes1} for $D^* = 0.2$. The inflow of liquid from above and the inflow from the sides neutralize each other to just give an almost flat, but wrinkled bubble at collapse on the solid surface. 

Very near to the solid boundary, a new type of axial liquid jet develops. 
Two examples, at $D^* =$ 0.1 and 0.05, are shown in figure\,\ref{fig:shapes1}. 
The transition region between $D^* = 0.3$ and $D^* = 0.1$ will be considered in a separate section below (Section \ref{subsec:sf}).
At $D^* = 0.1$ and 0.05 the inflow from the bubble sides is
so fast that it arrives as annular liquid flow 
at the axis of symmetry before the top of the bubble has time to pass by (see figure\,\ref{fig:shapes1}, $D^* = 0.1$, the last two shapes at 112.28 and 112.37 $\mus$, 
and  $D^* = 0.05$, the last two shapes at 113.6  and 113.85 $\mus$). 
The fast jets are studied in more detail in the next section (Section \ref{subsec:fastjet}).

\subsection{Fast jet formation dynamics}\label{subsec:fastjet}

\begin{figure}  %fig6
\begin{journal}
    \centerline{
\includegraphics[width=0.45\textwidth]{figure6a.eps}
\includegraphics[width=0.45\textwidth]{figure6b.eps}
}%\vspace{-4ex}
\end{journal}
\begin{arXiv}
    \centerline{
\includegraphics[width=0.45\textwidth]{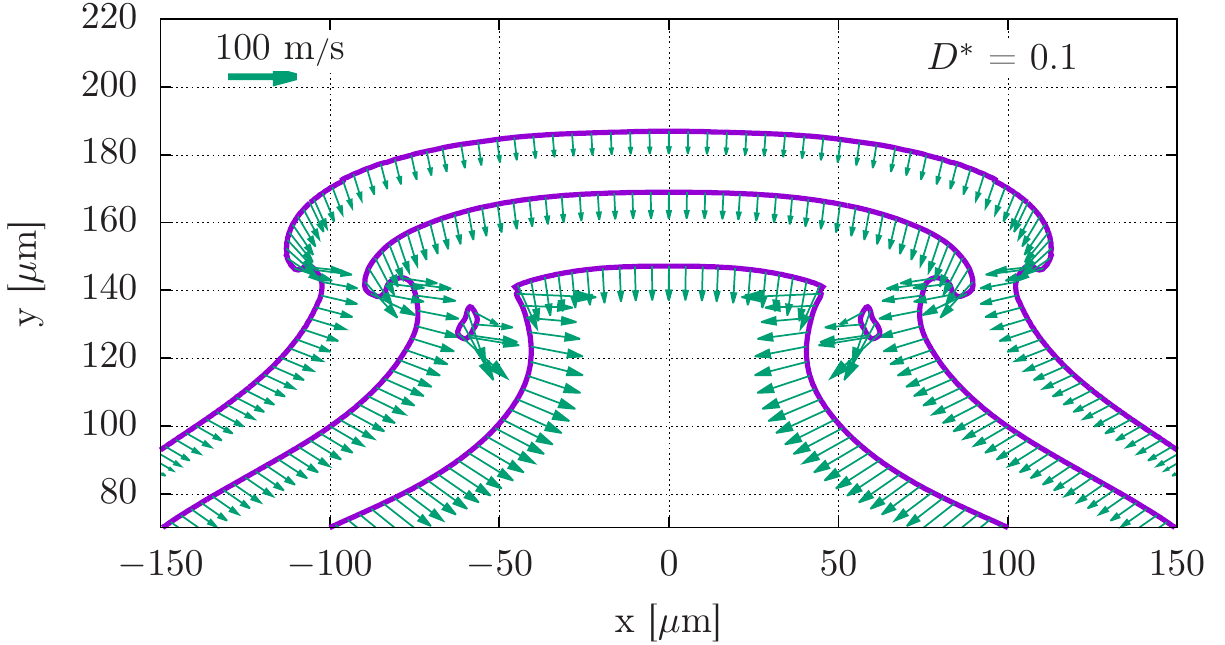}
\includegraphics[width=0.45\textwidth]{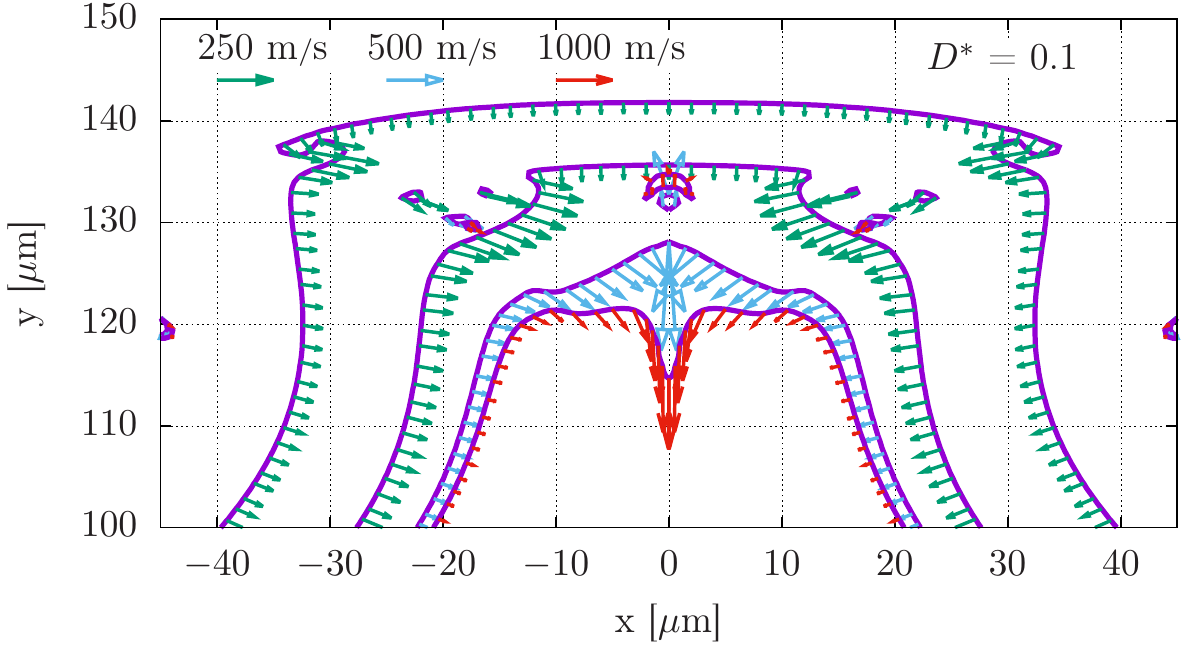}
}%\vspace{-4ex}
\end{arXiv}

\caption{(colour online)
Fast axial liquid-jet formation visualized by bubble shape development 
with bubble wall velocities. Neck formation and closing by annular inflow 
with jet initiation and small bubble separation. 
$\sigma = 0.0725$\, N\,m$^{-1}$ (water),
$D^* =$ 0.1,  
 time instants: 111, 111.5, and 112 $\mus$ (left), 112.10,
 112.20, 112.24, and 112.25 $\mus$ (right). 
 Note that the velocities (length of arrows) of the neck are larger than 
 the velocities of the head and strongly increase in time by cylindrical focusing. Three tiny torus bubbles are shed from the 
 neck--head transition region.
} 
\label{fig:wallvelo01}
\end{figure}

\begin{figure}  %fig7
\begin{journal}
%\vspace{-7ex}
    \centerline{
\includegraphics[width=0.45\textwidth]{figure7a.eps}
\includegraphics[width=0.45\textwidth]{figure7b.eps}
}%\vspace{-6ex}
\end{journal}
\begin{arXiv}
%\vspace{-7ex}
    \centerline{
\includegraphics[width=0.45\textwidth]{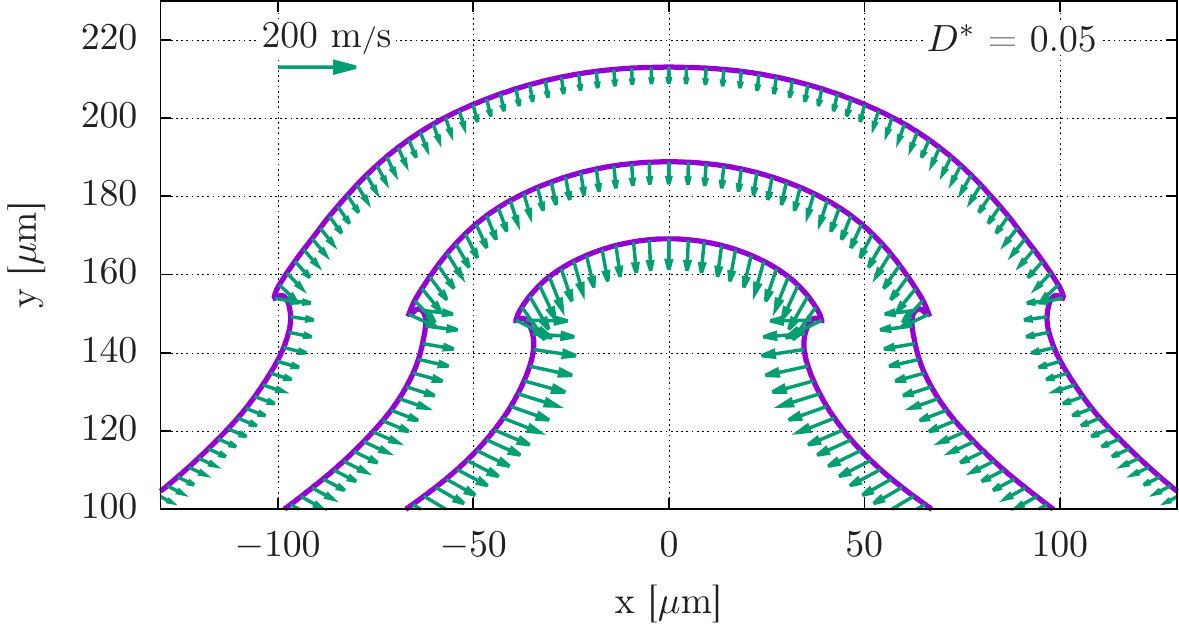}
\includegraphics[width=0.45\textwidth]{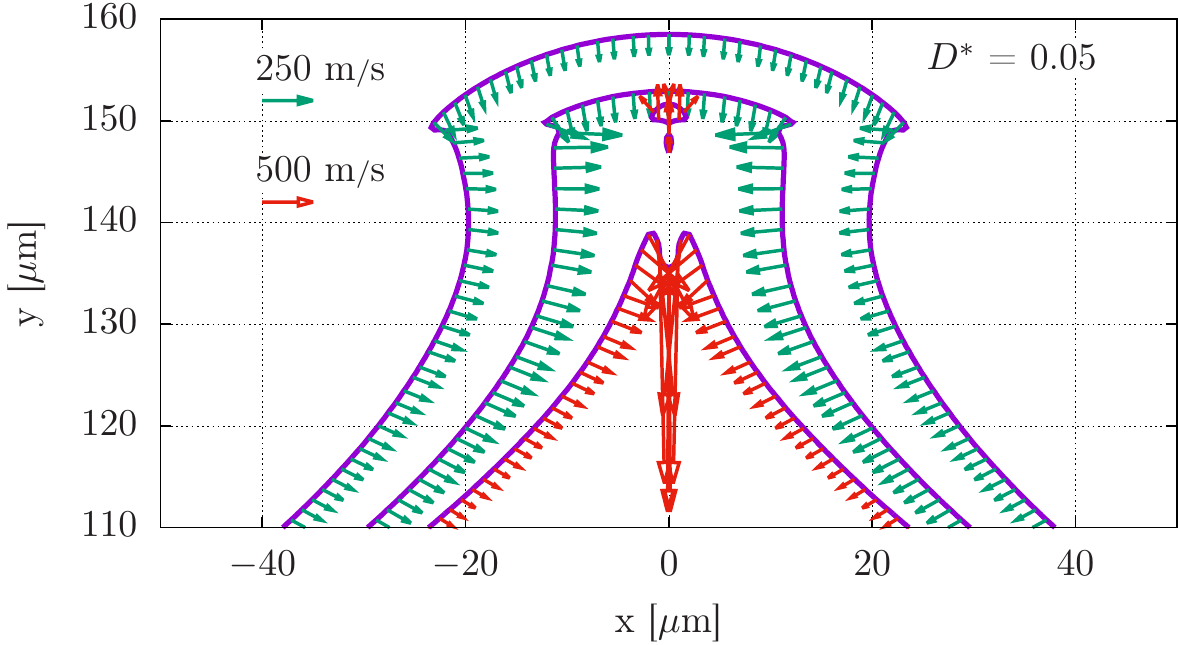}
}%\vspace{-6ex}
\end{arXiv}
\caption{(colour online)
Fast axial liquid jet formation visualized by bubble shape development with bubble wall velocities. 
Neck formation and closing by annular inflow with jet initiation and small bubble separation at the axis of symmetry. 
$\sigma = 0.0725$\, N\,m$^{-1}$ (water),
$D^* =$ 0.05,  
time instants: 112.8, 113.3, and 113.6 $\mus$ (left), 113.72, 113.77, and 113.80 $\mus$
(right).  
} 
\label{fig:wallvelo005}
\end{figure}

Two examples of fast jet formation are presented in figures \ref{fig:wallvelo01} and \ref{fig:wallvelo005}  in higher space and time resolution than in figure\,\ref{fig:shapes1} with the bubble shape evolution augmented by bubble wall velocities. This allows to anticipate the further development of the bubble shape by the length and direction of the velocity arrows and thus to figure out the formation of the fast axial liquid jet. A fast annular inflow (several 100 m\,s$^{-1}$) that has its origin in the outer rim of the bubble at maximum expansion constricts the bubble to eventually form a small head, a neck, a shoulder, and the main body of the bubble---a bell shape. 

The fast jet formation dynamics for $D^* = 0.1$ is shown in figure\,\ref{fig:wallvelo01} in seven shapes with bubble wall velocities that cover a time span of 1.25 $\mus$. 
In this case, the head of the bubble develops into an almost flat top with a sharp indentation at the head rim, from where a small torus bubble detaches (figure\,\ref{fig:wallvelo01}, left diagram). This process repeats until the further collapse of the bubble leads to neck closure with the formation of a fast jet (figure\,\ref{fig:wallvelo01}, right diagram). 
The head-on collision velocity of the neck in this case is about 660 m s$^{-1}$ water onto water at the axis of symmetry. 
The detached small bubbles (torus or simply connected ones) do not play a noticeable role in the formation of the fast jet.  
The cusp of the main bubble formed at self-impact of the annular inflow at the axis of symmetry will retract downwards to form the fast jet. 

The fast jet formation dynamics for $D^* = 0.05$ is shown in figure\,\ref{fig:wallvelo005} in six shapes with bubble wall velocities that cover a time span of 1 $\mus$. This time, the bubble head is not that flat as with $D^* = 0.1$, no small torus bubbles detach from the head rim, and the neck is more pronounced. From the length of the arrows it is evident that the velocity of the neck is higher than the velocity of the head. Again, a small, simply connected bubble is formed at the axis of symmetry as a remnant of the self-impact of the annular inflow at the axis.

%
%++++++++++++++++++++++++++++++++++++++++++++
%
% 
%-----------------------

\subsection{The limiting case $D^*=0$}\label{subsec:D0}
The case $D^* = 0$ cannot be treated as the previous cases, because a spherical bubble can not be placed with its centre 
directly at the solid surface. 
It would partly penetrate the solid.
However, when approaching the surface with spherical bubbles inserted, upon expansion a more and more hemispherical bubble shape is soon reached. A natural extension is therefore to place a small hemispherical bubble of high pressure directly at the boundary, giving $D^* = 0$. 
To comply with the previous cases as a kind of limiting case, a hemispherical bubble with the same initial volume and internal pressure (i.e., energy) as the previous bubbles, is placed directly at the solid boundary. 
The initial radius of the hemisphere then amounts to ${2}^{1/3} \times 20\,\mum$.
Without viscosity, and in the absence of instabilities, the bubble 
would expand and collapse in exactly hemispherical shape. 
The maximum radius of the bubble would be ${2}^{1/3} R_{\max}$, and the 
corresponding Rayleigh collapse time would amount to ${2}^{1/3}T_c = 1.26 \, T_c$, 
with $T_c$ the Rayleigh collapse time of a spherical (empty) bubble with maximum
radius $R_{\max}$ (the small amount of gas only marginally influences the collapse time): 
\begin{equation}
T_c = 0.915 R_{\rm max} \sqrt{\frac{\rho_l}{p_\infty}}\, .
\label{eq:Tc}
\end{equation}

\begin{figure}  %fig8 
%\vspace{-4ex}
\begin{journal} 
\centerline{
\includegraphics[width=0.45\textwidth]{figure8a.eps}
\includegraphics[width=0.45\textwidth]{figure8b.eps}
}
\end{journal}
\begin{arXiv} 
\centerline{
\includegraphics[width=0.45\textwidth]{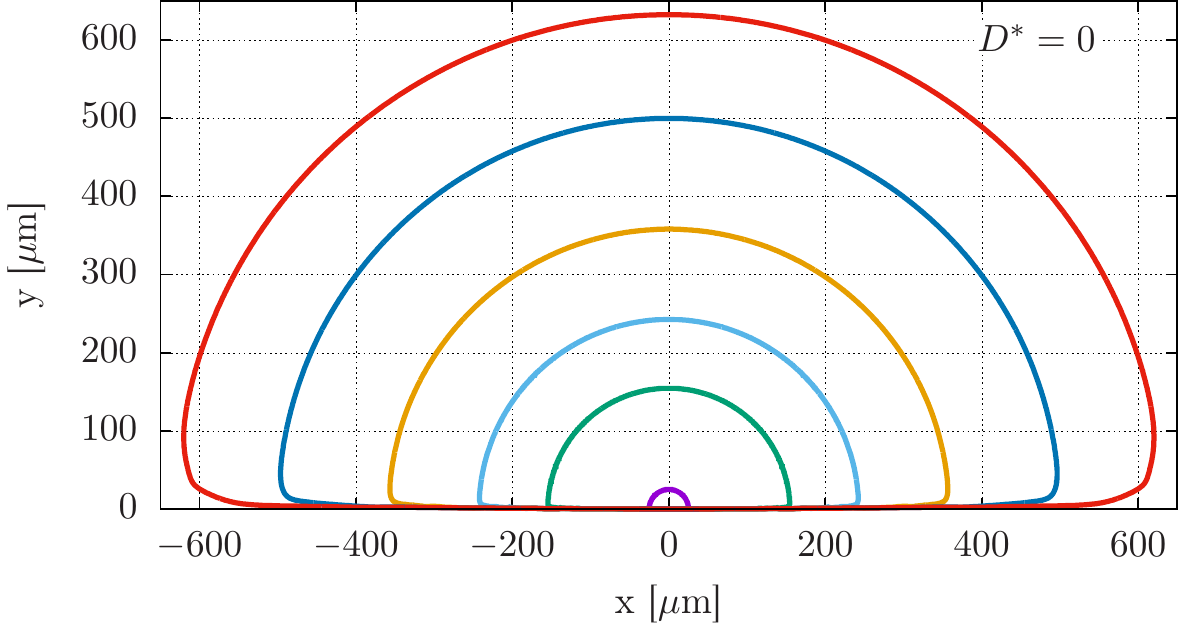}
\includegraphics[width=0.45\textwidth]{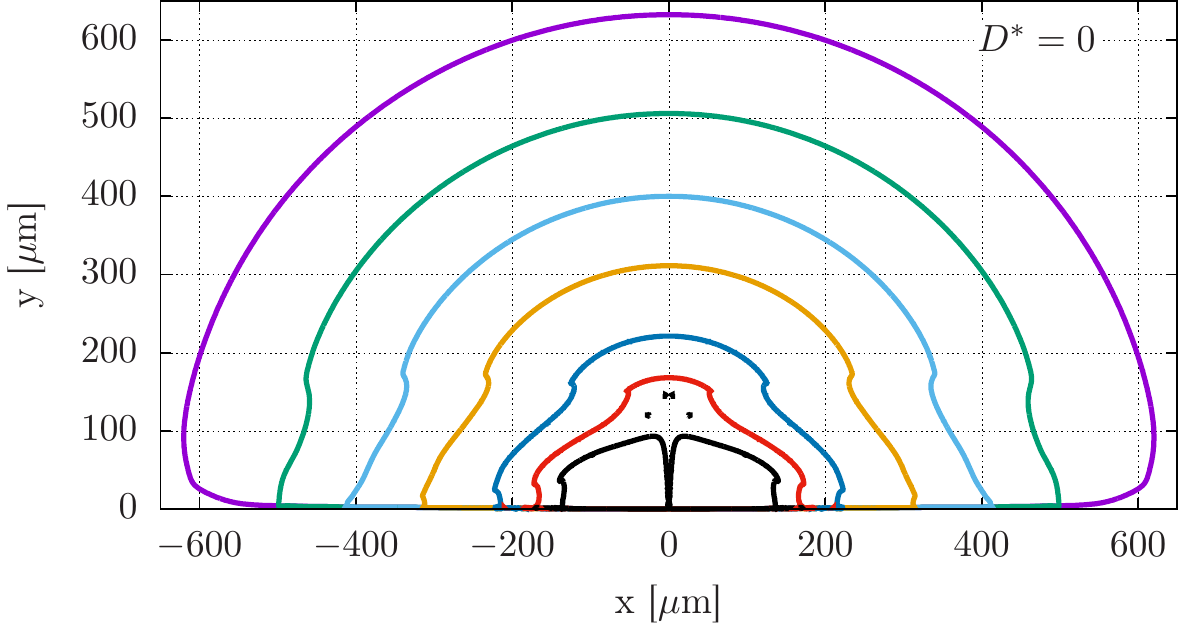}
}
\end{arXiv}
%\vspace{-4ex}
\caption{(colour online)
Bubble shapes at expansion and collapse of an initially hemispherical bubble attached to the
solid boundary. 
$\sigma = 0.0725$\, N\,m$^{-1}$ (water),
$D^* = 0$. 
Left: expansion, inner to outer curves, 
$t$ = 0, 1, 3, 8, 20, and 57.4 $\mus$. 
Right: collapse, outer to inner curves,  
$t$ = 57.4, 95, 105, 110, 113, 114, and 114.46 $\mus$. 
}
\label{fig:shapes_zero}
\end{figure}

\begin{figure} %fig9 
%\vspace{-4ex}
\begin{journal}
\centerline{
\includegraphics[width=0.45\textwidth]{figure9a.eps}
\includegraphics[width=0.45\textwidth]{figure9b.eps}
}
\end{journal}
\begin{arXiv}
\centerline{
\includegraphics[width=0.45\textwidth]{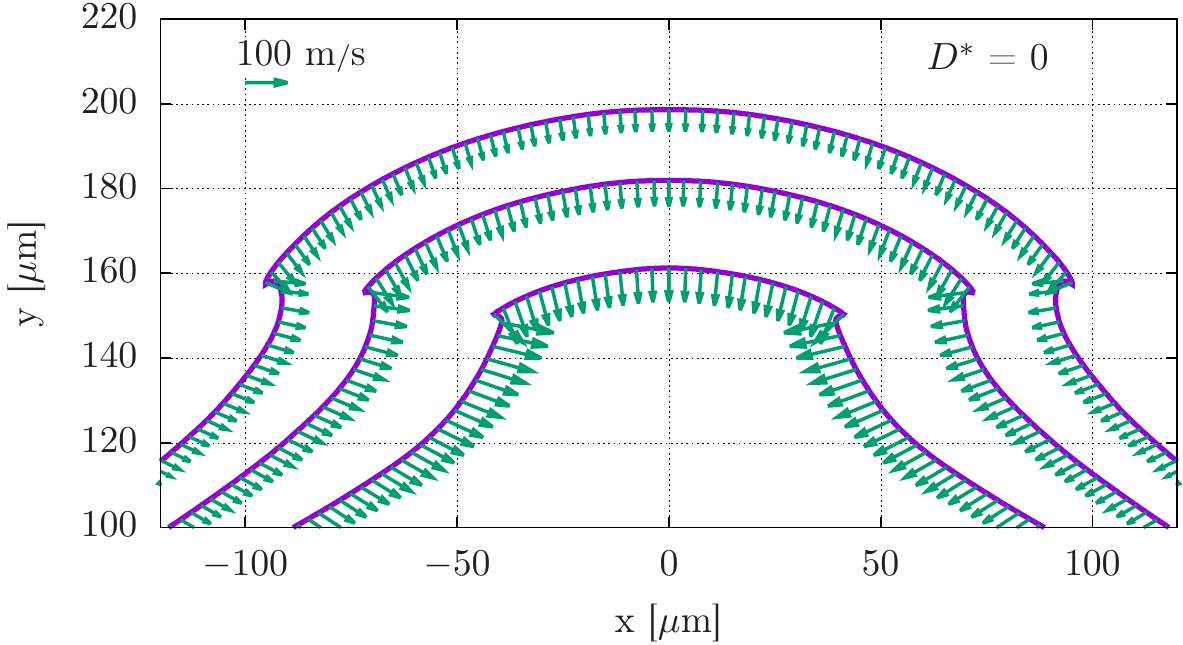}
\includegraphics[width=0.45\textwidth]{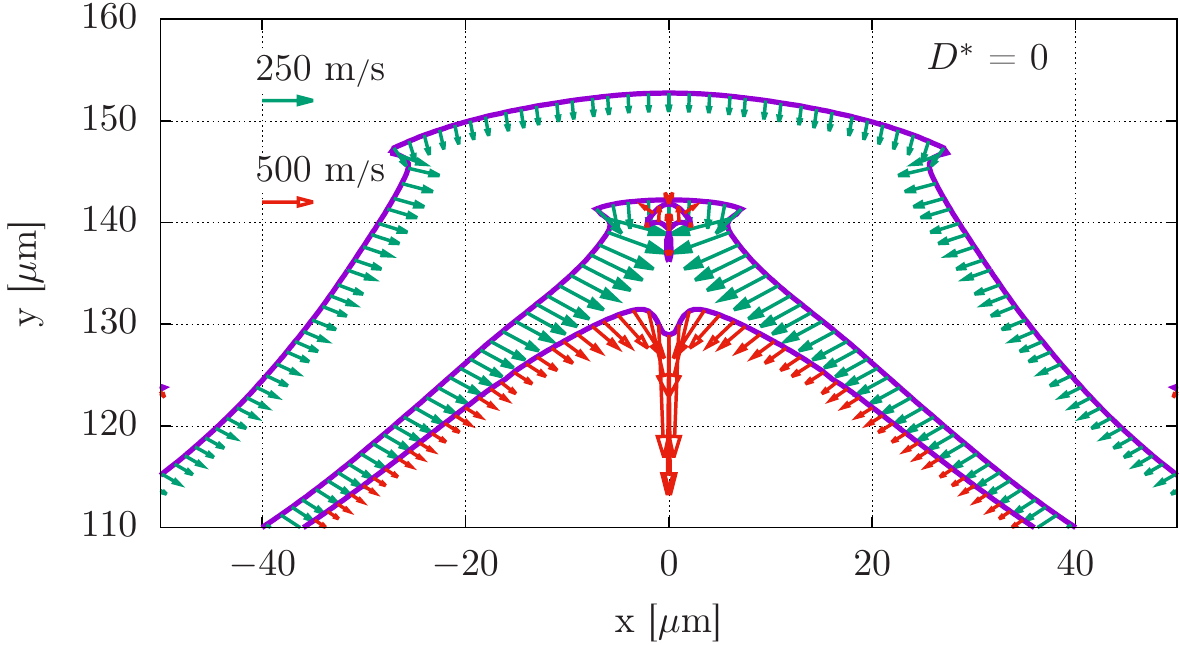}
}
\end{arXiv}
%\vspace{-4ex}
\caption{(colour online) 
Fast axial liquid-jet formation visualized by bubble shape development with bubble wall velocities. 
Neck formation and closing by annular inflow with jet initiation and small bubble separation at the axis of symmetry. 
$\sigma = 0.0725$\, N\,m$^{-1}$ (water),
$D^* = 0$. Jet formation phase (outer to inner curves),
$t$ = 113.5, 113.8, and 114.1 $\mus$ (left), $t$ = 114.2, 114.3, and 114.32 $\mus$ (right). 
}
\label{fig:wallvelo0}
\end{figure}

Including viscosity (and thereby the no-slip boundary condition), 
the flow upon bubble expansion from its initially hemispherical shape 
leads to a bubble shape with again a strongly curved outer rim at some distance from the solid boundary
(figure\,\ref{fig:shapes_zero}). 
The bubble shape
at maximum extension closely resembles the shape of an initially spherical
bubble with $D^*=0.05$ (figure\,\ref{fig:shapes1}). 
As in the latter case, the region of high curvature of the bubble wall close
to the outer rim leads to the typical bell-shaped form during collapse and
the formation of a fast jet (figure\,\ref{fig:wallvelo0}). 
The details of fast jet formation, however, are somewhat different as the dynamics of neck closure proceeds in a smoother way.  
But still a tiny, simply connected bubble is separated at the axis of symmetry 
from the top of the bubble following neck closure. 

\subsection{Slow--fast jet transition}\label{subsec:sf} 
The transition from the slow to the fast jet between $D^* = 0.3$ and $D^* = 0.1$ is a gradual and complicated process. It proceeds via strongly deformed bubble shapes. 
At $D^* = 0.3$ a clearly visible slow, broad jet is formed, and at $D^* = 0.1$ a clearly visible fast, thin jet is present (figure\,\ref{fig:shapes1}).  The region in between is explored in more detail in this Section. To save simulation time, the region $[0.1, 0.3]$ that contains the slow--fast jet transition is scanned with the 
help of a bisection-like search. 

\begin{figure} %fig10
%\vspace{-15ex}  
\begin{journal}
\centerline{
\includegraphics[width=0.64\textwidth]{figure10a.eps}
}%\vspace{-25ex}
\centerline{
\includegraphics[width=0.6\textwidth]{figure10b.eps}
}%\vspace{-15ex}
\end{journal}
\begin{arXiv}
\centerline{
\includegraphics[width=0.64\textwidth]{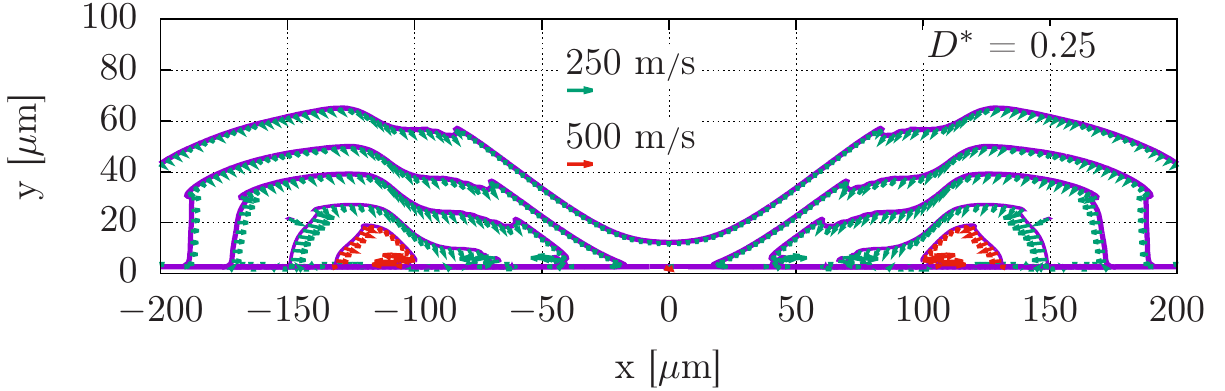}
}%\vspace{-25ex}
\centerline{
\includegraphics[width=0.6\textwidth]{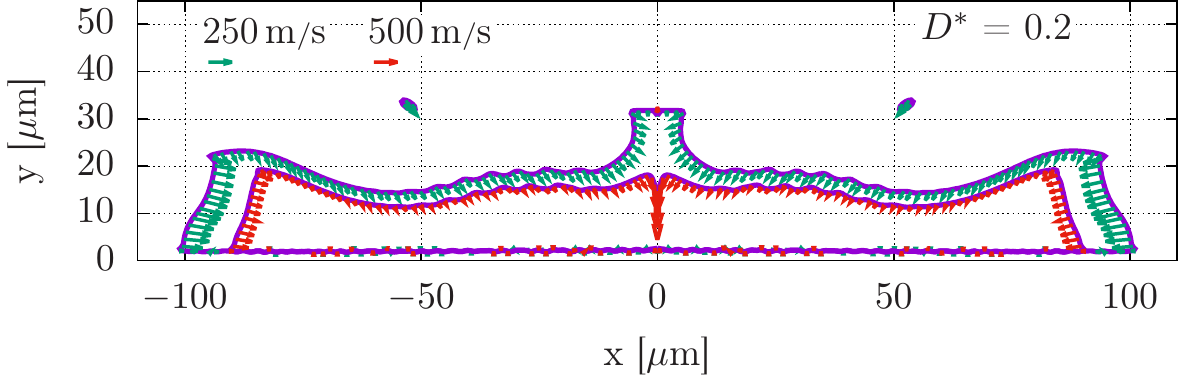}
}%\vspace{-15ex}
\end{arXiv}
\caption{(colour online)
Bubbles shapes at collapse with bubble wall velocities (outer to inner curves) 
for $D^* = 0.25$, $t = $ 110.2, 110.5, 110.7, 110.9, and 111 $\mus$
(upper  diagram), and $D^* = 0.2$, $t = $ 111.23 and 111.265 $\mus$ 
(lower  diagram).
} 
\label{fig:Dstar025}
\end{figure}

When looking for the type of jet formation at the intermediate value $D^* = 0.2 \in [0.1,0.3]$ in high resolution, a fast jet is found. And taking the intermediate value $D^* = 0.25 \in [0.2,0.3]$  a slow jet is found. This is demonstrated in figure\,\ref{fig:Dstar025}. At $D^* = 0.25$, the at first present annular inflow has been bent downwards 
by the (at this time) faster axial microjet and only produces a small dip at the basis of the microjet (see, e.g., $y \approx 60\,\mum$ at $t = 110.5\,\mus$ and its development at later times).
The standard microjet impact leads to an outbound annular liquid nanojet into the bubble, as reported before for this type of impact 
\citep{Lechner-2017}, and the subsequent collapse of the torus bubble formed. 
The velocity of the slow jet at impact amounts to about 35 m\,s$^{-1}$ ($D^* = 0.25$). 

At $D^* = 0.2$, on the other hand, the annular inflow is
so fast that it arrives at the axis of symmetry before the slow jet is able to pass by. Thereby the ``tip'' of the axial microjet is turned into a flat top of the bubble (figure\,\ref{fig:Dstar025},
lower  diagram). Actually, the self-impact of the annular inflow occurs almost at the ``tip'' of the slow jet and only leaves a very tiny bubble split off from the contact site (the tiny bubble at the flat top in figure\,\ref{fig:Dstar025},
lower  diagram). 

\begin{figure}  %fig11 
%\vspace{-14ex}
\begin{journal}
    \centerline{\hspace{-3ex}
  \includegraphics[width=0.6\textwidth]{figure11a.eps}
}%\vspace{-27ex}
    \centerline{
  \includegraphics[width=0.62\textwidth]{figure11b.eps}
}%\vspace{-13ex}
\end{journal}
\begin{arXiv}
    \centerline{\hspace{-3ex}
  \includegraphics[width=0.6\textwidth]{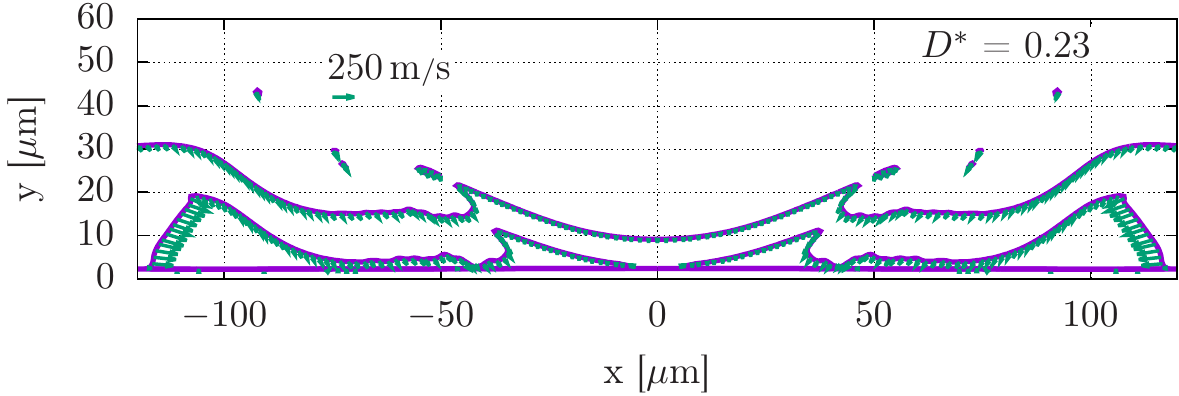}
}%\vspace{-27ex}
    \centerline{
  \includegraphics[width=0.62\textwidth]{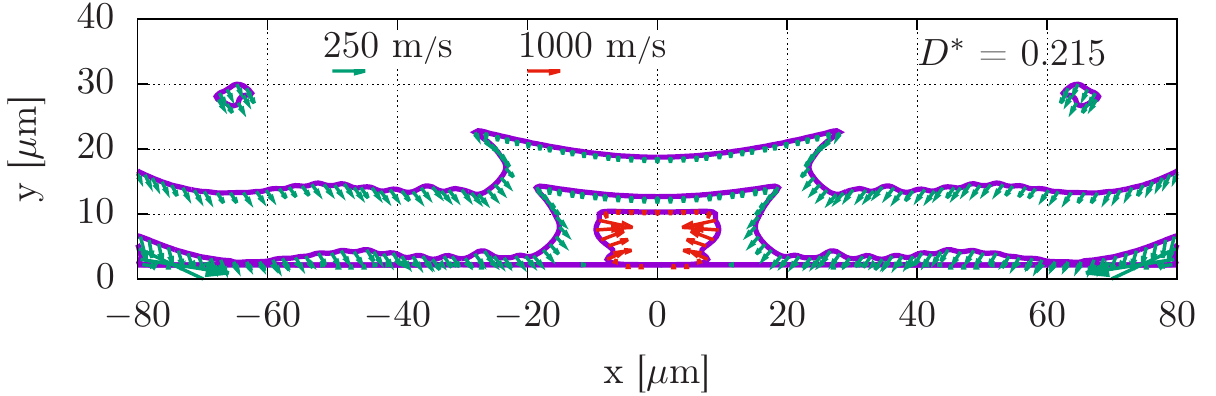}
}%\vspace{-13ex}
\end{arXiv}
\caption{ (colour online)
Bubble shapes upon collapse (outer to inner curves) after expansion to maximum volume  
for two different normalized distances $D^*$ with different jet formation dynamics.
Upper diagram: $D^* = 0.23$, $t$ = 110.9 and 111.05 $\mus$. 
Lower diagram: $D^*=0.215$, $t$ = 111, 111.13, and 111.17 $\mus$. 
}
\label{fig:shapes2} 
\end{figure}

When checking $D^*= 0.23 \in [0.2,0.25]$ for its jet type, 
it is seen that the slow jet just will succeed to impact the lower bubble wall first, before the annular flow arrives at the axis of symmetry (figure\,\ref{fig:shapes2}, upper diagram). The annular jet, however, makes contact with the lower bubble wall in the outskirts of the almost flat bubble about simultaneously or even before (not shown). 

At $D^* = 0.215 \in [0.2,0.23]$, the indentation of the bubble head (figure\,\ref{fig:shapes2}, lower diagram) is progressively flattened by the annular inflow and proceeds with a relatively low velocity.  
The shoulder hits the lower bubble wall and generates an additional inflow along the lower bubble wall towards the axis of symmetry. This leads to a scrumbled bubble surface. Thus neither a discernable slow jet nor a fast jet are present. Both jets are prevented by the fast annular inflow along the lower bubble wall.  

This finding points to a complicated transition process in a finite region of $D^*$, 
an interval $I_{\rm sf} = [D^*_{\rm f}, D^*_{\rm s}]$, with $D^*_{\rm s}$ the slow-jet transition threshold, $D^*_{\rm f}$ the fast-jet transition threshold, and $D^* = 0.215 \in [D^*_{\rm f}, D^*_{\rm s}]$. 
The task now 
is rather
the determination of two thresholds, $D^*_{\rm f}$ and $D^*_{\rm s}$. These thresholds will depend on their definition. There are several possibilities: just
slow or fast jet \textit{formation}, or including the \textit{impact} onto the lower bubble wall without being disturbed on the way to the solid boundary. Here, the definition is adopted that $D^*_{\rm f}$ is the largest $D^*$, where the fast jet hits the opposite bubble wall and leaves a torus bubble to collapse. Similarly, $D^*_{\rm s}$  is defined as the lowest value of $D^*$, where the slow jet hits the opposite bubble wall and leaves a torus bubble to collapse. With this definition of the transition region, $D^* = 0.215$ lies inside the transition region, as the impact of the shoulder makes a different 
collapse scenario (a ring impact). The same applies to $D^* = 0.23$. 

\begin{figure} %fig12 
%\vspace{-5ex}
%\vspace{-8ex}
 \begin{journal}
\centerline{
  \includegraphics[width=0.63\textwidth]{figure12.eps} 
 }
 \end{journal}
  \begin{arXiv}
\centerline{
  \includegraphics[width=0.63\textwidth]{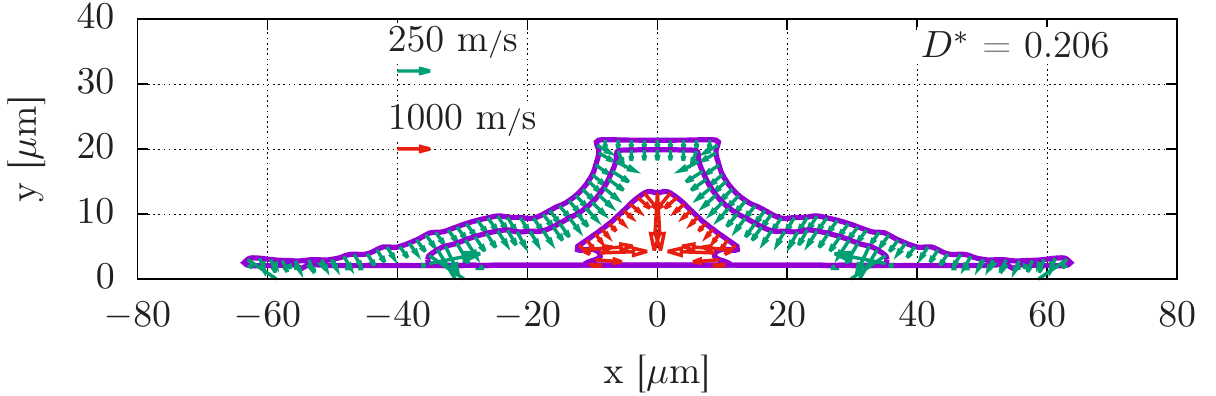} 
 }
 \end{arXiv}
\caption{(colour online) 
Fast jet formation dynamics (outer to inner curves) for $D^* = 0.206$, $t$ = 111.26, 111.28, and 111.302 $\mus$.  
}
\label{fig:shapeveltrans}
\end{figure}

When looking for the fast-jet transition threshold, $D^*_{\rm f}$, the already studied value of $D^* = 0.2$ is found, as the only slightly larger value of $D^* = 0.206$ (figure\,\ref{fig:shapeveltrans}) already shows a different collapse dynamics. Here, the contact of the shoulder with the lower bubble wall generates a very acute angle. The fast closure generates a second fast annular inflow, this time along the lower bubble wall (similarly to the outbound nanojet mentioned above, but inbound). It gains an exceedingly high speed by cylindrical convergence, more than 1000 m\,s$^{-1}$. The impact of the second annular jet onto the fast axial jet or itself leads to a
complicated flow on a very small scale, differently from the standard collapse behaviour. 
Thus, the fast-jet transition threshold can be located at $D^*_{\rm f} \approx 0.2$. Below this threshold, down to $D^* = 0$, the fast jet is present. 

%
%-----------------------
\begin{figure} %fig13 
%\vspace{-13ex}  
 \begin{journal}
\includegraphics[width=0.64\textwidth]{figure13.eps}
 \end{journal}
 \begin{arXiv}
\includegraphics[width=0.64\textwidth]{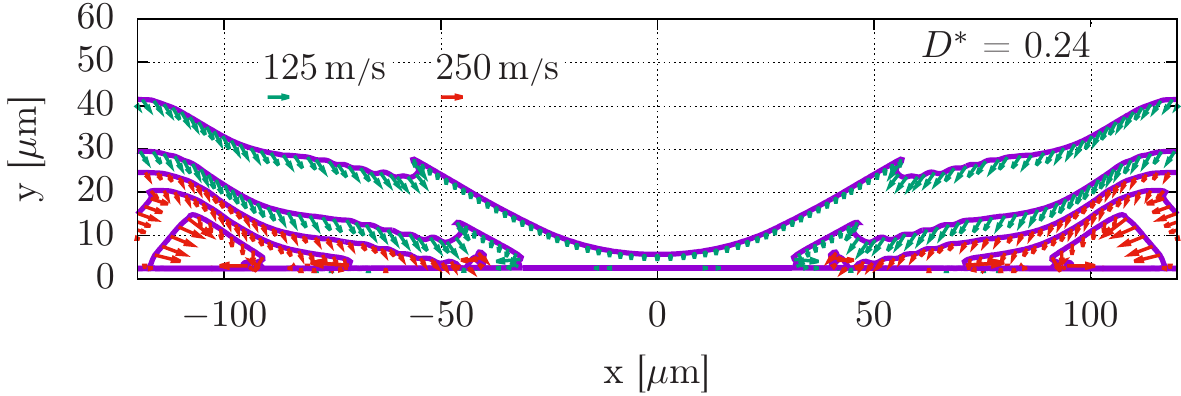}
 \end{arXiv}
\caption{(colour online)
Bubbles shapes at collapse with bubble wall velocities (outer to inner curves) for $D^* = 0.24$, $t = $ 110.7, 110.9, 110.97, 111.02, and 111.06 $\mus$. A standard, slow axial jet is formed. 
}
\label{fig:Dstar024}
\end{figure}

When looking for the slow-jet transition threshold, $D^*_{\rm s}$, a new search must be made. According to the results already present, $D^*_{\rm s}$ should be $\in [0.23,0.25]$. 
Figure \ref{fig:Dstar024} shows the intermediate value $D^* = 0.24$. There is a slow-jet impact with outbound annular nanojet formation. 
The annular inflow has been bent downwards by the standard axial microjet and impacts with the lower bubble wall and the outbound annular nanojet. A tiny (torus) bubble is split off that way. The subsequent dynamics is tricky in that the bent-down inflow 
now \textit{cum grano salis} resumes the slow-jet impact with the formation of a second outbound annular nanojet (figure\,\ref{fig:Dstar024}, curves at 111.02 and 111.06 $\mus$). The further dynamics very much resembles the torus-bubble collapse at larger $D^*$. 
The impact velocity of the slow jet onto the lower bubble wall is about 37 m\,s$^{-1}$. 
When comparing the collapse scenario at $D^* = 0.24$ with the one at $D^* = 0.25$ (figure\,\ref{fig:Dstar025},
upper diagram) and separate simulations for $D^* = 0.245$ (not shown), a gradually increasing influence of the bubble shoulder on the formation of the slow jet is observed. Very near below $D^* = 0.24$ the slow jet can no longer dominate the collapse. Thus, the slow-jet transition threshold can be located at $D^*_{\rm s} \approx 0.24$.

The transition region has been determined to $[D^*_{\rm f}, D^*_{\rm s}] = [0.2, 0.24]$. 
A total reorganization of the bubble shape dynamics occurs in this region.
It is characterized by additional jets. Outside that region,  a clear distinction between the slow and the fast jet can be made. The transition region is dominated by the appearance of a second inbound annular jet that forms along the lower bubble wall and gains high speed by cylindrical convergence (figure\,\ref{fig:shapeveltrans}). Its optimal formation conditions are found at about $D^* = 2.1$. To both higher and lower values of $D^*$ it fades away.

\subsection{Jet--bubble relationships}\label{subsec:jetbubble}
The changes in the jet-forming process, when the bubble is inserted closer and closer 
to the solid boundary, are clearly visible in figures \ref{fig:shapes} to  \ref{fig:Dstar024}. 
To be quantitative, specific quantities connected with the jet-forming process in dependence on $D^*$  are defined: \\
(a) $T(V_{\rm min})$, the bubble collapse time in the presence of the boundary: the time from the maximum bubble volume to the time of the main torus-bubble collapse; \\
(b) $T_{\rm ji}$, the jet impact time: the time from the maximum bubble volume (jet initiation reference time) to the time of the microjet impact onto the lower bubble wall; \\
(c) $V_{\rm ji}$, 
the volume of the bubble at the time of the microjet impact onto the lower bubble wall; \\
(d) $\Delta T_{\rm ji} = T(V_{\rm min}) - T_{\rm ji}$, the time difference between bubble collapse and jet impact onto the lower bubble wall. It is a measure of how early the jet develops in the process of bubble collapse.  $\Delta T_{\rm ji}/T_{\rm c}$ is called the jet earliness parameter.

\begin{figure}  %fig14
\begin{journal}
\centerline{
\includegraphics[width=0.45\textwidth]{figure14a.eps}
\hspace{2ex}
\includegraphics[width=0.45\textwidth]{figure14b.eps}
}
\end{journal}
\begin{arXiv}
\centerline{
\includegraphics[width=0.45\textwidth]{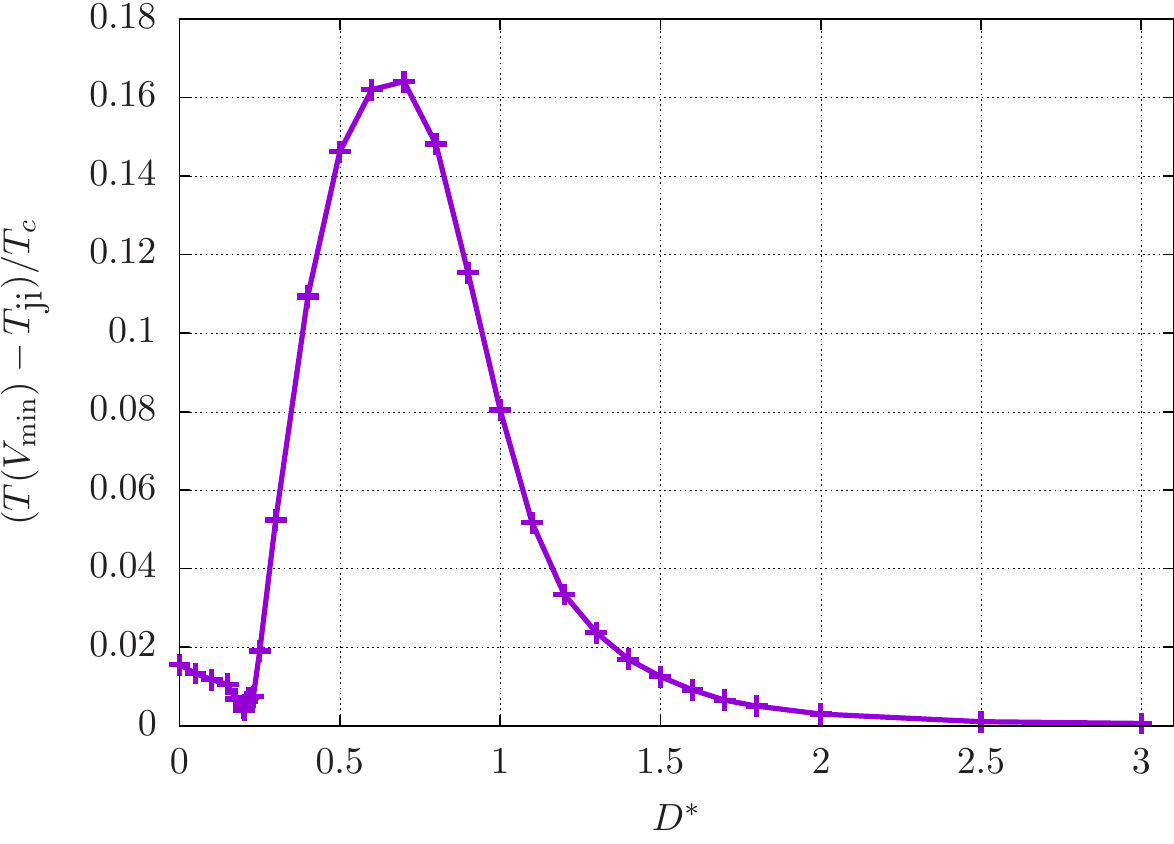}
\hspace{2ex}
\includegraphics[width=0.45\textwidth]{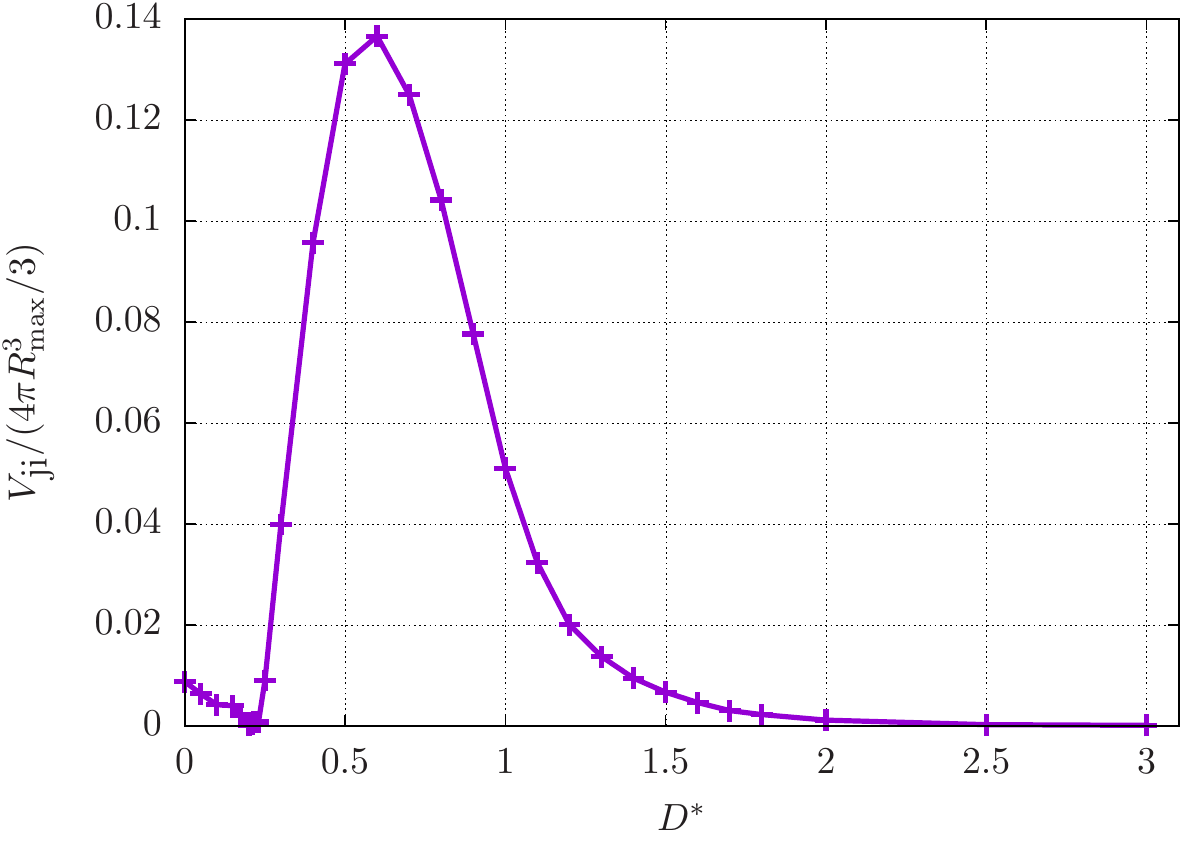}
}
\end{arXiv}
\caption{
(colour online)  
Jet--bubble relationships. Left: Normalized time difference between the time from the maximum volume of the bubble to the main torus-bubble collapse, $T(V_{\rm min})$,  and the time of microjet impact onto the lower bubble wall, $T_{\rm ji}$, in dependence on $D^*$ between 0 and 3. 
Right: Normalized volume of the bubble at jet impact onto the lower bubble wall. 
Surface tension 
$\sigma = 0.0725$\, N\,m$^{-1}$ (water)
for $D^*$ up to 0.4. 
}
\label{fig:TimediffDstar_ed}
\end{figure}

Figure \ref{fig:TimediffDstar_ed}, left diagram, shows the jet earliness parameter $\Delta T_{\rm ji}/T_{\rm c}$ versus $D^*$ and  
figure \ref{fig:TimediffDstar_ed}, right diagram, shows 
 $V_{\rm ji}$, normalized to the maximum bubble volume of the corresponding spherical bubble, versus $D^*$. 
Both curves, for $\Delta T_{\rm ji} = T(V_{\rm min}) - T_{\rm ji}$  and for $V_{\rm ji}$, show a similar behaviour. 
For large distances, the time difference $T(V_{\rm min}) - T_{\rm ji}$ is small. 
At $D^* =$ 3, the difference is positive, but already almost zero, when normalized to the Rayleigh collapse time, $T_c = 0.915 R_{\rm max} \sqrt{\rho_l/p_\infty}$ = 45.4 $\upmu$s  ($R_{\rm max} = 500\,\upmu$m,\ $\rho_l =
998\ {\rm kg}\,{\rm m}^{-3}$, 
$p_\infty = 101315$ Pa), i.e., jet impact and bubble collapse occur almost simultaneously. The
corresponding bubble volume, 
$V_{\rm ji}$, then must also be small, again 
almost zero, when normalized to the bubble volume at its spherical maximum, $V_{\rm max,sphere} = 4\pi R_{\rm max}^3/3 =$ 0.524 mm$^3$.
This points to a strong collapse for large $D^*$.

The difference between the time of jet impact and bubble collapse first grows with decreasing $D^*$ from large values ($D^* = 3$), and also the bubble volume at jet impact increases. 
However, at $D^* \approx 0.6$, there is a maximum in these quantities, and from thereon both quantities get smaller. 
It is conjectured that the reason may be found in the change from the collapse of a nearly spherical bubble at maximum volume to one from an essentially hemispherical one, as seen from a comparison of the bubble shapes in figures \ref{fig:shapes} and \ref{fig:shapes1}. 
Indeed, below $D^* \approx 0.6$ to 0.7, 
the fast annular inflow is starting  that, additionally to the spherical inflow, 
diminishes the bubble volume, so that the bubble volume at jet impact gets smaller again. 
This correlates with $T(V_{\rm min}) - T_{\rm ji}$ getting smaller and also 
with $T(V_{\rm min})/T_{\rm c}$ starting to decrease.  

The bubble volume at jet impact, $V_{\rm ji}$, 
attains a (local) minimum 
at the boundary of the transition region $[D^*_{\rm f}, D^*_{\rm s}] = [0.2, 0.24]$ with respect to the slow and the fast jet. 
$V_{\rm ji}$ gets larger again below $D^*_{\rm f}$, because the annular inflow is then faster, arrives earlier for self-impact, and leaves a larger bubble volume at impact of the now fast jet. Again this correlates with $T(V_{\rm min}) - T_{\rm ji}$ getting larger. 

\begin{figure}  %fig15
\begin{journal}  
    \centerline{
\includegraphics[width=0.45\textwidth]{figure15a.eps}
\hspace{2ex}
\includegraphics[width=0.45\textwidth]{figure15b.eps} 
}
\end{journal}
\begin{arXiv}  
    \centerline{
\includegraphics[width=0.45\textwidth]{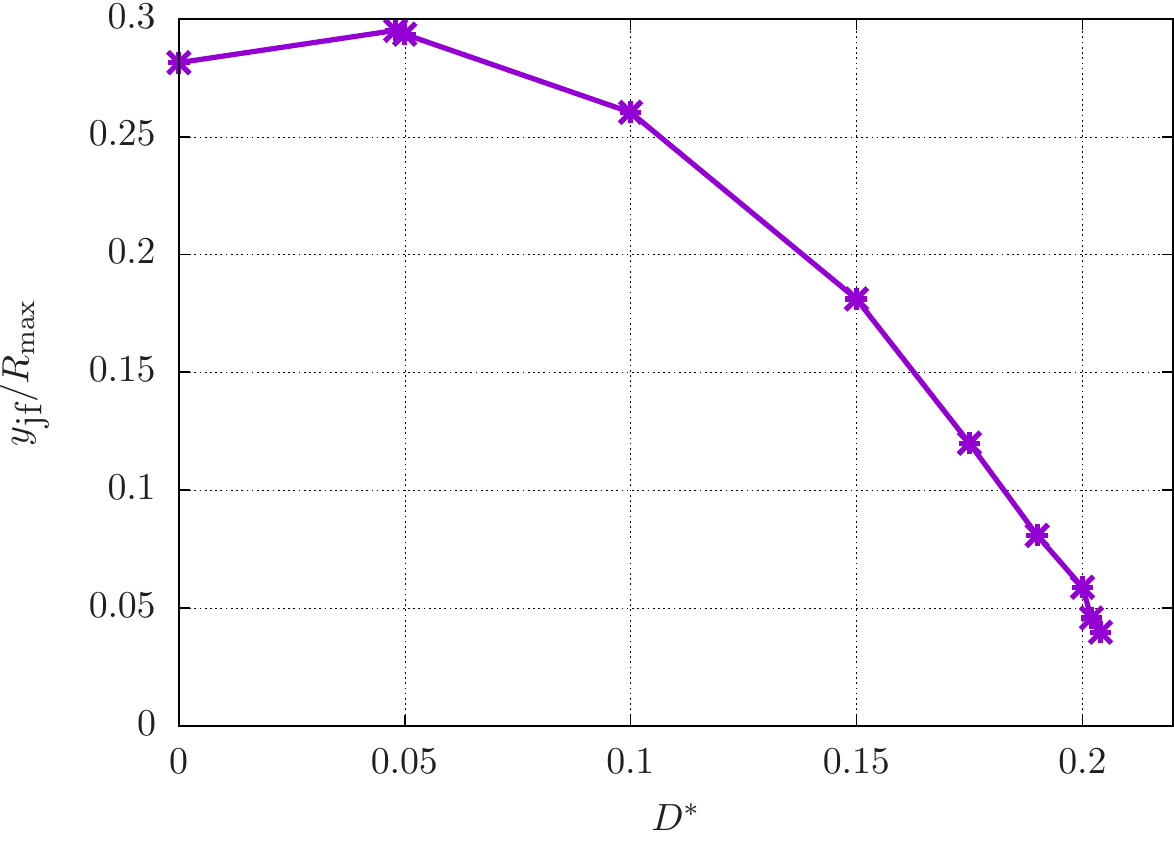}
\hspace{2ex}
\includegraphics[width=0.45\textwidth]{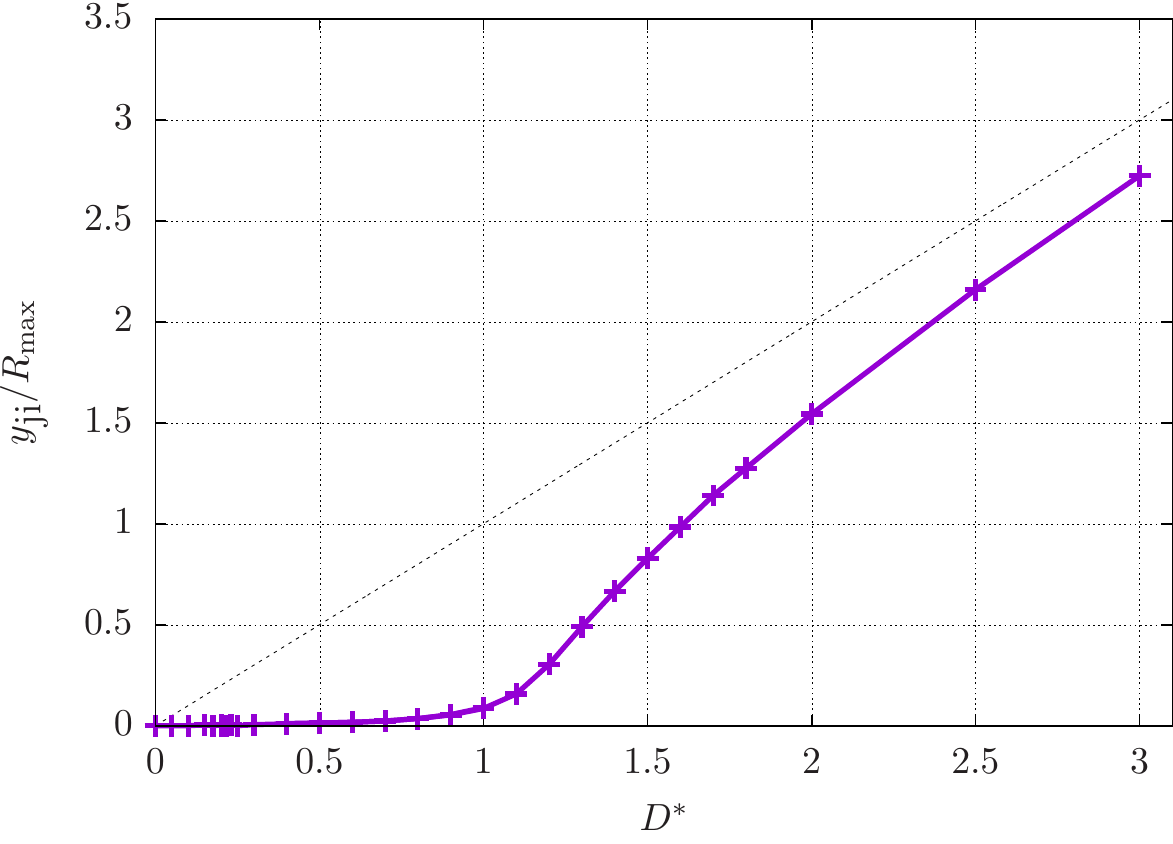} 
}
\end{arXiv}
\caption{(colour online) 
Jet--bubble relationships. 
   Left: Normalized distance from the solid boundary of the impact site of the annular inflow at the axis of symmetry versus $D^*$ (fast jet formation point).
   Right: Normalized distance from the solid boundary of the impact site of the
 axial jet (slow or fast) onto the opposite bubble wall (jet impact point). The dashed line indicates the
 initial distance of the bubble centre from the solid boundary. 
Surface tension 
$\sigma = 0.0725$\, N\,m$^{-1}$ (water)
for $D^*$ up to 0.4. 
}
\label{fig:yaji}
\end{figure}

The fast jet starts, when the annular inflow hits the axis of symmetry before the flow along the axis towards the solid boundary arrives there. This can be shown quantitatively, as given in figure\,\ref{fig:yaji}. In the left diagram, the distance $y_{\rm jf}$ from the solid boundary of the impact site of the annular inflow at the axis of symmetry is given, normalized with $R_{\rm max}$. There is no annular self-impact for values of $D^*$ larger than about 0.204. For smaller $D^*$, the self-impact occurs farther and farther away from the solid boundary, i.e., also earlier and earlier before the final collapse of the bubble. $D^* = 0$ is a special case, as the initial conditions had to be adjusted. The slight deviation from an extrapolation of the values for $D^* > 0$ may find its explanation in this fact. 

On the right side of figure\,\ref{fig:yaji}, the normalized impact distance, $y_{\rm ji}$, of the jets (fast or slow) onto the lower bubble wall is given. Up to $D^* \approx 1$  the impact occurs near to the solid boundary, beyond, the curve rises grossly with the distance of bubble generation, up to $D^* \approx 2$ somewhat faster, then only marginally faster. 
Erosion from axial jet impact is to be expected only up to $D^* = 1$. For larger $D^*$, the cushioning effect of the liquid layer between impact site and solid boundary together with the spherical spreading of the impact pressure wave will become increasingly larger. This is valid only for the axial jet of the first bubble collapse. Torus bubble collapse would need a separate study. 

% 
%
%-----------------------

\begin{figure}  %fig16
\begin{journal}
\includegraphics[width=0.64\textwidth]{figure16.eps}
\end{journal}
\begin{arXiv}
\includegraphics[width=0.64\textwidth]{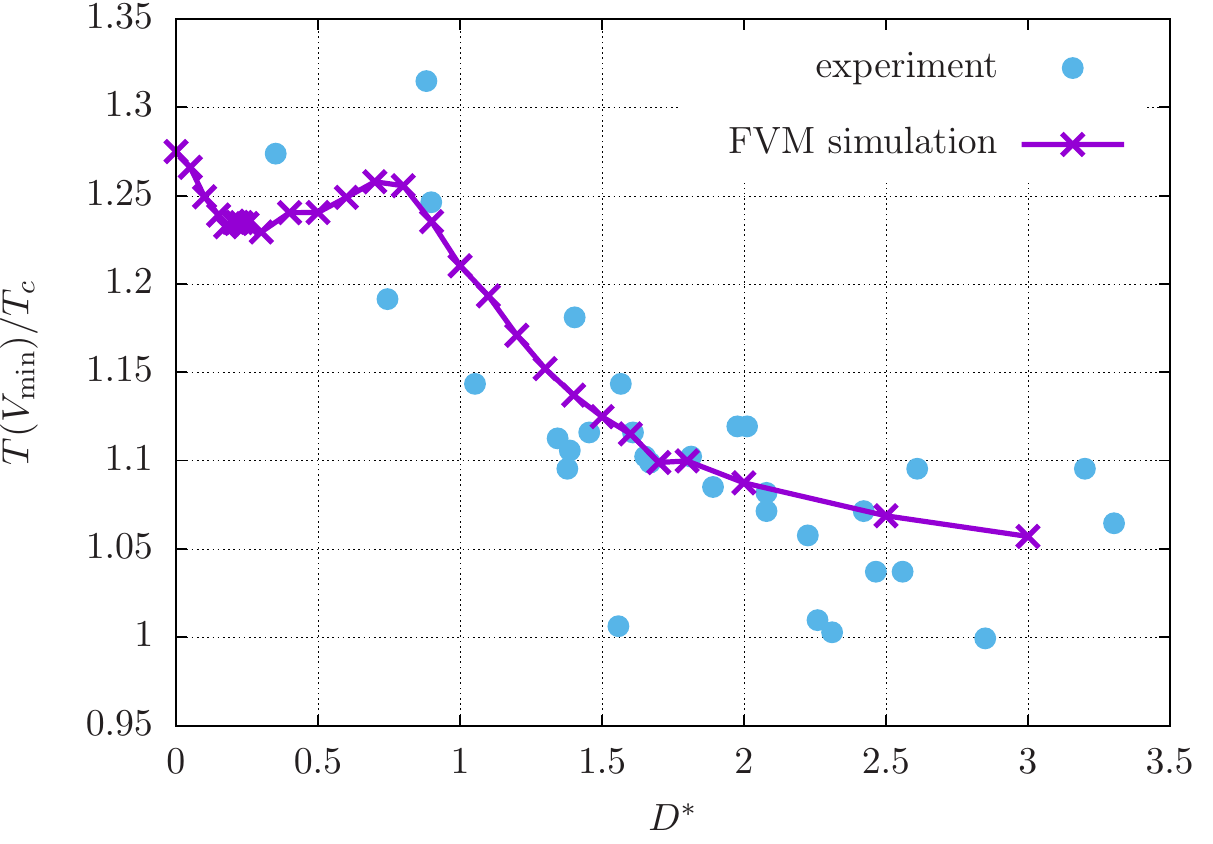}
\end{arXiv}
\caption{
(colour online)  
Normalized collapse time, $T(V_{\rm min})/T_{\rm c}$, versus $D^*$ in the interval $[0,3]$ in comparison with experimental data \citep{Vogel-1988a}.
Surface tension 
$\sigma = 0.0725$\, N\,m$^{-1}$ (water)
for $D^*$ up to 0.4. 
}
\label{fig:TimecollDstar_ed}
\end{figure}

Figure \ref{fig:TimecollDstar_ed} gives a comparison of the computed normalized bubble collapse time, $T(V_{\rm min})/T_{\rm c}$, with experimental data from laser-induced bubbles \citep{Vogel-1988a}. Similar experimental data exist with spark-induced bubbles 
\citep{Krieger-2005}. In view of the experimentally very scattered data base, the numerical data fit into the general experimental trend and give a refined view on the non-monotonous dependence of bubble collapse properties on $D^*$. It is interesting to note that the curve for $D^*$ oscillates around 1.26 ($=2^{1/3}$). This fact can be understood as a result of the relation $R_{\rm max,hemisphere} = 2^{1/3}\,R_{\rm max,sphere}$ and therefore $T_{\rm c,hemisphere} = 2^{1/3}\,T_{\rm c}$) for a bubble with the same volume, i.e., the same potential energy at maximum expansion (see Section \ref{subsec:D0}).

In figures  \ref{fig:TimediffDstar_ed} to \ref{fig:TimecollDstar_ed} 
the surface tension has only been included in the simulations up to $D^* = 0.4$ 
because the differences are marginal for larger $D^*$.

\subsection{Jet velocities}\label{subsec:jetvelo}
\begin{figure}  %fig17  
 \begin{journal}
\includegraphics[width=0.64\textwidth]{figure17.eps} 
 \end{journal}
 \begin{arXiv}
\includegraphics[width=0.64\textwidth]{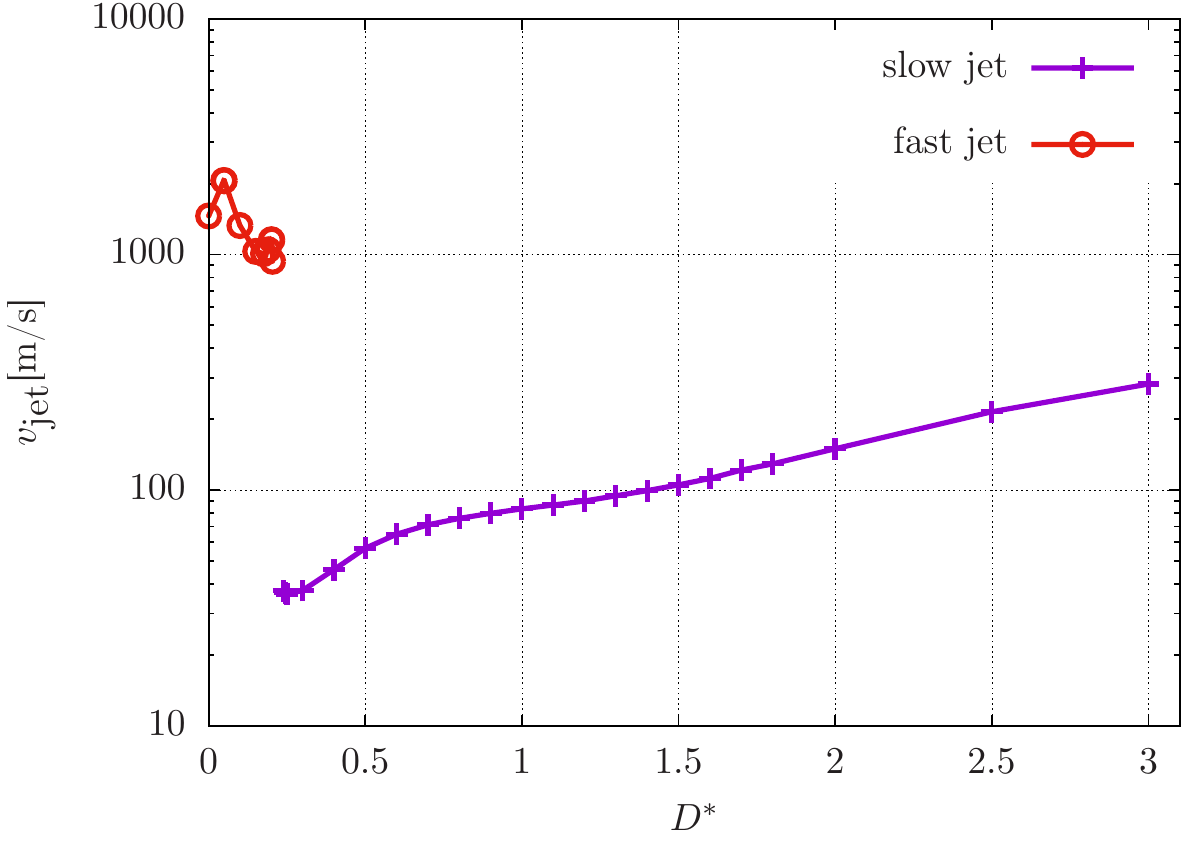} 
 \end{arXiv}
\caption{(colour online) 
Axial jet velocity, $v_{\rm jet}$, 
versus the normalized distance, $D^*$. Spherical 
flow focusing prevails down to $D^* \approx 0.24$ (slow jet). 
Below $D^* \approx 0.2$, the axial jet is generated by annular-inflow collision at the axis of symmetry 
(fast jet). 
No velocities are given for the transition region $0.2 < D^* < 0.24$, owing to the complicated, varying dynamics 
(see Section \ref{subsec:sf}). 
}
\label{fig:jetveloc_D}
\end{figure}

Results on the normalized distance dependence of the axial jet velocity 
are given in figure\,\ref{fig:jetveloc_D} for the range of $D^*$ from 0 to 3. 
%
%-----------------------
As the axial jet velocity is not constant 
along its total path through the bubble, but is starting slowly (standard axial jet) 
or exceedingly fast (fast axial jet), it must be defined how ``the jet velocity'' 
is to be determined. In the case of the standard axial jet by involution of the top 
of the bubble 
the velocity of the jet tip
is taken shortly before impact onto the opposite bubble wall. 
That is, because already shortly before impact the gas in the gap 
decelerates the jet by compression. 
As to the fast jet, the definition has been adopted to take the 
average velocity 
from annular jet collision up to jet impact.
This quantity is unambiguously defined. 
Time and location of the two impact events are determined 
from their respective pressure maxima. 

The (average) velocities of the fast jet given in figure\,\ref{fig:jetveloc_D}
are obtained with a resolution of $\Delta x_{\min} = 0.5\,\mum$ in the region of the
fast jet (mesh size 500\,000 cells). 
Additional local refinement shows that the 
velocity of the jet tip 
increases further with higher resolution. 
Therefore, the velocities of the fast jet given in figure 
\ref{fig:jetveloc_D} 
should be understood as lower bounds. 
This is considered a direct consequence of the nearly singular nature of the self-impact of the annular inflow at the axis and is discussed in Appendix \ref{app:jet}.

The velocity curve in figure\,\ref{fig:jetveloc_D} thus obtained consists of two separate parts, one for the slow axial jet for $D^* > 0.24$ and one for the fast axial jet for $D^* <0.2$. The range in between develops jets of different kinds, as described in
Section \ref{subsec:sf}, and must be left as a gap for the time being. It should be realized that the standard axial jet gets slower and slower, when the bubble is approaching the solid boundary to just below 40 m\,s$^{-1}$. These velocities are not sufficient to erode harder materials. The sudden jump to a more than 20-fold higher velocity is unexpected and owes its existence to the different jet formation mechanism as described in
Section \ref{subsec:fastjet}. These velocities are well apt to erode even strong materials.

\subsection{Influence of the viscosity on the fast jet at $D^* = 0$}\label{subsec:visc}
The viscosity of water can be considered to be small in many cases, but it is decisive for the fast jet. This finding will be studied in more detail in the present Section. The case $D^* = 0$ is chosen, because it is of interest in other areas, too, e.g., ophthalmology 
\citep{Vogel-2003}
and laser ablation with nanoparticle formation 
\citep*{Zhang-2017}.
The case of water with its dynamic viscosity of 1.002 $\times 10^{-3}$ Pa\,s has been treated in Section \ref{subsec:D0}. There, it has been stated that without viscosity no fast jet is generated. It may be added that without viscosity no jet is possible at all (at $D^* = 0$), because the motion is perfectly (hemi)spherical. Viscosity, however, is damping any flow, and thus the role of viscosity in this case of enabling extremely fast flows surely is strange and of utmost interest. 
For some more insight as to the role of viscosity, simulations with increased viscosity have been done. The outcome is as expected: in a liquid with higher and higher viscosity the fast jet gets slower and slower. However, quite high viscosities with respect to water are needed for the fast jet to slow down and disappear. 
%-----------------------
\begin{figure}  %fig18   
%\vspace{-4ex}
 \begin{journal}
    \centerline{
\includegraphics[width=0.45\textwidth]{figure18a.eps} 
\includegraphics[width=0.45\textwidth]{figure18b.eps} 
}%\vspace{-8ex}
    \centerline{
\hspace{2ex}\includegraphics[width=0.45\textwidth]{figure18c.eps} 
\hspace{0ex}\includegraphics[width=0.46\textwidth]{figure18d.eps} 
}%\vspace{-5ex}
 \end{journal}
 \begin{arXiv}
    \centerline{
\includegraphics[width=0.45\textwidth]{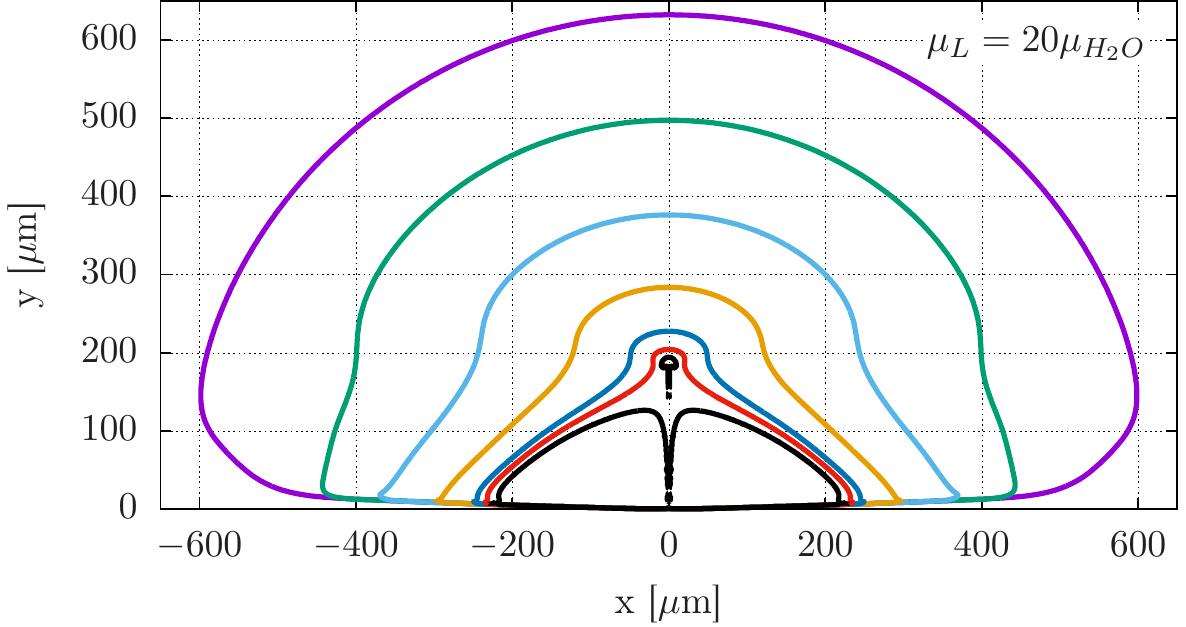} 
\includegraphics[width=0.45\textwidth]{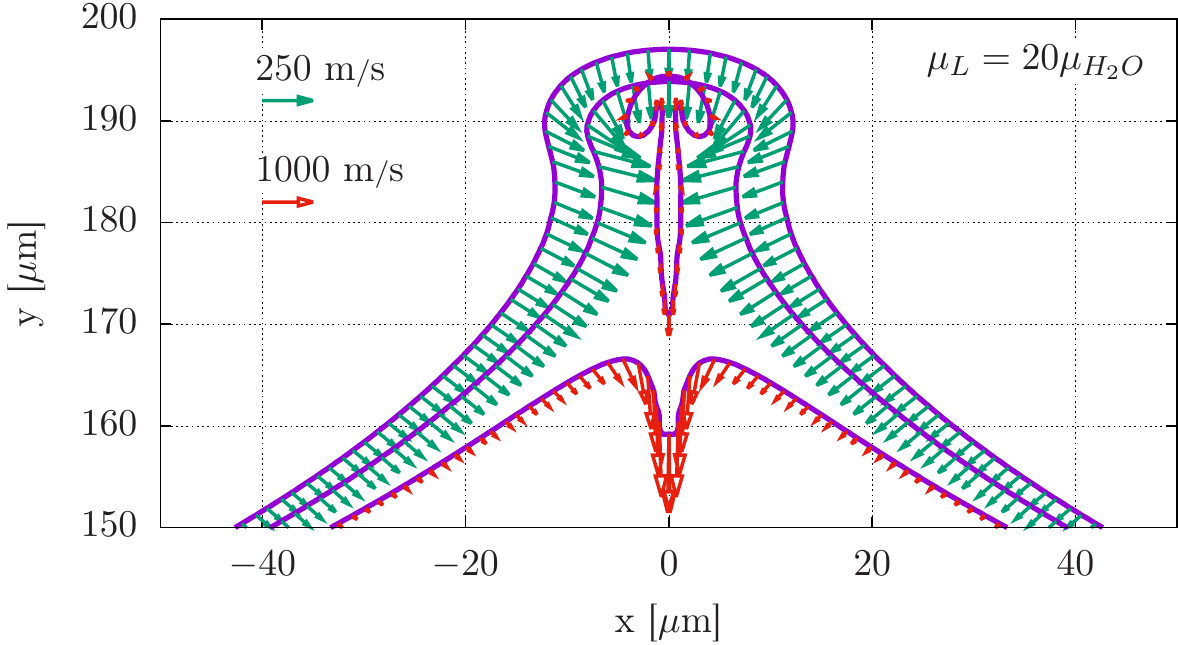} 
}%\vspace{-8ex}
    \centerline{
\hspace{2ex}\includegraphics[width=0.45\textwidth]{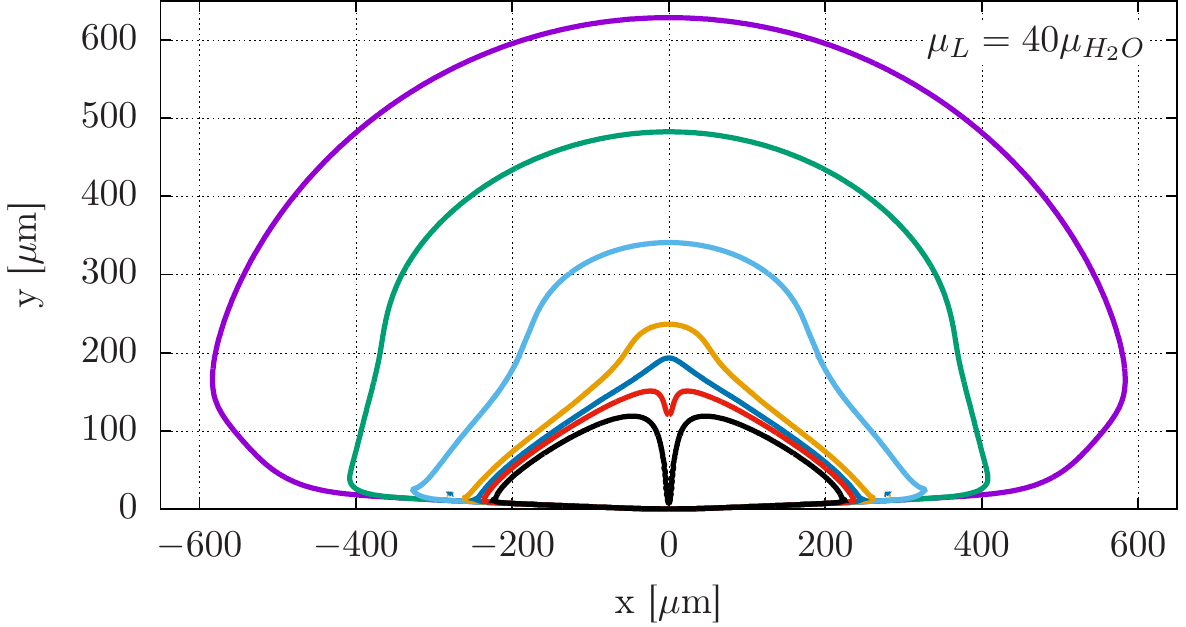} 
\hspace{0ex}\includegraphics[width=0.46\textwidth]{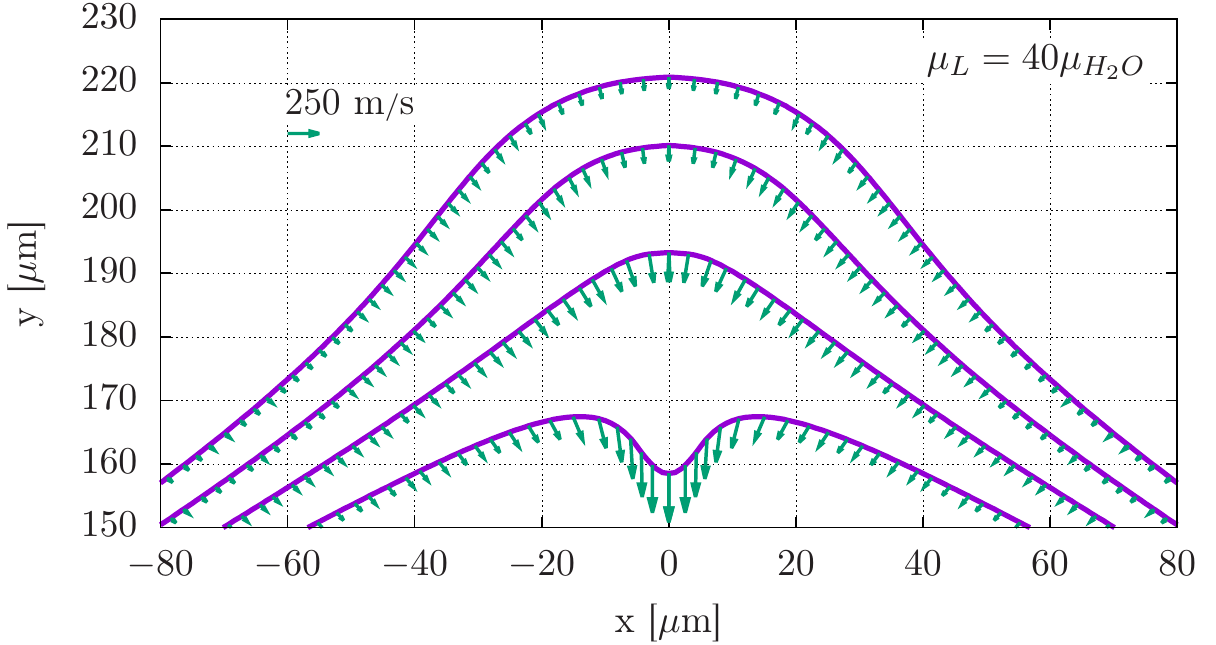} 
}%\vspace{-5ex}
 \end{arXiv}
\caption{(colour online) 
Bubble shape dynamics with higher viscosities than water. $D^* = 0$,
$\sigma = 0.0725$ N\,m$^{-1}$. 
Upper row: $\mu_{\rm l} = 20\,\mu_{\rm water}$, bubble shapes at $t =$ 55, 95, 105, 109, 110.3, 110.6, and 110.9
$\mus$ (left) 
and bubble shapes with bubble wall velocities at $t =$ 110.65, 110.67, and 110.7
$\mus$ (right).  
Lower row: $\mu_{\rm l} = 40\,\mu_{\rm water}$, bubble shapes at $t =$ 55, 95, 105, 108, 108.4, 108.6, and 108.9
$\mus$ (left) 
and bubble shapes with bubble wall velocities at $t =$ 108.2, 108.3, 108.4, and 108.5
$\mus$ (right). 
}
\label{fig:viscmultwater}
\end{figure}
In figure\,\ref{fig:viscmultwater}, bubble shapes upon collapse from maximum expansion and selected bubble wall velocities near (fast) jet formation are given for a liquid with a viscosity of 20 times that of water (figure\,\ref{fig:viscmultwater}, upper row) and 40 times that of water (figure\,\ref{fig:viscmultwater}, lower row). At a viscosity of $\mu_{\rm l} = 20\,\mu_{\rm water}$, still a fast jet is formed by fast inflow of an annular jet with self-impact. When doubling the viscosity to $\mu_{\rm l} = 40\,\mu_{\rm water}$, the annular inflow is not fast enough for self-impact and a typical slow jet is formed by involution of the top of the bubble. 
The example is for $D^* = 0$. A thorough investigation of the influence of viscosity (different liquids) for the whole range of normalized bubble distances $D^*$ is beyond the scope of the present study. 
However, some more thoughts about the influence of viscosity on jet formation 
are given in the discussion section (Section \ref{sec:discussion}) together with a discussion 
of slow- and fast-jet formation.

\subsection{Pressure and velocity fields} \label{subsec:fields}

The simulations also deliver pressure and velocity fields throughout the liquid and the gas that give additional insight, why the bubble changes its shape in the characteristic ways. Typical examples are again chosen from the range 0 $\le D^* \le$ 3. 
%
%++++++++++++++++++++++++++++++++++++++
\begin{figure}  %fig19
 \begin{journal}
\includegraphics[width=0.9\textwidth]{figure19.eps} 
 \end{journal}
 \begin{arXiv}
\includegraphics[width=0.9\textwidth]{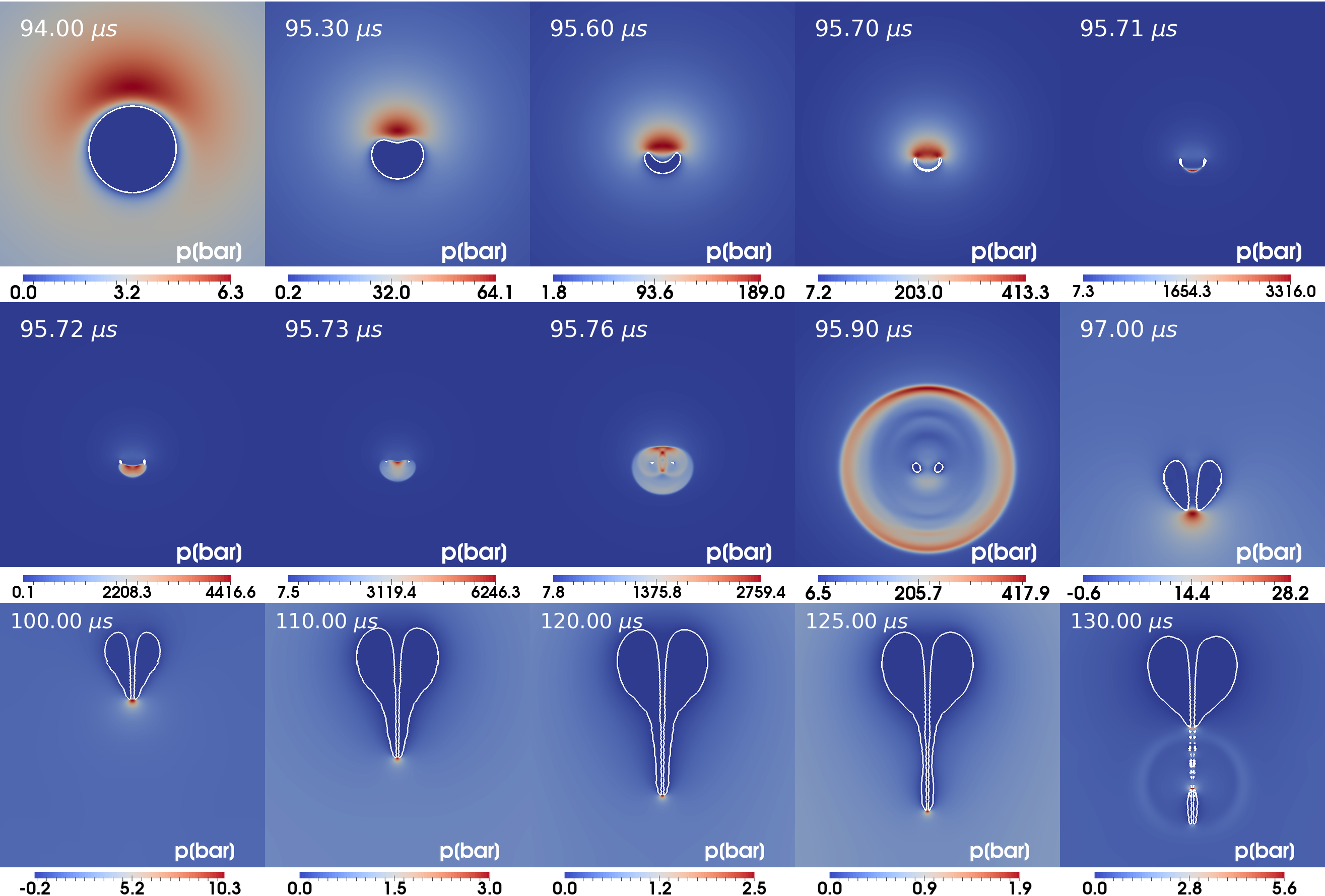} 
 \end{arXiv}
\caption{(colour online) 
Pressure fields for $D^* = 3$. Overview of the first collapse and rebound  with jet formation, 
jet impact with shock-wave radiation, torus-bubble collapse with torus-bubble shock-wave emission and rebound up to jet disintegration. 
The jet  is broad in relation to the bubble dimension at the time of impact, but thin in relation to the maximum rebound radius. 
In the upper two rows the frame size is 1 mm $\times$ 1 mm and the solid boundary is located 1 mm below the lower border of each frame.
In the bottom row the frame size is 1.6 mm $\times$ 1.6 mm
and the solid boundary coincides with the lower border of each frame.  
}
\label{fig:ed_p_Dstar3jet}
\end{figure} 

Figure \ref{fig:ed_p_Dstar3jet} shows pressure fields for the collapse and rebound phase of a bubble 
at $D^* =$ 3. 
The value of $D^*$ has been chosen for validation of the code in figure\,\ref{fig:valcompD3} by comparison with experiments. The time resolution in the experiment was limited to 1 $\mus$ interframe time. Therefore, the fast collapse phase is not resolved. The simulations do not suffer from this restriction and an almost arbitrary time resolution can be reached. That way, a kind of interpolation of the dynamics between the frames can be done. At least a time resolution of 10 ns was found to be necessary to about resolve the very collapse phase in this case. 
The series in figure\,\ref{fig:ed_p_Dstar3jet} starts some time after expansion to maximum volume of the bubble, when a high-pressure concentration on top of the bubble has set in and the bubble top is starting to involute. This high-pressure region persists up to at least jet impact at 95.71 $\upmu$s. 
The jet impact and the immediate post-impact phase  are covered in steps of 10 ns from 95.70 to 95.73 $\mus$. 
Torus-bubble collapse takes place only 50 ns after jet impact, at about 95.76 $\upmu$s, where both the pressure wave from jet impact and the torus pressure wave from the (torus) bubble collapse are visible. Due to the short time difference of only 50 ns between both events essentially 
one shock wave is finally propagating into the liquid (see frame at 95.9 $\upmu$s in figure\,\ref{fig:ed_p_Dstar3jet}). Upon re-expansion (rebound) of the bubble after collapse the jet advances towards the solid boundary with a gaseous hull. The long, thin jet disappears, when the gaseous hull finally disintegrates into tiny (torus) bubbles.

\begin{figure}  %fig20
\begin{journal}
\includegraphics[width=0.9\textwidth]{figure20.eps} 
 \end{journal}
\begin{arXiv}
\includegraphics[width=0.9\textwidth]{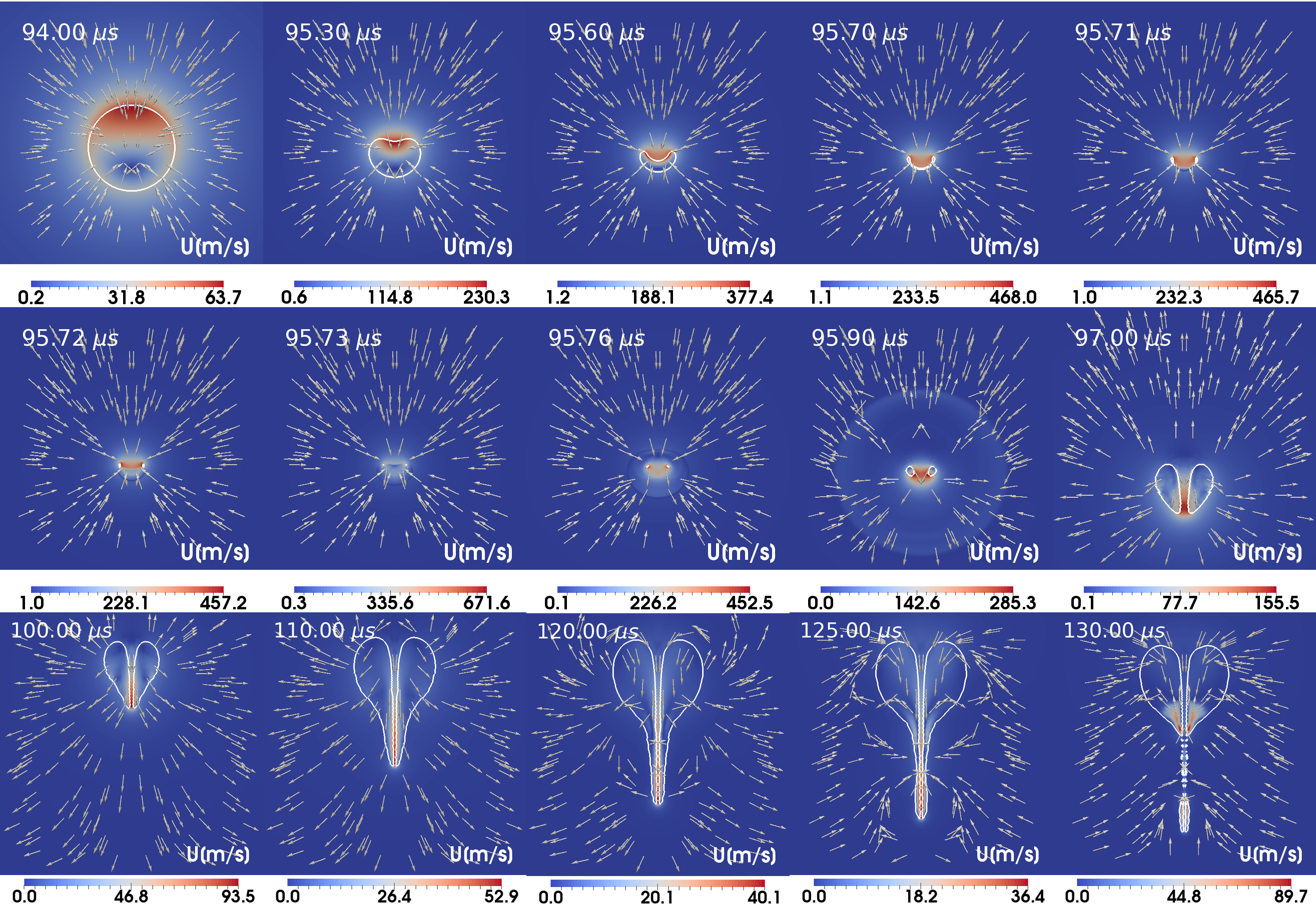} 
 \end{arXiv}
\caption{(colour online) 
Velocity fields for $D^* = 3$. Overview of the first collapse and rebound  with jet formation, 
jet impact, torus-bubble collapse and rebound up to jet disintegration. 
Frame sizes are the same as in figure\,\ref{fig:ed_p_Dstar3jet}. 
Note that the maximum velocity on the scale bars are not necessarily the maximum liquid jet velocity in the respective frame.   
}
\label{fig:ed_U_Dstar3jet}
\end{figure}

Figure \ref{fig:ed_U_Dstar3jet} shows velocity fields for the collapse and rebound phase of a bubble 
at $D^* =$ 3. The series matches the time instants of the pressure fields in figure\,\ref{fig:ed_p_Dstar3jet}. 
%
%-----------------------
The velocity scales show very high maximum velocities, up to 671.6 m\,s$^{-1}$ (at 95.73 $\upmu$s). 
These are not necessarily 
the maximum jet velocity at the time instant of the frame (for instance at 100 $\upmu$s), but may be also 
either gas velocities inside the bubble (as at 94 $\upmu$s) or bubble collapse velocities (as at 95.76 $\upmu$s). 
Careful inspection of the velocity field at 95.9 $\upmu$s reveals that the shock wave on the respective frame in 
figure\,\ref{fig:ed_p_Dstar3jet} is also visible in the velocity field by the change of the arrow direction. 
In front of the shock wave, the flow direction points towards the bubble, behind the shock wave away from it.
The jet extends almost down to the solid boundary and carries a gaseous hull with it that develops surface undulations. 
The maximum jet velocity occurs at the tip of the jet (actually shortly before the tip and as long as it is coherent) and continuously decreases 
as the jet advances through the liquid. The collapse of the individual tiny torus bubbles 
formed upon disintegration of the hull again generates higher velocities in the diagrams (and pressures, 
as can be read from the respective frames in figure\,\ref{fig:ed_p_Dstar3jet}). 

%
%-----------------------
\begin{figure}  %fig21  
 \begin{journal}
\includegraphics[width=0.9\textwidth]{figure21.eps} 
 \end{journal}
 \begin{arXiv}
\includegraphics[width=0.9\textwidth]{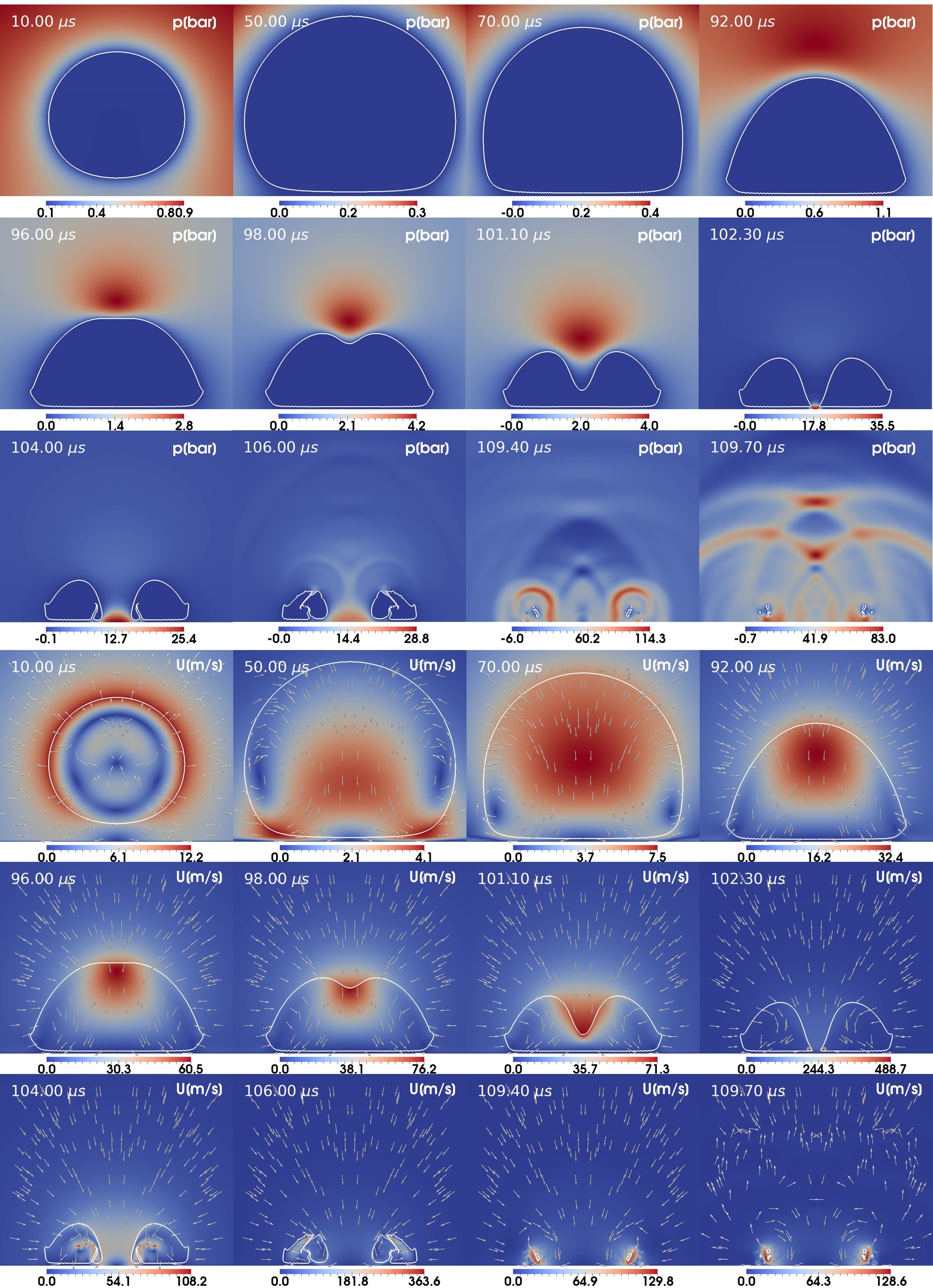} 
 \end{arXiv}
\caption{(colour online)  
Pressure (upper three rows) and velocity fields (lower three rows) for $D^* =$ 0.7. 
Bubble expansion, standard axial jet formation, torus bubble collapse with torus splitting by nanojet formation and the Blake splash, shock wave emission, and the very beginning of rebound with vortex formation. 
Frame size is 1.1 mm $\times$ 0.9 mm (width $\times$ height), the solid boundary coincides with the lower border of each frame. 
} 
\label{fig:dyn07}
\end{figure}
Figure \ref{fig:dyn07} gives an example of bubble expansion, axial jet formation, torus bubble collapse, and rebound 
for $D^* = 0.7$ with pressure (upper three rows) and velocity fields (lower three rows). This value of $D^*$ is special in that the normalized bubble volume at jet impact has about a maximum (figure\,\ref{fig:TimediffDstar_ed})  
and the bubble wall already gets significantly distorted during expansion to a more hemispherical shape. 
Nevertheless, still the standard axial jet by involution of the top of the bubble prevails, owing to the early pressure concentration ($\sim$4 bar) on top of the bubble (96 to 101.1 $\upmu$s). The velocity of the jet is about 70 m\,s$^{-1}$ on its way to the solid surface (101.1 $\upmu$s in the velocity diagram series). 
The high velocity of near 500 m\,s$^{-1}$ at 102.3 $\upmu$s can be ascribed to the nanojet 
\citep{Lechner-2017}
and the upward jet inside the bubble at 104 $\upmu$s to the Blake splash 
\citep{Tong-1999,Lauterborn-2018b}.
The Blake splash with the nanojet on top is shifted by the jet flow along the surface of the solid to the interior of the bubble (frames at 104 and 106 $\upmu$s). The impact of the axial jet onto the opposite bubble wall---besides the impact pressure---also generates a negative pressure zone of $-6$ bar (frame at 109.4 $\upmu$s) in a small region around the symmetry axis above the now torus bubble. 
The subsequent torus bubble collapse proceeds in three steps. At first, the torus bubble splits into mainly two torus bubbles of slightly different sizes through the action of the nanojet and Blake splash. Then the smaller torus bubble collapses first with emission of a torus pressure wave (109.4 $\upmu$s), finally the larger torus bubble collapses, again with emission of a torus pressure wave (109.7 $\upmu$s). The self-intersection and mutual intersection of the two torus pressure waves prominently shape the frame at 109.7 $\upmu$s. In the respective velocity diagram the formation of a vortex ring is visible. 

\begin{figure}  %fig22  
 \begin{journal}
\includegraphics[width=0.9\textwidth]{figure22.eps} 
 \end{journal}
 \begin{arXiv}
\includegraphics[width=0.9\textwidth]{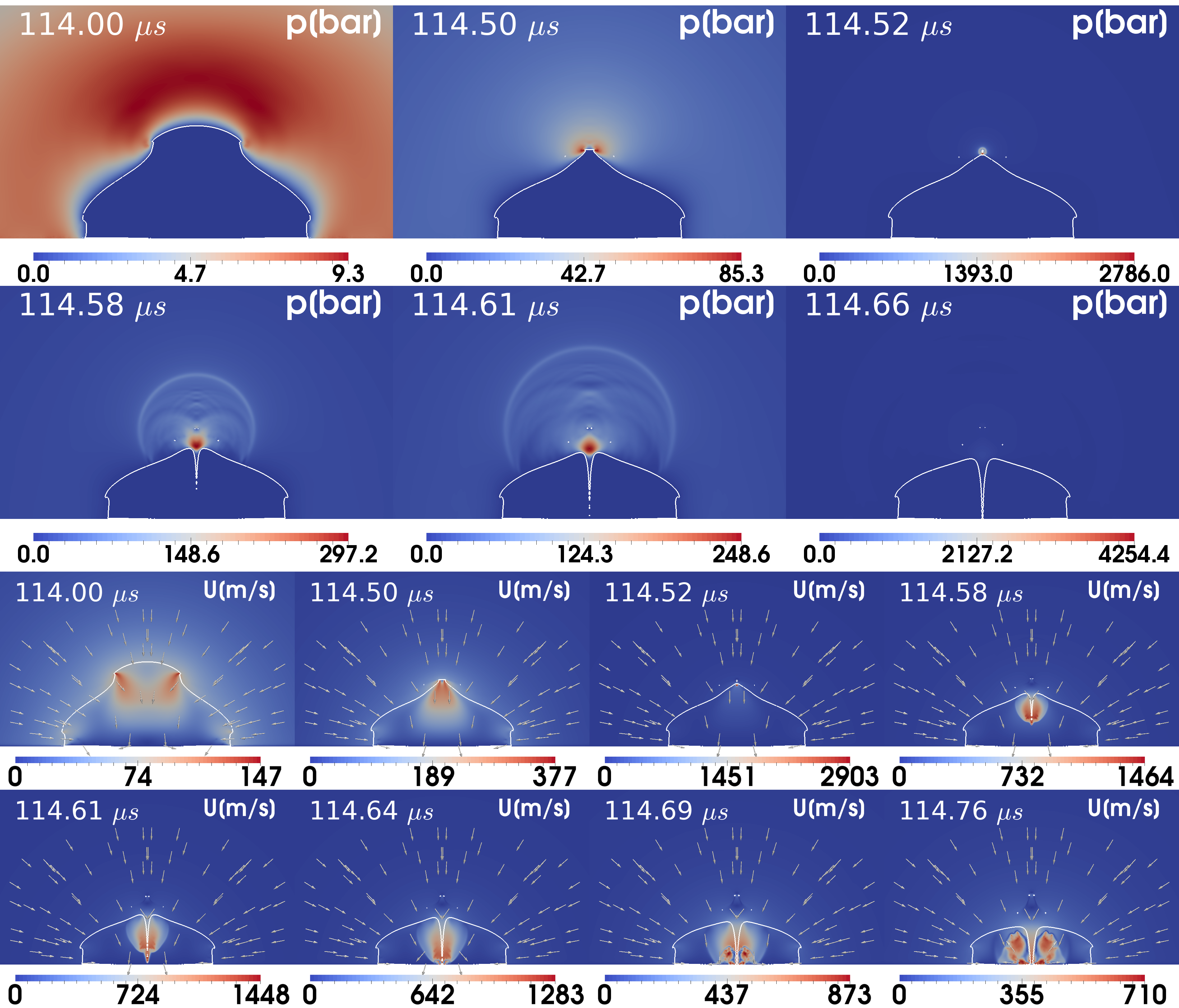} 
 \end{arXiv}
\caption{(colour online)  
Final phase of bubble collapse for $D^* =$ 0. Upper two rows: pressure fields. 
Note the high pressure above and aside the bubble that grows until the fast axial jet is formed around $t = 114.52\,\upmu$s. 
Lower two rows: velocity fields. Note the fast flows at the sharp curvatures that lead to the fast axial inflow with self-collision at the top of the bubble. 
Frame size is 0.64 mm $\times$ 0.38 mm (width $\times$ height), 
the solid boundary coincides with the lower border of each frame.  
} 
\label{fig:dyn00}
\end{figure}

Figure \ref{fig:dyn00}  shows pressure and velocity fields for $D^* =$ 0, i.e., for a bubble that develops the fast, thin jet to be seen in figure\,\ref{fig:wallvelo0}. 
%
%-----------------------
Only 660 ns of the collapse phase are presented from the beginning of the pressure concentration above the bubble to jet
impact.  The special bell shape is a result of annular inflow of liquid and is the precursor form that leads to the fast jet. 
The annular inflow leads to a very high pressure at the time of self-impact (2786 bar at 114.52 $\upmu$s). A strong pressure wave is radiated from the impact site that is best visible in the two subsequent frames. At this time, the jet has formed and is already visible. Driven by the high pressure at its base and the continuous inflow of liquid from the further collapsing bubble it heads for the solid boundary. When looking into the velocity fields, the high velocities at the sharp curvatures are noticed (frame at 114 $\upmu$s), later the fast (supersonic) shock wave in the gas inside the bubble (frame at 114.52 $\upmu$s and subsequent frames).

\section{Summary and discussion}\label{sec:discussion} 
The most prominent case of a bubble in front of a solid boundary has been taken for the present numerical study to cover the wide range of normalized distances $D^*$ between 0 and 3. 
As the simulations allow for easy parameter changes and experimentally difficult to access quantities (for instance pressure and velocity fields), numerical calculations may substantially enhance the understanding of the phenomena that show up upon bubble collapse, in particular jet formation. An example has been given here for $D^* = 3$. In figure\,\ref{fig:valcompD3}, the comparison between experimental and numerical shape dynamics is given with excellent agreement. In figures \ref{fig:ed_p_Dstar3jet} and \ref{fig:ed_U_Dstar3jet}, pressure and velocity fields for $D^* = 3$ are given in both higher temporal and higher spatial resolution than in the experiment. These diagrams may be considered as a kind of interpolation between the experimental frames in figure\,\ref{fig:valcompD3}. Such interpolations are always possible, when the respective governing equations are known 
\citep{Lauterborn-2018a},
in the present case the Navier-Stokes equations.  

Successful comparisons also give confidence for extension to other parameter values, here the range of $D^*$ below $D^* = 3$ down to zero. Bubble shapes upon expansion and collapse have been given for this range with remarkable results. With the change of an essentially spherical bubble collapse at larger distances from the solid boundary to a near hemispherical collapse at very small distances the type of the jet formation mechanism alters from involution of the bubble top to self-impact of an annular inflow. How this transition comes about, is discussed in the 
following.

In Section \ref{subsec:D0}, where the case $D^* = 0$ has been treated, 
it has been found that it is the viscosity of the liquid ($\mu_l = \mu_{\rm water}$) 
that initiates 
the sequence of events that leads to  
the fast jet, because at $\mu_l = 0$ the bubble collapses perfectly hemispherically without any jet. 
This is valid for both the energy-deposit boundary case as studied here and the Rayleigh boundary case 
(initial half sphere at maximum volume, see also \cite{Lauer-2012}).
To discern between the two different cases that generally give different results, the above nomenclature has been introduced by 
\citet{Lauterborn-2018b}.
It is the viscous boundary layer that the bubble generates upon expansion that makes the difference to the case of $\mu_l = 0$. 
Including viscosity, the bubble wall gets strongly different curvatures. 
 As a side effect, this leads to strongly different surface tension forces along the bubble wall. 
This fact evokes the question, whether it is the surface tension that subsequently generates the fast jet. 
This can be proven numerically by switching off the surface tension.
The result of the calculation is unequivocal. Only slight differences are observed between the calculations with and without surface tension. 
Thus, surface tension does not play a decisive role.

The only further forces involved (gravity omitted) are inertial forces. To probe the action of the inertial forces, the following numerical experiment has been conducted. The shape of the fully expanded bubble with action of the viscosity via the boundary layer 
is taken and placed in a free liquid
(without solid boundary). 
The subsequent dynamics (viscosity included) delivers the fast jet. 
Thus it can be stated that the fast jet 
is a cooperative phenomenon of viscosity and inertia, viscosity shaping the bubble via its boundary layer and inertia delivering the energy for the jet via subsequent flow focusing. 

But how does inertia proceed to concentrate the energy into a jet? For that, the curvature of the bubble interface with the liquid comes into play, but not via surface tension (see the numerical experiment mentioned above), instead via spherical and/or cylindrical convergence of the flow, i.e., flow focusing. \textit{Cum grano salis} the principle holds: More highly curved parts of a bubble collapse faster, i.e., attain a higher velocity faster than less curved parts 
\citep{Lauterborn-1976, Lauterborn-1982}. 
This principle is elaborated in Appendix \ref{app:flow}. 
Axial flow focusing directly explains the slow jet and the decrease of its velocity with a decrease of $D^*$. That is, because with a decrease in $D^*$ the jet develops earlier during the collapse of the bubble, i.e., at a weaker curvature (larger radius of curvature,) and therefore attains a slower velocity. 
Quantitatively, the velocity of the axial, flow-focused standard jet drops from 282 m\,s$^{\rm -1}$ at $D^* = 3$ to 35 m\,s$^{\rm -1}$ at $D^* = 0.25$ (figure\,\ref{fig:jetveloc_D}). 

The formation of the fast jet needs a more involved explanation, as it consists of two steps. The rim of the bubble has a higher curvature than the other parts of the bubble. Therefore it collapses faster than the other parts of the bubble surface. The cylindrical convergence accelerates the flow on its way towards the axis of symmetry. When the flow arrives there earlier than the also concentrated flow along the axis of symmetry from the top of the bubble, the fast jet is born from the self-impact of the annular inflow. 
The birth of the fast jet gives rise to a shock wave radiated into the liquid and into the interior of the bubble.  
An example of shock-wave generation and propagation has been given in figure\,\ref{fig:dyn00} for $D^* = 0$. 

The slow and the fast jet occupy distinct ranges on the $D^*$-axis, the slow jet the range $D^* > 0.24$ 
(checked for $D^*$ up to 10), 
the fast jet the range $D^* < 0.2$ down to $D^* = 0$. 
The intermediate range $0.2 \le D^* \le 0.24$ is characterized by complicated flow structures with fast inbound and outbound annular jets. The gross picture has been presented in
Section \ref{subsec:sf}. 

It is astonishing that the fast jet has been detected numerically only recently 
\citep{Lechner-2019},
as its existence is known experimentally since the work of 
\citet{Benjamin-1966}
more than fifty years ago. The reason may be that the standard axial jet was considered as the right description of the experimental jet found by 
Benjamin \& Ellis. This would have obviated the need to search for another explanation. Moreover, numerical difficulties still prevent detailed results on the fast jets. 
The head-on collision of the cylindrically converging 
bubble wall
presents a numerically challenging problem, despite the collision singularity is removed by the compressibility and viscosity of the liquid. Therefore, only lower limits of the velocity of the fast jets could be presented (figure\,\ref{fig:jetveloc_D}). 
These lower limits, however, are already more than tenfold higher than the heretofore known jet velocities and will strongly alter the view on the erosion problem.

It has been found in the present study that compressibility and viscosity strongly 
influence the flow and cannot be neglected.
In particular, the role of viscosity in enabling the fast jet is surprising, but has been made comprehensible by a set of additional simulations as described above in this Section. Compressibility enters the fast axial jet forming process in the annular inflow collision that determines how the axial jet is formed by redirecting the flow from annular to axial. Again, the curvature of the colliding flow is conjectured to play an essential role. It is reminiscent of drop impact 
\citep{Visser-2015},
where the curvature and the speed of the droplet and, moreover, the gas in between drop and impacted surface play a role. Enclosed small gas bubbles upon drop impact have  been observed between drop and impact site. The same phenomenon may occur in the case of fast jet formation upon annular jet collision (figure\,\ref{fig:viscmultwater}). They additionally cushion the impact when appearing 
and influence the initial jet velocity.
The experimental  measurement of the fast jet is a challenging task. The bubble is very small when the fast jet is formed, and a look into the interior is difficult. In this context, the approach to bore a little hole into the solid plate representing the solid boundary and squeeze the jet through it to an observable space may be an approximation to the jet velocity at impact. This has been done with an electric spark across the plate at the hole entrance from the bubble side 
\citep*{Gonzalez-2015}.
This experiment corresponds to $D^* = 0$. A hemispherical bubble is generated as with the laser-induced bubbles here. As there is liquid in the hole, the expansion of the 
bubble shoots the liquid in the hole outwards in the form of a jet 
(the first jet in the indicated reference). More interesting for the present study is the second jet exiting the hole in the collapse phase of the bubble. 
The jet velocity increases with the bubble radius. It
grows from about 200 m\,s$^{-1}$ for bubbles with a maximum radius as in the
present study to
about 400 m\,s$^{-1}$ for a bubble of $\approx$1\,mm in radius. These values point to a fast-jet-formation mechanism. However, no indications of the type of jet formation can be found in the pictures. The interframe times are too low. Because in the bubble collapse phase the bubble contents (gas and vapour including the sucked-in gas) are blown out through the hole, 
the collapse mechanism may be different from the one found here. 
The bubble collapse cannot be stopped by compression of the gas and/or vapour, and the rushing-in liquid finally escapes through the hole at high velocity. This would be a separate mechanism that delivers fast jets, one with a hole in the solid needed to generate them. Only further experiments or simulations can decide on the real jet formation mechanism. 

High jet velocities of $\sim$1000 m\,s$^{\rm -1}$ are also obtained in shock-compressed bubbles \citep{Hawker-2012}. The shock-wave--bubble case, however, is a topic of its own with a large literature that will not be discussed here further. Just to mention the subclass of shock-wave lithotripsy or, generally, the shock-wave--boundary-bubble case,  
where the bubbles are located near or on boundaries. The jet formation then is determined by both the shock wave and the boundary 
\citep[see, e.g.,][]{Wolfrum-2002, Wolfrum-2003, Johnsen-2008, Sankin-2006}.
It must be noted, however, that the study of the shock-wave--boundary-bubble case is still in its starting pits.

\section{Conclusions}\label{sec:conclusions} 
The distance dependence of bubble dynamics, in particular jet formation, has been studied numerically by solving the Navier-Stokes equations for a model of a laser-induced bubble in water in front of a flat solid boundary. Bubble shapes, pressure and velocity fields, and jet velocities are given for the range of the normalized distance $D^*$ between 0 and 3. Numerically obtained bubble shapes are compared with experimental ones including bubble rebound for $D^* = 3$. Very good agreement is found validating the model and the numerical code written with the help of the open software environment OpenFOAM. Two types of axial liquid jets are found, a relatively slow, broad one ($\sim$100 m\,s$^{\rm -1}$) for 0.24 $\le D^* \le$ 3 (upper limit of the present study) and a fast, thin jet ($\sim$1000 m\,s$^{\rm -1}$) for 0 $\le D^* \le 0.2$, with $D^* =0$ being the lower limit of the present study. In the transition region ($0.2 <D^* < 0.24$), the jet formation mechanism changes from 
axial flow focusing (slow, broad jet)
to an annular-liquid-flow collision one (fast, thin jet). The transition region is characterized by additional inbound and outbound annular jets with velocities reaching ~1000 m\,s$^{\rm -1}$. The inclusion of the viscosity of water is essential for the fast jet. 
High viscosities of more than thirty times the viscosity of water are needed to change the fast jet back to a jet by axial flow focusing. The high jet velocities found for bubbles very near to the solid boundary are expected to alter the view on the erosion problem in cavitation. \\  

\begin{acknowledgments}
The authors thank the Cavitation Bubble Dynamics Group at the Institute 
for many inspiring discussions. The work was supported in part by the Deutsche Forschungsgemeinschaft 
(German Research Foundation) under contract Me 1645/8-1. 
C.\ L. thanks H.\ C.\ Kuhlmann and the Institute of Fluid Mechanics and Heat Transfer, TU Wien, 
for their hospitality. 
Computational resources on the cae cluster of 
TU.it,
TU Wien, are gratefully acknowledged. 
\end{acknowledgments}

\appendix
\section{Grid independence studies}\label{appA}

The code has been validated in 
\citet{Koch-2016},
both in the spherically
symmetric case (for a bubble in an unbounded liquid) and the axially symmetric case 
(for a bubble collapsing next to
a solid boundary, $D^* \simeq 1.4$). In both cases excellent agreement 
was found 
concerning the bubble shape and location as functions of time when comparing to
experimental data. For the present investigations, 
grid independence 
is demonstrated separately for both the Cartesian and polar grids 
(figure~1) using 
the examples of $D^*=1.2$ (Cartesian grid) and $D^* = 0$ (polar grid).

\begin{figure}%[!h]  %fig23
\begin{journal}
\centerline{
\includegraphics[width=0.46\textwidth]{figure23a.eps}
\raisebox{-1ex}{\includegraphics[width=0.45\textwidth]{figure23b.eps}}
}
\end{journal}
\begin{arXiv}
\centerline{
\includegraphics[width=0.46\textwidth]{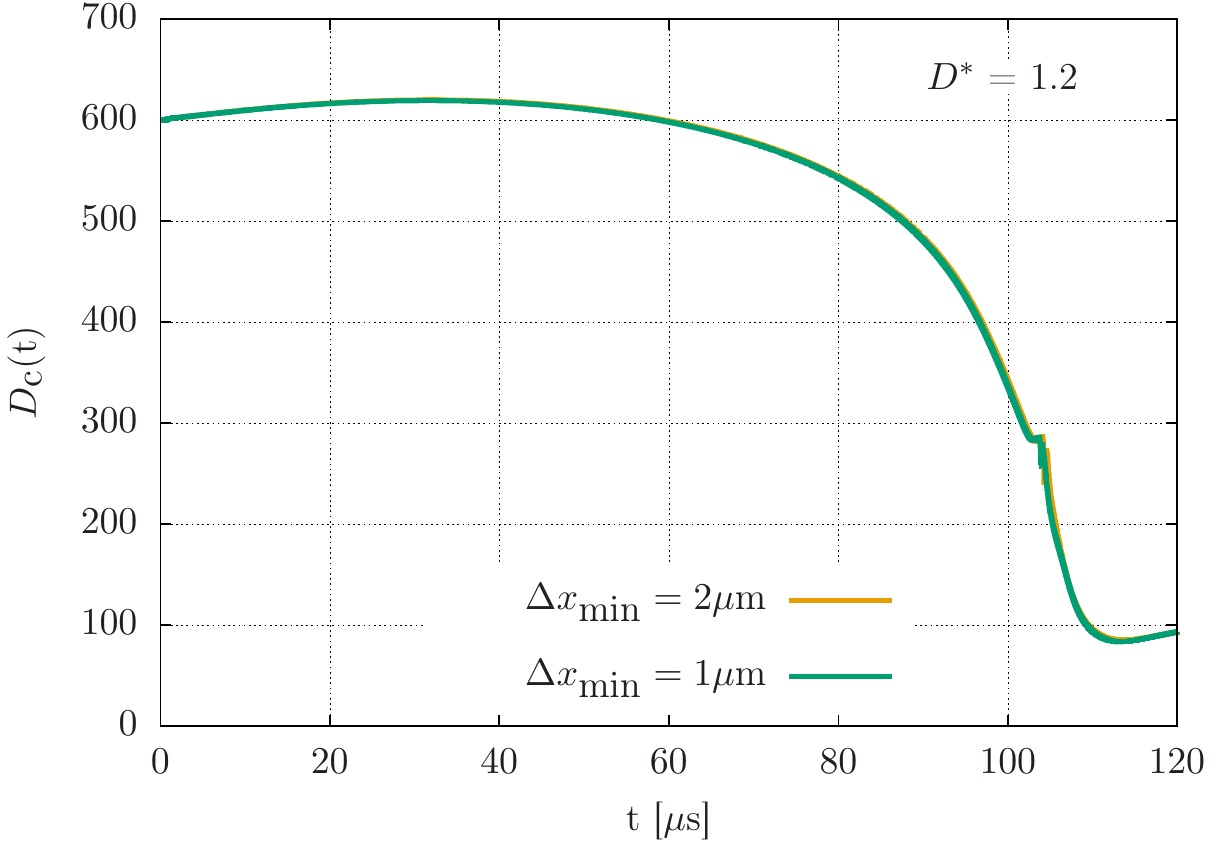}
\raisebox{-1ex}{\includegraphics[width=0.45\textwidth]{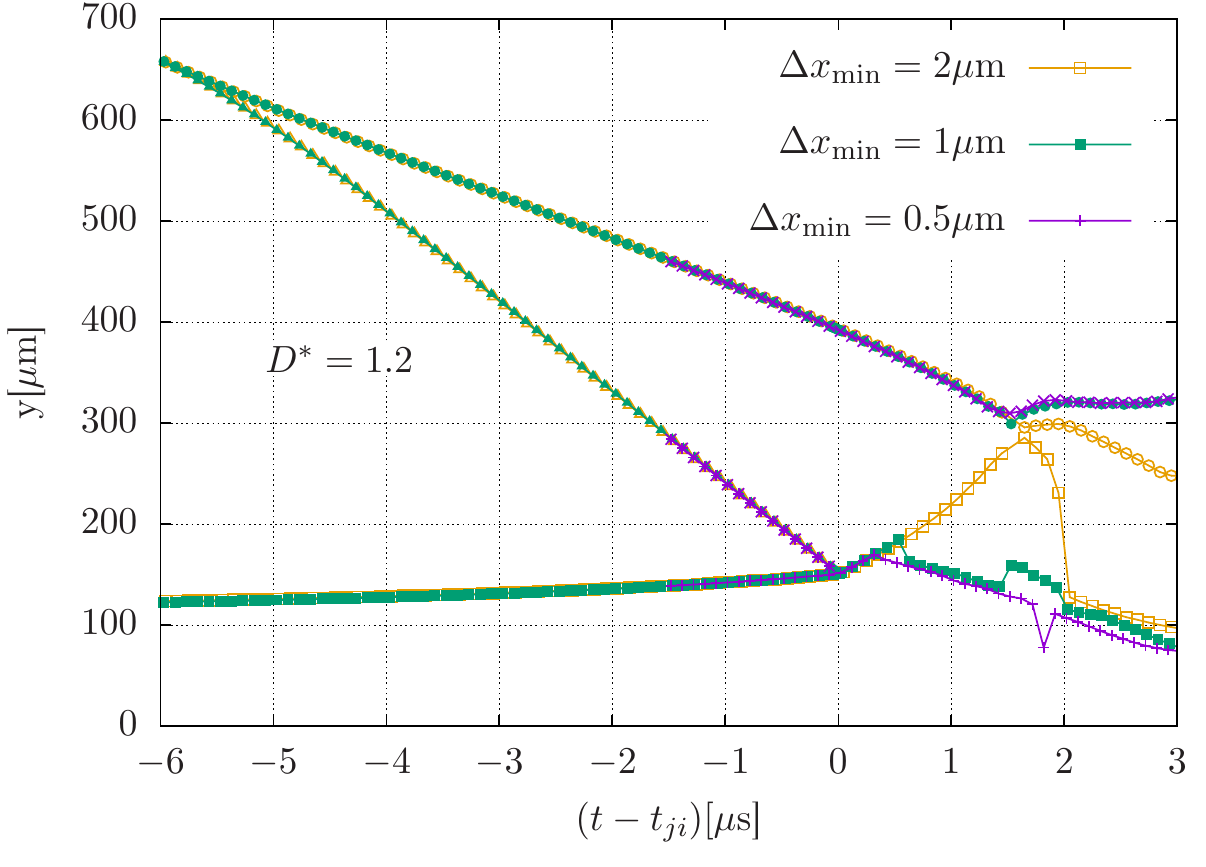}}
}
\end{arXiv}
\caption{(Colour online)
Simulations performed on a Cartesian grid for a bubble with
$D^*=1.2$, $\rmax = 500\,\mum, \sigma = 0$. Left: distance of the 
geometric bubble center, $D_{\rm c}$, from the solid boundary as a function of time.
Resolutions $\Delta x_{\min} = 2\,\mum$ 
($\cdot \!\! \cdot \!\! \cdot \!\! \cdot \!\! \cdot$)
and $\Delta x_{\min} = 1\,\mum$ (---). 
Right: bubble skeleton, i.e., the distance of the lowermost 
({\small $\square$}, {\small $\blacksquare$}, $+$)
and uppermost
(\protect\raisebox{-0.3ex}{{\Large $\circ$}},\protect\raisebox{-0.3ex}{{\Large $\bullet$}},$\times$)
points on the bubble wall as well as of the jet tip 
($\vartriangle$,$\blacktriangle$,$*$)
from
the solid boundary. The time interval shown is around the moment of 
jet impact onto the lower bubble wall,
$t_\tx{ji}$, 
and the subsequent 
first bubble collapse. 
Resolutions $\Delta x_{\min} = 2\,\mum$
(open symbols) and $\Delta x_{\min} = 1\,\mum$ (filled symbols) and 
$\Delta x_{\min} = 0.5\,\mum$ ($+,\times,*$). 
\label{fig:Cart_conv}
}
\end{figure}

Figure \ref{fig:Cart_conv} shows grid indepence on the Cartesian grid
(see figure\,\ref{fig:inited}, left) for a bubble with $D^* = 1.2$ using two
resolutions with $\Delta x_{\min} = 2\,\mum$ and $\Delta x_{\min} = 1\,\mum$. The left
diagram in figure\,\ref{fig:Cart_conv} shows the distance of the geometric
bubble center from the solid boundary as a function of time. The
translational motion of the bubble is captured equally well by both 
resolutions. 
A more differentiated view is given in
figure\,\ref{fig:Cart_conv}, right, which displays the ``bubble skeleton'',
i.e., the distances of the uppermost and lowermost points on the 
bubble wall and the jet tip 
from the solid boundary as functions of time. 
Excellent agreement is found for all three resolutions 
up to the moment of jet impact onto the lower bubble wall.  
Around the first bubble collapse about 1.5\,$\mus$ later
the higher resolutions ($\Delta x_{\min} = 1\,\mum$
and $\Delta x_{\min} = 0.5\,\mum$) 
show more structure. 
The variations in the curves
after jet impact visible in figure\,\ref{fig:Cart_conv}, right,
are due to the splitting of the bubble into several torus bubbles that collapse individually 
\citep*[see, e.g.,][]{Ohl-1995, Philipp-1998}.  
Although the details
around the first bubble collapse (i.e., after jet impact) are not subject of the present investigations, 
the simulations on the Cartesian grid are performed with the medium resolution, 
i.e., with a minimum grid spacing of $\Delta x_{\min} = 1\,\mum.$

\begin{figure}%[!h]  %fig24
\begin{journal}
\centerline{
\includegraphics[width=0.45\textwidth]{figure24a.eps}
\hspace{1ex}
\protect\raisebox{0ex}{\includegraphics[width=0.46\textwidth]{figure24b.eps}}
}
\end{journal}
\begin{arXiv}
\centerline{
\includegraphics[width=0.45\textwidth]{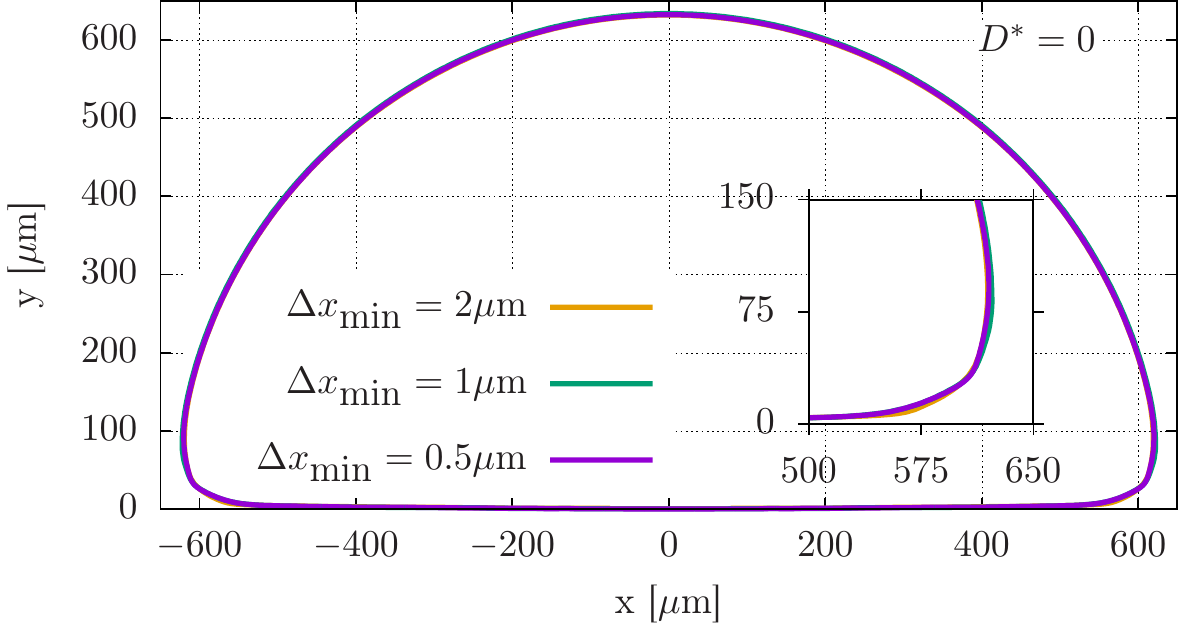}
\hspace{1ex}
\protect\raisebox{0ex}{\includegraphics[width=0.46\textwidth]{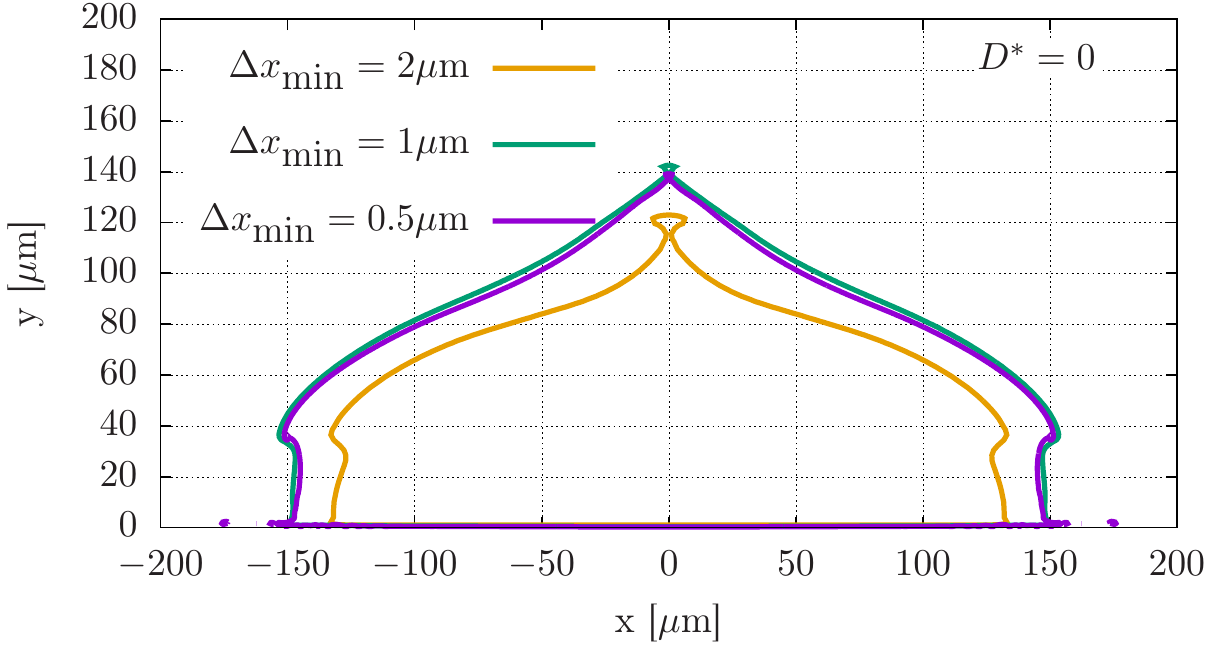}}
}
\end{arXiv}
\caption{(Colour online) Simulations performed on a polar grid with
Cartesian center region for a bubble with
$D^*=0$, $\rmax = 500\,\mum, \sigma = 0.0725$\,N\,m$^{-1}$ (water). 
The Cartesian center regions have a uniform grid spacing of 
$\Delta x_{\min} = 2\,\mum$ ($\cdot \!\!\, \cdot \!\!\, \cdot \!\!\, \cdot \!\!\, \cdot$), 
$\Delta x_{\min} = 1\,\mum$ (---) 
and $\Delta x_{\min} = 0.5\,\mum$ ($-\, \!\! -\, \!\! -$). Left: bubble shape at maximum
extension.
Right: bubble shape directly before the impact of the annular inflow at the
axis. 
\label{fig:polar_conv}
}
\end{figure}

For bubbles close to the wall, $D^* \le 0.4 $, simulations are performed on a
polar grid matched to a Cartesian region in the center (see
figure\,\ref{fig:inited}, right). Figure \ref{fig:polar_conv} shows cuts through
a bubble with $D^*=0$ from simulations with $\Delta x_{\min} = 2, 1$, and $0.5\,\mum$ 
The bubble shape at maximum extension shown in the left diagram of
figure\,\ref{fig:polar_conv}---in particular the location of the region of high curvature 
at the outer rim---is reproduced very well by all three resolutions. 
It follows that the boundary layer during expansion is sufficiently resolved
for our purpose. The right diagram in figure\,\ref{fig:polar_conv} shows the
bubble shape right before the moment of the formation of the fast jet. 
The two higher resolutions display a similar bubble shape and location of
impact of the annular inflow at the axis. 
In the study, a minimum grid spacing 
of $\Delta x_{\min} = 1\ \rm{or}\ 0.5\,\mum$ 
is taken for this grid configuration.

\section{Convergence studies on the speed of the fast jet}\label{app:jet}

While the overall bubble dynamics leading to the fast jet 
is sufficiently resolved with a resolution of $\Delta x_{\min} = 1\,\mum$, 
the self-impact of the annular inflow leading to the fast jet, in general  
is a nearly singular process and thus cannot be resolved properly. 
In axial symmetry, the annular inflow is forced to converge 
towards a single line, i.e., the axis. 
In the absence of any damping (empty bubble, incompressible and inviscid
liquid) this would lead to an unbounded impact velocity at the axis 
\citep{Voinov-1976}.
In the present case, the process is damped by the viscosity and mitigated by the 
compressibility of the liquid as well as the gas in the
bubble. 
Since the gas can easily evade into the main body of the bubble, its
cushioning effect is small. It is expected to depend on 
the shape of the closing neck and thus will vary with $D^*$ (see 
figures \ref{fig:wallvelo01}, \ref{fig:wallvelo005}, and \ref{fig:wallvelo0}).
For the case of $D^*=0$, local refinement has been pursued 
down to $\Delta x_{\min} = 7.8125$ nm without obtaining converged values of the
key quantities around the self-impact of the annular inflow.
In particular, the 
speed of the fast axial jet still increases with
increasing resolution.

Computer time limitations inhibited 
further grid refinement.
In figure\,\ref{fig:jetveloc_D}, the average velocity of the jet tip
obtained with a resolution of $\Delta x_{\min} = 0.5\,\mum$ is given,
because the jets found are 
already a factor of more than ten higher in velocity than anything known
before 
for bubble jets near solid boundaries.
Thus, the velocity values of the fast jet given in 
figure\,\ref{fig:jetveloc_D} are to be understood as lower bounds.  
The 
challenge to reach out for the limit in these velocities is left for 
further studies.

\section{Flow focusing and jet formation}\label{app:flow}

There exists a connection between jet formation and the curvature of the surface of aspherical bubbles. It is mediated by the concept of local flow focusing.

Flow focusing in its purest form appears irrespective of surface tension or gravity. It is always present when there are curved surfaces with a pressure difference across, as, e.g., with Rayleigh bubbles or overexpanded laser-induced bubbles, and is most pronounced in an initially zero flow field. With a flow present, there is an overlay of both, flow and flow focusing.

The basic example of flow focusing is given by the Rayleigh bubble 
\citep{Rayleigh-1917}.
An empty, spherical bubble of initial radius $R_{\rm max}$ in an incompressible, inviscid and unbounded liquid of density $\rho_l$ with positive pressure at infinity of $p_{\infty}$ and flow field zero will collapse by spherical flow focusing. The collapse ends in a singularity at radius $R=0$ of the bubble. The time between $R_{\rm max}$ and $R=0$ is called the Rayleigh-bubble collapse time $T_{\rm c} = 0.915\, R_{\rm max}\sqrt{\rho_l/p_{\infty}}$. There is no surface tension present to drive the flow, it is solely driven by the pressure difference between the empty interior of the bubble and the exterior liquid. The curvature of the bubble surface is constant at any time, but increasing with time. The collapse of a spherical Rayleigh bubble is the basic example of \textit{spherical flow focusing}.

\begin{figure}  %fig25
 \centering
\begin{journal}
\includegraphics[width=0.35\textwidth]{figure25a.eps}
\hspace{0.1\textwidth}
 \protect\raisebox{16ex}{\includegraphics[width=0.18\textwidth]{figure25b.eps} }
\end{journal}
\begin{arXiv}
\includegraphics[width=0.35\textwidth]{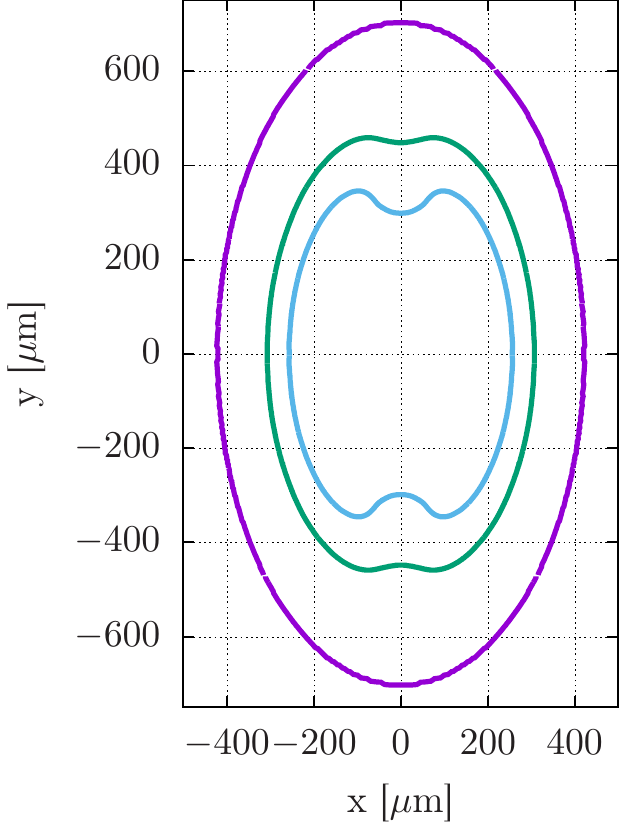}
\hspace{0.1\textwidth}
 \protect\raisebox{16ex}{\includegraphics[width=0.18\textwidth]{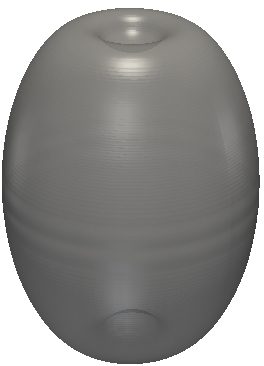} }
\end{arXiv}
\caption{(colour online).
Example of spherical flow focusing leading to jet formation. The higher curved parts develop the jets.
Prolate spheroidal Rayleigh bubble of eccentricity $e=0.8$, 
$R_{\max}^{\tx{equiv}} = 500\,\mum$, $\sigma = 0$. 
Left: Central cut through the bubble at  $t=0, 33, 38 \, \mus$.  
Right: 3D-rendering with a viewing angle from slightly above for $t=38\,\mus$. 
\label{fig:prolate_spheroid}
}
\end{figure}
Flow focusing is also present in bubbles with non-uniform curvature. For demonstration, a non-spherical Rayleigh bubble is taken, i.e., a bubble with all properties of a spherical Rayleigh bubble except for its shape. For definiteness, a prolate spheroid of eccentricity $e= 0.8$ is taken and its collapse calculated with OpenFOAM (the boundary integral method would also do). Figure \ref{fig:prolate_spheroid}, left, gives the result of the first steps of collapse. Figure \ref{fig:prolate_spheroid}, right, shows the 3D appearance of this type of jet formation by flow focusing. Axial jet formation is noticed from the involution of the most strongly curved parts of the bubble. 
This fact leads to the statement: More highly curved parts of a (Rayleigh or overexpanded) bubble collapse faster, i.e., attain a higher velocity faster than less curved parts. This is a general statement or principle of flow focusing, not a specialty of the example. It can be derived from the Rayleigh 
equation  $R \ddot R + \frac{3}{2} \dot R^2 = -p_\infty/\rho_l$. 
When starting with $\dot{R} = 0$, the equation reads $\ddot{R} = -p_\infty/(\rho_l R_{\rm max})$, and it is detected that a bubble with a smaller $R_{\rm max}$ 
(higher curvature) 
experiences a higher acceleration than a larger bubble with its weaker curvature and attains an initially higher velocity as stated. This situation prevails in the next step and so on until bubble collapse. It finds its analytical expression in the Rayleigh collapse time.
The Rayleigh collapse time scales with the maximum radius of a spherical bubble. On a fixed time scale, a bubble with an initially smaller maximum radius (higher curvature) collapses faster than a different bubble with an initially larger maximum radius (weaker curvature).
Applying this argument locally to the respectively curved parts of a non-spherical bubble surface 
yields the above statement.

It has been found that this type of jet soon reaches an almost constant velocity. 
The fact has been noted in the numerical study by Plesset \& Chapman already in 1971.
However, no explanation was given.
The almost constant velocity has been visualized in skeleton diagrams, e.g., experimentally in 
\cite{Lauterborn-1975}
and lately numerically in \cite{Lauterborn-2018b}.
A further example is given in Appendix \ref{appA} in connection with the convergence studies 
and can be seen in figure\,\ref{fig:Cart_conv}, right diagram. The almost straight line (middle curve) is the jet with almost constant velocity. 
An explanation may run as follows. 
The faster velocity of the highly curved parts of the bubble 
 leads to an involution of the respective surface part from concave to convex (seen from the liquid, 
 see figure\,\ref{fig:prolate_spheroid}). 
 Consequently, axial flow
 focusing is gone locally and therefore the acceleration by flow focusing stops in the jet funnel. Thus the jet flow can proceed at an almost constant velocity 
owing to the inertia of the liquid---almost, because of surface tension and viscosity. 
But both quantities are of minor influence as the inertial forces by far 
 outperform the surface tension forces and the viscous forces in the present case. 

The example presented is for spherical flow focusing and axial jet formation, applicable to the standard jet. 

\begin{figure}  %fig26
 \centering
\begin{journal}
   \includegraphics[width=0.45\textwidth]{figure26a.eps}
\hspace{7ex}
\protect\raisebox{10ex}{
\includegraphics[width=0.23\textwidth]{figure26b.eps}}
\end{journal}
\begin{arXiv}
   \includegraphics[width=0.45\textwidth]{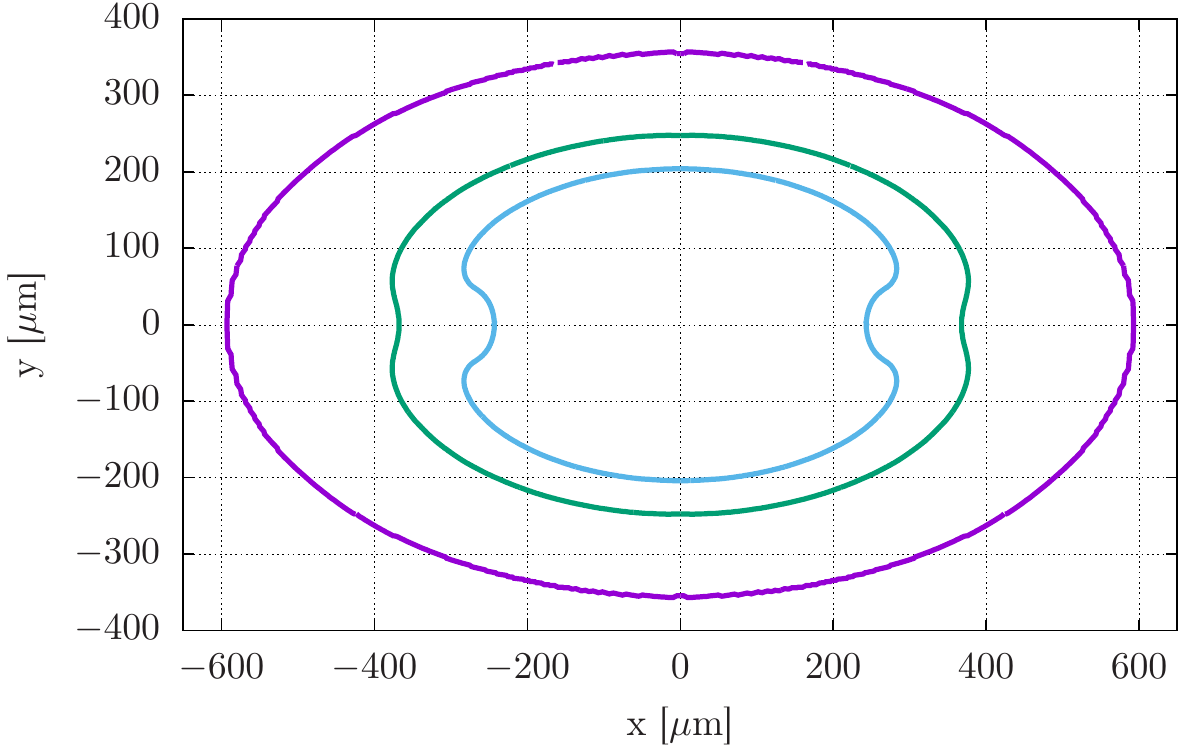}
\hspace{7ex}
\protect\raisebox{10ex}{
\includegraphics[width=0.23\textwidth]{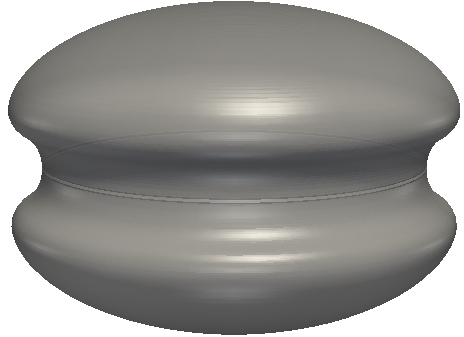}}
\end{arXiv}

\caption{(colour online). 
Example of annular flow focusing leading to jet formation. The higher curved parts develop the jet. 
Oblate spheroidal Rayleigh bubble of eccentricity $e=0.8$, 
$R_{\max}^{\tx{equiv}} = 500\,\mum$, $\sigma = 0$. 
Left: Central cut through the bubble at  $t=0, 35, 40 \, \mus$.  
Right: 3D-rendering with a viewing angle from slightly above for $t=40\,\mus$ to better appreciate the annular jet. 
\label{fig:oblate_spheroid}
}
\end{figure}

In the study, a second type of flow focusing appears that finally---after self-impact of the flow---leads to the thin, fast jet. An example of this type of flow focusing is given by an oblate spheroidal Rayleigh bubble (figure \ref{fig:oblate_spheroid}). The same eccentricity of $e=0.8$ is chosen and the same initial volume as in the prolate case. Again the more highly curved parts of the bubble collapse faster and develop a jet. The type of the jet, however, is different from the prolate case, where two oppositely directed axial jets develop from two separated more highly curved parts of the bubble. Instead, in the oblate case, one annular jet is formed (albeit in the central cut in figures \ref{fig:prolate_spheroid} and \ref{fig:oblate_spheroid} both jets look similar). The difference is better seen in the 3D-renderings. 
A further difference is that the annular jet does not diminish its acceleration as the axial jet does, because the annular flow focusing proceeds owing to the 
ongoing cylindrical shrinking of the flow in the 
equatorial plane of the initially oblate spheroidal bubble.

The inevitable self-impact of the annular jet at the axis of symmetry will give birth to two oppositely directed axial jets of the same speed along the axis of symmetry (not shown here).
A numerical example can be found in \cite{Blake-1997} and an experimental 
example in \citep[][figure 6]{Lauterborn-1976}. 
Recent examples with respect to solid boundaries can be found in 
\cite{Lechner-2019} and \cite{Pishchalnikov-2019}.

Annular flow focusing can be considered a necessary precursor to the fast, thin jet as found for bubbles expanding and collapsing very near to a solid boundary.

\bibliography{Manuscript}

\end{document}